\newif\ifemulate
\emulatetrue

\ifemulate
   \documentclass{emulateapj}
   \slugcomment{Accepted for publication in ApJ}
   \usepackage{graphicx}
   \usepackage{lscape}
\else
   \documentclass[12pt, preprint]{aastex}
\fi

\lefthead{Rudnick et al.}
\righthead{Red Sequence Cluster Galaxy LF at $z<0.8$}

\newcommand\lrest{${\rm L^{rest}_\lambda}$}

\newcommand\zp{$z_{phot}$}

\newcommand\ha{${\rm H\alpha}$}

\newcommand\lstar{${\rm L}^\star$}

\newcommand\phistar{${\rm \phi^\star}$}

\newcommand{\mllam}[1]{${\rm M}/{\rm L}_{#1}$}

\newcommand{\lmtwo}{${\rm j_{\rm crs}}/{\rm M_{200}}$}
\newcommand{\mtotl}{${\rm M_{\rm tot}}/{\rm L}$}
\newcommand{\mtwo}{${\rm M_{200}}$}

\newcommand{\mlstarlamav}[1]{$\langle {\cal M_{\star}}/{\rm L}_{V}\rangle$}

\newcommand\chisq{$\chi^2$}

\newcommand\msol{${\cal M_{\odot}}$}

\renewcommand\bv{$(B-V)$}
\newcommand\vi{$(V-I)$}

\newcommand{\jsdss}{$j_{crs,SDSS}$}

\newcommand\mitot{$I_{\rm tot}$}
\newcommand\miauto{$I_{AUTO}$}

\newcommand\mb{$B$}

\newcommand\mv{$V$}
\newcommand\mr{$R$}
\newcommand\mi{$I$}
\newcommand\mj{$J$}

\newcommand\mk{$K$}
\newcommand\mks{$K_s$}
\newcommand\gs{$g$}
\newcommand\rs{$r$}
\newcommand\is{$i$}
\newcommand\zs{$z$}

\newcommand{\mstar}{${M^{\star}}$}

\newcommand\pclust{$P_{clust}$}
\newcommand\pz{$P(z)$}
\newcommand\pthresh{$P_{thresh}$}
\newcommand\zclust{$z_{clust}$}
\newcommand\rtwo{$R_{200}$}
\newcommand{\magrest}[1]{$#1_{rest}$}
\newcommand{\absmag}[1]{$M_{#1}$}
\newcommand{\colrest}[2]{$(#1-#2)_{rest}$}
\newcommand{\lfzp}{LF$_{\rm zp}$}
\newcommand{\lfss}{LF$_{\rm ss}$}
\newcommand{\jcrs}{$j_{\rm crs}$}
\newcommand{\jbg}{$j_{bg}$}

\newcommand{\jpred}{$j_{crs,pred}$}
\newcommand{\nlf}{$N_{\rm lum}/N_{\rm faint}$}

\begin{document}
\title{The Rest-Frame Optical Luminosity Function of Cluster Galaxies at $z<0.8$ and the Assembly of the Cluster Red Sequence \altaffilmark{1}}

\author{Gregory Rudnick\altaffilmark{2,3}, 
Anja von der Linden\altaffilmark{4,5}, 
Roser Pell\'o\altaffilmark{6}, 
Alfonso Arag\'on-Salamanca\altaffilmark{7}, 
Danilo Marchesini\altaffilmark{8}, 
Douglas Clowe\altaffilmark{9}, 
Gabriella De Lucia\altaffilmark{4,10}, 
Claire Halliday\altaffilmark{11}, 
Pascale Jablonka\altaffilmark{12}, 
Bo Milvang-Jensen\altaffilmark{13,14}, 
Bianca Poggianti\altaffilmark{15},
Roberto Saglia\altaffilmark{16}, 
Luc Simard\altaffilmark{17}, 
Simon White\altaffilmark{4}, 
\& Dennis Zaritsky\altaffilmark{18}}

\altaffiltext{1}{Based on observations collected at the European Southern
    Observatory, Chile, as part of large programme 166.A-0162
    (the ESO Distant Cluster Survey).}

\altaffiltext{2}{NOAO, 950 N. Cherry Ave., Tucson, AZ 85719, USA}
\altaffiltext{3}{Currently at The University of Kansas, Department of Physics and Astronomy, Malott room 1082, 1251 Wescoe Hall Drive, Lawrence, KS, 66045, USA\texttt{grudnick@ku.edu}}
\altaffiltext{4}{Max-Planck-Institut f\"ur Astrophysik, Karl-Schwarzschild-Str. 1, D-85741, Garching, Germany}
\altaffiltext{5}{Currently at Kavli Institute for Particle Astrophysics and Cosmology,
 Stanford University, 452 Lomita Mall, Stanford, CA 94305-4085, USA}
\altaffiltext{6}{Laboratoire d'Astrophysique de Toulouse-Tarbes, CNRS, Universit\'e de Toulouse, 14 Avenue Edouard Belin, 31400-Toulouse, France}
\altaffiltext{7}{School of Physics and Astronomy, University of Nottingham, University Park, Nottingham NG7 2RD, United Kingdom}
\altaffiltext{8}{Astronomy Department, Yale University, P.O. Box 208101, New Haven, CT 06520-8101 USA }
\altaffiltext{9}{Department of Physics \& Astronomy, Clippinger Labs 251B, Athens, OH 45701 USA}
\altaffiltext{10}{Currently at INAF - Astronomical Observatory of Trieste, via Tiepolo 11, I-34143 Trieste, Italy}
\altaffiltext{11}{Osservatorio Astrofisico di Arcetri, Largo E.Fermi, 5. 50125 Florence, Italy}
\altaffiltext{12}{Observatoire de Gen\`eve, Laboratoire d'Astrophysique Ecole Polytechnique Federale de Lausanne (EPFL), CH-1290 Sauverny, Switzerland}
\altaffiltext{13}{Dark Cosmology Centre, Niels Bohr Institute, University of Copenhagen, Juliane Maries Vej 30, 2100 Copenhagen \O, Denmark }
\altaffiltext{14}{The Royal Library / Copenhagen University Library, Research Dept., Box 2149, 1016 Copenhagen K, Denmark }
\altaffiltext{15}{Osservatorio Astronomico di Padova, Vicolo dell'Osservatorio 5, 35122 Padova, Italy}
\altaffiltext{16}{Max-Planck Institut fur extraterrestrische Physik, Giessenbachstrasse, D-85748, Garching, Germany}
\altaffiltext{17}{Herzberg Institute of Astrophysics, National Research Council of Canada, Victoria, BC V9E 2E7, Canada}
\altaffiltext{18}{Steward Observatory, University of Arizona, 933 North Cherry Avenue, Tucson, AZ 85721 USA}

\begin{abstract}
We present the rest-frame optical luminosity function (LF) of red
sequence galaxies in 16 clusters at $0.4<z<0.8$ drawn from the ESO
Distant Cluster Survey (EDisCS).  We compare our clusters to an
analogous sample from the Sloan Digital Sky Survey (SDSS) and match
the EDisCS clusters to their most likely descendants.  We measure all
LFs down to $M\sim M^\star + (2.5 - 3.5)$.  At $z<0.8$, the bright end
of the LF is consistent with passive evolution but there is a
significant build-up of the faint end of the red sequence towards
lower redshift.  There is a weak dependence of the LF on cluster
velocity dispersion for EDisCS but no such dependence for the SDSS
clusters.  We find tentative evidence that red sequence galaxies
brighter than a threshold magnitude are already in place, and that
this threshold evolves to fainter magnitudes toward lower redshifts.
We compare the EDisCS LFs with the LF of co-eval red sequence galaxies
in the field and find that the bright end of the LFs agree.  However,
relative to the number of bright red galaxies, the field has more
faint red galaxies than clusters at $0.6<z<0.8$ but fewer at
$0.4<z<0.6$, implying differential evolution.  We compare the total
light in the EDisCS cluster red sequences to the total red sequence
light in our SDSS cluster sample.  Clusters at $0.4<z<0.8$ must
increase their luminosity on the red sequence (and therefore stellar
mass in red galaxies) by a factor of $1-3$ by $z=0$.  The necessary
processes that add mass to the red sequence in clusters predict local
clusters that are over-luminous as compared to those observed in the
SDSS.  The predicted cluster luminosities can be reconciled with
observed local cluster luminosities by combining multiple previously
known effects.

\end{abstract}
\section{Introduction}
\label{Intro}

 Most of the stellar mass in the local universe is contained in "red
 and dead" galaxies, i.e. galaxies which have stopped forming stars at
 an appreciably level and whose light is thus dominated by old, red
 stars\citep{Hogg02}.  To understand how stars form and galaxies are
 assembled, we therefore need to determine how the red galaxy
 population evolves through time.  Red galaxies are located on a tight
 sequence in color and magnitude, the ``red sequence'' (e.g. de
 Vaucouleurs 1961; Visvanathan \& Sandage 1977), and the very small
 intrinsic scatter in color implies that the red colors result from
 uniformly old stellar ages (e.g. Bower, Kodama \& Terlevich 1998).
 Old ages for red sequence galaxies are also found by studies of their
 stellar indices (e.g. Trager et al. 1998).  Some studies even find a
 stellar mass dependence in the mean stellar age, such that lower mass
 galaxies formed their stars at later epochs than those that are more
 massive (e.g. Thomas et al. 2005), but this result is still
 controversial as \citet{Trager08} find no such trend in their studies
 of Coma cluster early types.

 At face value direct lookback observations may support these local
 archaeological studies as the total stellar mass on the red sequence
 may have doubled since $z\sim 1$ \citep{Bell04,Faber07,Brown07}.
 \citet{Cimatti06} and \citet{Brown07} concluded that this mass growth
 comes primarily from the addition of low mass galaxies to the red
 sequence at late times, with the most luminous red sequence galaxies
 ($L>4$\lstar) appearing to have been in place since $z>1$.  As
 \citet{Trager08} point out, however, it may be hard to relate the
 direct lookback results to studies of local galaxies, as the latter
 may be susceptible to very small amounts (a few percent) of late star
 formation.

 It is impossible to study the evolution of red galaxies without
 examining the influence of environment.  Going all the way back to
 \citet{Hubble31} it has been known that there are significant
 correlations between color and environment, star formation rate (SFR)
 and star formation history (SFH) and environment, and morphology and
 environment (e.g. Dressler 1980), such that dense environments,
 e.g. the centers of galaxy clusters, have much higher fractions of
 red sequence galaxies than the field.  Local studies suggest that
 luminous ellipticals in galaxy clusters have older stellar ages than
 those in the field \citep{thomas05} but studies at high redshift
 detected no difference in the ages of field and cluster elliptical
 galaxies \citep{vanderwel05, dokkum07}.  Nonetheless, the large
 differences between clusters and the field even at intermediate
 redshift, which are measured in terms of the morphological fraction
 (e.g. Postman et al. 2005; Smith et al. 2005; Desai et al. 2007) and
 the fraction of star forming galaxies (e.g. Poggianti et al. 2006)
 implies that a galaxy's evolutionary path might be strongly affected
 by the environment in which it lives as it evolves through cosmic
 time.  \citet{Poggianti06} postulate that massive ellipticals in
 clusters may have been formed at high redshift but that lower
 luminosity red galaxies are added to the cluster at $z<1$.

 From a theoretical standpoint, some models (e.g. De Lucia et
 al. 2006) predict that stars in red galaxies were formed at high
 redshift and that the formation epoch of the stars is earlier for
 higher mass galaxies.  It is nonetheless not clear if these models
 can be reconciled in detail with the observed evolution in the
 increase of mass on the red sequence at $z<1$.  Also not clear is if
 the properties of galaxies as a function of environment are being
 properly treated in some models as none of the commonly implemented
 processes, e.g. ram-pressure stripping, harassment, strangulation,
 can reproduce the observed dependence of the red and blue galaxy
 fraction on e.g. halo mass and central halo galaxy type at low redshift
 (e.g. Weinmann et al. 2006) or at $z\sim1$ \citep{Coil08}.

 One way to study the evolving galaxy population is to use the
 luminosity function (LF; see Binggeli, Sandage, \& Tammann 1988 for a
 review), which describes the number of galaxies per unit luminosity.
 The LF encodes information about the efficiency of star formation and
 feedback in galaxies and how galaxies populate their parent dark
 matter halos.

 Enabled by large surveys at low redshift such as 2dF \citep{Folkes99}
 and the Sloan Digital Sky Survey \citep{York00} it is now possible to
 construct the detailed LF of low redshift galaxies in a range of
 environments.  For example, using the 2dFGRS, \citet{Propris03}
 measured the composite LF in a set of local galaxy clusters and found
 that clusters have a brighter characteristic luminosity and steeper
 faint-end slope than the field, with the largest difference being
 found for spectroscopically identified non-starforming galaxies.  The
 availability of these well-characterized local LF determinations
 provide well established reference points against which to measure
 evolution in the cluster galaxy population.  Simultaneously, the
 recent availability of deep multi-color photometry of intermediate
 and high redshift clusters with extensive spectroscopic follow-up
 have allowed the galaxy population to be studied out to $z\sim 1$ in
 the Universe's densest regions.

De Lucia et al. (2004; hereafter DL04) were the first to measure the
evolution of the red sequence LF in clusters at high redshift by
studying the ratio of luminous-to-faint red sequence galaxies \nlf\ in
four clusters at $z\sim 0.7$ drawn from the ESO Distant Cluster Survey
(EDisCS).  They found that this ratio was significantly higher in the
high redshift clusters than in the Coma cluster.  Subsequently, this
redshift trend in \nlf\ was confirmed by \citet{Goto05} and
\citet{Tanaka05} in a few clusters, and by De Lucia et al. (2007;
hereafter DL07), \citet{Stott07}, and \citet{Gilbank08} in
significantly larger samples.  \citet{Tanaka05}, DL07, and
\citet{Gilbank08} also found that the evolution of \nlf\ depends
weakly on cluster velocity dispersion and DL07 and \citet{Gilbank08}
find that poorer systems evolve marginally slower than richer systems
at $0.4<z<0.8$.  The behavior in \citet{Tanaka05} is based on only one
cluster and is harder to generalize.  Tracing the evolution to $z=0$,
however, there is some disagreement between DL07 and
\citet{Gilbank08}.  In DL07 it appears that the low-dispersion systems
have converged to the \nlf\ value of the Coma cluster while the
high-dispersion systems require significant evolution to reach the
value from SDSS or Coma.  On the other hand, the poor systems of
\citet{Gilbank08} have systematically higher \nlf\ values than rich
systems at $0.4<z<0.6$ and therefore need to evolve more at $z<0.4$ to
come into agreement with the local value.  The origin of this apparent
discrepancy is hard to track down since DL07 and \citet{Gilbank08} use
different effective velocity dispersion cuts and different magnitude
limits defining the split between faint and luminous galaxies.  At the
same time \citep{Andreon06,Andreon08} claim little trend
in \nlf\ with redshift and no trend with velocity dispersion.  In
their Figure 4, however, the amount of redshift evolution appears
similar to that from DL07.  It is also not easy to compare the trends
with velocity dispersion between the two works since the
\citet{Andreon08} sample contains no clusters below 600~km/s, which
comprises a large fraction of the DL07 and \citet{Gilbank08} samples.

This paper makes a series of advances over previous works by computing
the full red-sequence LFs from EDisCS and comparing them to the local
red sequence cluster LF as determined from the SDSS.  The EDisCS
sample is the largest sample that probes well past $z=0.5$, all the
way out to $z=0.8$, has deep multi-band photometry that enables the
construction of rest-frame optical luminosity functions, and has a
large range in cluster velocity dispersion.  In this paper we extend
the work of DL07 significantly by measuring the non-parametric LF,
fitting Schechter functions, and measuring the detailed evolution of
red sequence galaxies.  In doing so we pay specific attention to the
ability to determine membership from galaxies with only photometry.
Our large range in velocity dispersion permits us to study how
evolution in the LF depends on velocity dispersion and our deep
photometry make us complete well below \mstar.  We also make the first
comparison of the composite cluster red sequence LF to that in the
field and measure their comparative evolution.  This test is crucial
as it spans the full range of galaxy environment and speaks directly
as to whether the cluster and field red galaxy populations are built
up at different rates.  Finally we measure the evolution of the total
light on the red sequence in clusters and discuss its implications for
how mass is added to the cluster red sequence over time.  We do not
address in detail the total LF or that of blue galaxies as we show in
\S\ref{methodcomp_sec} that LFs from photometric data can only be
robustly computed for red galaxies.

 In this paper we examine the rest-frame optical LF of the red
 galaxies in EDisCS clusters.  The rest-frame near infrared LF and
 stellar mass function will be presented in Arag\'on-Salamanca et
 al. (in preparation).  In \S\ref{datasec} we discuss the survey
 strategy and describe the data.  In \S\ref{membsel} we discuss our
 techniques for determining cluster membership.
 In \S\ref{lf_meas_sec} we describe our estimation of rest-frame
 luminosities and present our construction of the rest-frame optical
 LF.  We present our results in \S5, discuss them in \S6, and
 summarize and conclude in \S7.  Throughout we assume ``concordance''
 $\Lambda$-dominated cosmology with
 $\Omega_\mathrm{M}=0.3,~\Omega_{\Lambda}=0.7,~\mathrm{and~H_o}=70~{\rm
 h_{70}~km~s^{-1}~Mpc^{-1}}$ unless explicitly stated otherwise.  All
 magnitudes are quoted in the AB system.

\section{Observations and data}
\label{datasec}

\subsection{Observations and Survey Description}

 The survey strategy and description are presented in detail in White
 et al. (2005; hereafter W05) who also present the optical photometry
 and the construction of photometric catalogs.  The near Infrared
 (NIR) photometry will be presented in (Arag\'on-Salamanca et al. in
 preparation).  The spectroscopic data are presented in
 \citet{Halliday04} for the first five clusters with full spectroscopy
 and in \citet{Jensen08} for the full EDisCS
 sample.  The survey description and data will be summarized briefly
 below.

 The original goal of EDisCS was to study in detail a set of 10
 clusters at $z\sim 0.5$ and 10 at $z\sim 0.8$.  Our survey draws on
 the optically selected sample of clusters from the LCDCS
 \citep{Gonzalez01}.  After confirming the presence of a galaxy
 surface overdensity at the expected position and the presence of a
 red sequence using short images with the FORS2 instrument on the VLT,
 we initiated deep imaging of 10 clusters in each redshift bin.  We
 observed every field in either the \mb, \mv, \mi, and \mks-bands or
 in the \mv, \mr, \mi, \mj, and \mks-bands depending on whether the
 LCDCS redshift estimate of the cluster was at 0.5 or 0.8
 respectively.  The optical data were all obtained with FORS2/VLT and
 the NIR data were obtained with the SOFI instrument on the NTT.

 From the first reduction of our imaging data we computed photometric
 redshifts to get a more precise redshift estimate for the clusters
 \citep{Pello09}.  These redshifts were used to target objects for
 spectroscopic observations with FORS2/VLT.  Now complete, our
 extensive spectroscopic observations consist of high signal-to-noise
 (S/N) data for $\sim 30-50$ members per cluster and a comparable
 number of field galaxies in each field down to $I\sim22$.  As
 explained in W05, deep spectroscopy was not obtained for two of the
 EDisCS fields (CL1122.9--1136 and CL1238.5--1144), the former of
 which showed no evidence for a cluster in the initial, shallow
 spectroscopic observations.  These clusters will not be used in this
 study, leaving 18 of which one (CL1119.3--1129) does not have any NIR
 data.

\subsection{Catalog Construction and Total Flux Measurements}
\label{catsec}

 We measured two types of magnitudes for our galaxies, matched
 aperture magnitudes and SExtractor AUTO magnitudes.  The former are
 used for measuring colors and the spectral energy distributions
 (SEDs) used to fit the photometric redshifts.  The latter are used to
 estimate the total magnitude of the galaxies in question.  We
 describe each in turn.  All magnitudes have been corrected for
 galactic extinction from \citet{Schlegel98}.

 Before the measurement of matched aperture fluxes, all images with
 seeing better than FWHM=$0\farcs8$ were convolved to FWHM=$0\farcs8$.
 The seeing across all bands ranged from $0\farcs6$ to $1\farcs0$ with
 most observations having FWHM$\leq0\farcs8$.

 Flux catalogs were created using the SExtractor software
 \citep{Bertin96} in the two image mode, detecting in the unconvolved
 (i.e. natural seeing) \mi-band image and measuring fluxes in matching
 apertures in all other bands.  Colors were measured with the same
 aperture in all bands, using either isophotal apertures defined from
 the detection images for those galaxies that were not crowded or
 using circular apertures with $r=1\farcs0$ for those galaxies that
 were crowded.  With this dual choice of matched apertures we obtained
 a high S/N measurement of the color while minimizing the biases due
 to neighboring objects.

\begin{figure}
\epsscale{1.30}
\plotone{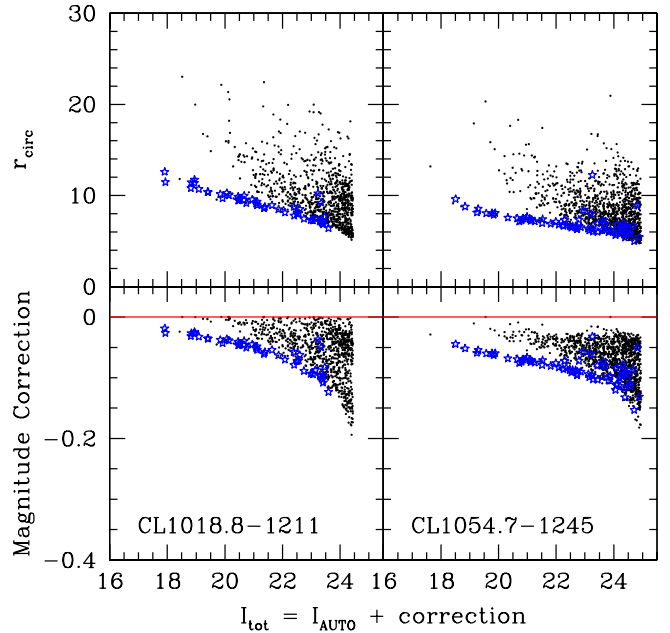}
\caption {
 An illustration of how our aperture correction depends on apparent
 magnitude for two clusters with the worst and best \mi-band image
 quality in our sample, CL1018.8-1211 and CL1054.7-1245 respectively.
 The x-axis in all plot is our ``total'' estimate of the \mi-band
 magnitude, \mitot, which is the AUTO \mi-band magnitude with a point
 source aperture correction.  The bottom row of panels shows how the
 correction depends on \mitot\ and the top panel shows how the
 circularized AUTO aperture radius depends on \mitot.  Only objects
 with no evidence of crowding have been used.  Stars are indicated by
 blue stars.  At every magnitude, the objects with the smallest
 apertures receive the largest correction.  The smallest apertures
 correspond to those for stars. }
\label{magcorr_emp_fig}
\end{figure}

 Obtaining accurate total magnitudes is important when characterizing
 the LF.  A true total magnitude estimate is not possible, however,
 due to uncertainties in the galaxy profile at large radii coupled
 with an uncertain knowledge of the sky level.  As described in W05,
 therefore, we attempted to measure pseudo-total magnitudes (called
 ``total'' magnitudes throughout) in the \mi-band using the AUTO
 magnitude from SExtractor.  These magnitudes were measured on the
 images at their natural seeing.  The SExtractor AUTO measurement is
 executed within a Kron like aperture (Kron 1980) and measures the
 flux within a radius corresponding to 2 times the first moment of the
 light distribution.  The AUTO magnitudes for each object have a
 minimum aperture radius of 3.5 pixels (or $0\farcs7$).  The AUTO
 aperture is quite large for bright objects but for faint objects the
 AUTO aperture shrinks its size to the minimum allowable limit.  In
 this regime, light will be lost out of the aperture even for point
 sources, since the stellar PSF throws significant amounts of light
 beyond this minimum aperture.  Such an effect was also noted in the
 absolute magnitude estimates of \citet{Labbe03} and we adopt their
 approach for correcting for this effect, which we summarize here.
 Correcting for this offset explicitly is difficult because we don't
 know the intrinsic profile of the galaxies whose photometry we wish
 to measure.  However, a conservative and necessary correction can be
 made by accounting for the light that would be missed assuming that
 the object is a point source.  While the amount of light lost may be
 larger for extended objects, this robust correction must be made
 regardless of the intrinsic object shape.  Since we only define the
 total magnitude consistently from the \mi-band image, and use this to
 scale our rest-frame luminosities (as measured in the matched
 apertures) to total luminosities, we only calculated the aperture
 correction for the \mi-band image.  This neglects the effects of
 large color gradients, but the resultant error in the total
 magnitudes should not dominate our uncertainties.  We determined an
 empirical stellar curve of growth for each image using a set of
 bright, unsaturated, and isolated stars.  Using the curve of growth,
 we computed the correction as a function of AUTO aperture area and
 apply it to the AUTO magnitudes.  The corrected magnitudes become our
 ``total'' magnitudes, \mitot.  For the two clusters with the worst
 and best seeing in the \mi-band we plot the dependence of these
 corrections and the AUTO aperture size on the \mitot\ in
 Figure~\ref{magcorr_emp_fig}.  The corrections range from median
 values of $\sim 0.04$ magnitudes at $20.4<I_{\rm tot}<22.4$ to $\sim
 0.09$ magnitudes at $24.4<I_{\rm tot}<24.9$.

 To check how well this aperture correction does in retrieving the
 true total magnitude, we compared the \mitot\ values to those derived
 from 2D profile fits to the \mi-band data using the GIM2D software
 (Simard et al. 2002; Simard et al. in prep).  We fit bulge$+$disk
 models to the galaxies and extrapolated the profiles to get total
 magnitude estimates $I_{\rm GIM2D}$.  For sources with no nearby
 neighbors, $I_{\rm GIM2D}$ should be relatively free of bias
 \citep{Haeussler07}.  At $20.4<I_{\rm tot}<22.4$ and $22.4<I_{\rm
   tot}<24.4$ we find a median difference $I_{\rm tot} - I_{\rm
   GIM2D}$ of 0.02--0.04 and 0.06--0.1 respectively, such that $I_{\rm
   GIM2D}$ is systematically brighter.  However, in \citet{Simard02}
 those authors used extensive simulations to show that the GIM2D
 magnitudes are biased brighter by the same order as our measured
 difference between \mitot\ and $I_{\rm GIM2D}$, implying that our
 \mitot\ magnitudes indeed are good approximations to the true total
 magnitude.

 We have verified that our results do not depend sensitively on the
 value of the correction, as it only significantly effects the very
 faintest galaxies in the sample, which do not dominate any of the
 observed effects.

\section{Determining Cluster Membership}
\label{membsel}

 At intermediate redshift, the contrast of a cluster against the
 background and foreground is very low and an estimation of the
 cluster galaxy LF necessitates that a sample of cluster members be
 assembled which has been cleaned of foreground and background
 interlopers.  Spectroscopy is obviously the most accurate method of
 accomplishing this and spectroscopic redshifts can be determined for
 single objects down to $I\sim24.5$ with the use of 8-meter class
 telescopes.  Nonetheless, determining redshifts for large numbers of
 cluster members in multiple clusters, even with large time
 allocations on 8-meter class telescopes, is limited to relatively
 bright magnitudes, e.g. $I\lesssim22-23$ \citep{Tran07,Halliday04}.
 To determine cluster membership for magnitude selected samples down
 to $I\sim25$ it is therefore necessary to use alternate techniques.
 We have employed two methods to accomplish this, one based on
 photometric redshifts \zp\ and one based on statistical background
 subtraction.  These membership techniques have also been used in
 previous works on the EDisCS clusters, e.g. DL07, DL07.  LFs computed
 from photometric redshifts and statistical background subtraction
 will hereafter be referred to as \lfzp\ and \lfss\ respectively.  We
 discuss these two methods in this section, while in
 \S\ref{methodcomp_sec} we compare \lfzp\ and \lfss\ to determine the
 robustness of our results.

\subsection{Photometric Redshifts}
\label{zp_membsel}

 In general, photometric redshift techniques estimate the redshift of
 a galaxy by modeling the broad-band SED with a set of template
 spectra (e.g. Fern\'andez-Soto, Lanzetta, and Yahil 1999; Rudnick et
 al. 2001).  The resulting \chisq\ of the template fit as a function
 of redshift gives an estimate of the redshift probability
 distribution \pz\ and hence the most likely redshift.  As an example
 application of photometric redshift techniques to cluster studies,
 \citet{Toft04} used their \zp\ estimates to determine membership by
 taking a very wide $\Delta z=\pm0.3$ slice in redshift and selected
 every galaxy within this slice as being a cluster member.  A slice
 this wide, however, is $\sim 100$ times larger in velocity than the
 expected velocity width of the cluster, implying a large
 contamination from field galaxies.  Also, the performance of
 photometric redshifts is expected to depend on galaxy SED shape,
 e.g. blue star-forming galaxies have weak Balmer/4000\AA\ breaks
 which result in weaker photometric redshift constraints and possible
 larger systematic errors.  This color dependence on the \zp\ accuracy
 can only be quantified by using a large number of spectroscopic
 redshifts that span a large range of SED shape/color in the desired
 redshift range, preferably with identical photometry.  Until now,
 such large spectroscopic samples in intermediate redshift cluster
 fields have not been available.

 We explore an alternative photometric-redshift-based interloper
 subtraction technique with EDisCS, which tries to mitigate the
 disadvantages mentioned above.  The photometric redshifts for the
 EDisCS sample, their performance, and their use to isolate cluster
 members, is described in detail in \citet{Pello09}.  Here we provide
 a brief summary.

 Photometric redshifts were computed for every object in the EDisCS
 fields using two independent codes, a modified version of the
 publicly available Hyperz code \citep{Bolzonella00} and the code of
 \citet{Rud01} with the modifications presented in \citet{Rudnick03}.
 The accuracy of both methods is $\sigma(\delta z)\approx 0.05-0.06$,
 where $\delta z= \frac{z_{spec}-z_{phot}}{1+z_{spec}}$.  By fitting
 stellar templates to the observed SEDs of stars we searched for
 zeropoint offsets and found no offsets except for a small one in the
 $B$-band of CL1353.0-1137.  We applied the offset for this one band
 when performing the photometric redshift fits.  We established
 membership using a modified version of the technique first developed
 in \citet{Brunner00}, in which \pz\ is integrated in a slice around
 the cluster redshift for the two codes.  The width of the slice
 around which \pz\ is integrated should be on the order of the
 uncertainty in redshift for the galaxies in question. In our case we
 use a $\Delta z=\pm0.1$ slice around the spectroscopic redshift of
 the cluster \zclust.  We reject a galaxy from our membership list if
 $P_{clust}<P_{thresh}$ for either code.  We calibrate \pthresh\ from
 our $\sim 1900$ spectroscopic redshifts.  Our values of
 \pthresh\ were chosen to maximize the efficiency with which we can
 reject spectroscopic non-members (down to $I=22$) while retaining at
 least $\approx90\%$ of the confirmed cluster members, independent of
 their rest-frame \bv\ color or observed \vi\ color.  In practice we
 were able to choose thresholds such that we satisfied this criterion
 while rejecting 45--70\% of spectroscopically confirmed non-members.
 Applied to the entire magnitude limited sample, our thresholds reject
 $75-93\%$ of all galaxies with $I_{\rm tot}<24.9$.  It is worth
 noting that it is very difficult to assess our absolute contamination
 for two reasons.  First, even the extensive spectroscopy we currently
 have was performed on a subsample of the photometric catalog that was
 designed to exclude objects with an extremely low probability of
 being at the cluster redshift.  Any estimates based on this
 spectroscopy may therefore not be entirely indicative of the true
 contamination down to the spectroscopic completeness limit.  Second,
 we do not have spectroscopy for galaxies down to the faint limit of
 the photometric catalog and it becomes impossible to definitively
 measure the contamination at these faint magnitudes without
 significantly deeper spectroscopy or highly model dependent
 assumptions.

 Our method establishes cluster membership using a redshift interval
 smaller than that employed in other photometric-based membership
 techniques (e.g. Toft et al. 2004) and therefore should suffer
 considerably lower field contamination.  As a check of how much more
 contamination we would have if we adopted the technique of
 \citet{Toft04} we have re-measured our membership requiring that each
 galaxy be within $\Delta z=\pm0.3$ of the cluster redshift.  The
 number of cluster members with this technique is typically 2-3 times
 larger than when using our membership technique, implying a
 correspondingly larger contamination.

 Despite the apparently good performance of the photometric redshift
 technique, the \zp\ estimates are only well tested at relatively
 bright magnitudes, e.g. $I\lesssim 22$.  Because the \zp-based
 membership technique is largely untested at $I\gtrsim 22$ it will be
 difficult to trust the faint end slope of the LF derived from such
 techniques.  For this reason it is desirable to use complementary
 photometric methods to establish membership.

\subsection{Statistical Background Subtraction}
\label{statsub_membsel}
 
 An independent method of establishing cluster membership is the
 statistical subtraction technique (e.g. Arag\'on-Salamanca et
 al. 1993; Stanford et al. 1998).  In this technique, number counts in
 the cluster field are compared to those in an ``empty'' field and the
 excess counts are used to assign a membership probability to each
 galaxy in the cluster field.  This method becomes increasingly
 inefficient at high redshift, where the contrast of the cluster
 against the background becomes increasingly low.  In addition, this
 method provides no membership probability for individual galaxies,
 but rather gives every galaxy in a given region of magnitude (and
 color) space an identical probability.  At the same time, it suffers
 from completely different uncertainties than the photometric redshift
 technique and is a useful complement to judge the robustness of our
 results.

 Ideally, the comparison catalog used to create the field counts
 should contain the same bands as used in the cluster fields and cover
 a large enough area to minimize cosmic variance.  For our
 statistical-background-subtraction-based membership we utilized a
 ``field'' catalog from the Canada France Deep Field (CFDF; McCracken
 et al. 2001)\footnote{This catalog has been kindly provided to use by
   H. McCracken}. This field has the advantages of having matched
 aperture photometry in \mv\ and \mi-bands and AUTO magnitudes in the
 \mi-filter, while also covering 0.25 square degrees, roughly 20 times
 the area of the optical coverage in an individual EDisCS field.  The
 depth of the CFDF is only $I=24.5$ and so all LFs computed via
 statistical background subtraction will be limited to $I<24.5$.  The
 CFDF is the only publicly available field that satisfies our
 requirements for a background field.  These were: 1) that it must
 have photometry in at least V and I since these filters are in common
 for both of the EDisCS filter sets (BVIK and VRIJK) and 2) that it
 must have a large enough area to overcome the effects of cosmic
 variance in the background subtraction estimate.  While there are
 other fields with deep multi-field photometry over a moderate area
 (e.g. Chandra Deep Field South, NOAO Deep Wide Field Survey), there
 are no publicly available surveys with both deep V and I at a depth
 comparable to EDisCS and with large enough area to overcome cosmic
 variance.  For example, The Chandra Deep Field South (CDF-S) that was
 targeted by the FIREWORKS survey (Wuyts et al. 2008) is known to be
 underdense at $z\sim0.7$ compared to the much larger Extended Chandra Deep
 Field South (ECDF-S; Taylor et al. 2009) and so is not a good sample
 of the mean background.  Also, the NOAO deep wide field survey (Brown
 et al. 2007), which we use in \S\ref{field_sec} has no $V$ filter and
 a very wide $B$-band filter (essentially $U+B$), which makes it
 impossible to use as a background field for the EDisCS clusters with
 only BVIK photometry.

 We use a method similar to the one presented by \citet{Pimbblet02}
 and refer to that paper for details, although we summarize it briefly
 here.  We bin the CFDF data and our own in observed \vi\ color
 and \miauto\ using bins of 0.5 in color and magnitude (using color
 bins of 0.3 results in nearly identical LFs).  Note that we do not
 use \mitot\ when performing the statistical subtraction, as the CFDF
 does not have aperture corrected magnitudes.  We assume that the AUTO
 magnitudes perform similarly for both surveys.  In a given bin we
 scale the number of field galaxies to the area of the cluster under
 consideration to derive the number of expected field galaxies.  We
 first retain all spectroscopically confirmed members and exclude all
 spectroscopically confirmed non-members.  Then we subtract off a
 random subset of the remaining galaxies equal in number to the
 expected number of field galaxies (minus the number of
 spectroscopically confirmed non-members) to obtain a realization of
 the cluster member population.  In bins where the number of expected
 field galaxies are greater than the number of member candidates, we
 merge adjacent bins in color until the number of expected field
 galaxies is greater than or equal to the number of member candidates
 in the expanded bin.  This is analogous to expanding the bins until
 the membership probabilities again lie between 0 and 1.  As explained
 in Appendix A of \citet{Pimbblet02} this method has an advantage over
 similar methods in that it preserves the original probability
 distribution, albeit smoothed over larger scales.

 The moderately large area of the CFDF gives an accurate
 representation of the mean density of field galaxies but on spatial
 scales similar to that of our clusters the number counts of field
 galaxies may vary and the true underlying field may be systematically
 different from the mean.  We use the entire CFDF area to calculate
 our best estimate of the membership sample for each cluster.  When
 calculating the uncertainty in the cluster membership, we split the
 CFDF into tiles, with each tile having the same area as the area of
 the cluster under consideration.  In practice this resulted in
 greater than 20 independent tiles in the CFDF.  We then performed 100
 Monte Carlo iterations of the subtraction, where each iteration uses
 a randomly chosen tile to derive the expected field population.

\section{Measuring the Luminosity Function}
\label{lf_meas_sec}

 In this section we will present our method for determining rest-frame
 luminosities, for measuring the LF of cluster galaxies as a whole and
 split by color, and for fitting Schechter (1976) functions to the
 measured LFs.  We will present a comparison of \lfzp\ and \lfss\ and
 discuss why robust LF determination of cluster galaxies can only be
 made for the red galaxy population.  

\subsection{Determining Rest-frame Luminosities}
\label{restlum_sec}

 Rest-frame luminosities \lrest\ and rest-frame colors were calculated
 using the technique described in \citet{Rudnick03} and assuming that
 every galaxy selected as a cluster member has $z=z_{clust}$.
 Our \lrest\ estimates were computed from the matched aperture
 magnitudes (see \S\ref{catsec}), which almost certainly miss flux
 compared to the \mitot\ estimate.  To scale our \lrest\ estimates to
 total values we therefore multiply every \lrest\ value by the ratio
 of the total \mi-band flux to that in the \mi-band matched aperture.
 The median correction ranges from a few percent at $I_{\rm tot} \sim
 20-21$ to $\sim 30-50\%$ at $I_{\rm tot} \sim 24.4-24.9$.

 Which rest-frame luminosities we are able to use depends on which
 technique we employ to determine cluster membership.  For the
 photometric redshift method the full range of rest-frame wavelengths
 are available, as the probability of each galaxy residing at the
 cluster redshift is computed directly from its SED.  Therefore the
 SED is by definition consistent with being at (or near) the cluster
 redshift and any interpolation between the observed bands based on
 the templates at that redshift should yield a robust estimate
 of \lrest.  We therefore can compute rest-frame magnitudes of cluster
 members in many rest-frame bands spanned by our observed filter sets,
 e.g. \magrest{g}, \magrest{r}, and \magrest{i}.  The rest-frame NIR
 luminosity functions will be presented in Arag\'on-Salamanca et
 al. (in preparation).

 The statistical background subtraction method, however, limits the
 rest-frame wavelengths for which luminosities can be robustly
 computed to those that are straddled by the observed subtraction
 filters.  The reasoning is as follows.  Recall that the photometric
 redshift technique uses the full SED information to determine
 membership on an individual basis.  With statistical subtraction,
 however, the membership probability is not known for each galaxy, but
 rather for all galaxies in a region of color-magnitude space based on
 their relative numbers with respect to those in an empty field image.
 This implies that some fraction of the galaxies classified as members
 will actually be at different redshifts than the cluster.  For
 rest-frame wavelengths straddled by the observed subtraction bands
 (in our case $\lambda_V<(1+z_{clust})\times\lambda_{rest}<\lambda_I$)
 this is not a problem, as the color of every candidate member is
 constrained to be similar to that of the very cluster galaxies that
 cause the overdensity in counts in that color-magnitude bin,
 regardless of whether or not that candidate truly is a member.
 Therefore the use of templates at \zclust\ can be used to determine
 \lrest\ without large systematic errors if the galaxy is truly a
 non-member.  However, this statistical subtraction method does not
 insist that the SED of the galaxy outside of the observed subtraction
 bands is consistent with one at the cluster redshift.  For this
 reason, rest-frame wavelengths outside of the subtraction bands will
 be subject to uncertain extrapolations and will not be robust.  For
 clusters at our redshift, the condition
 $\lambda_V<(1+z_{clust})\times\lambda_{rest}<\lambda_I$ is
 approximately satisfied for the \magrest{g}\ and \magrest{B}-bands,
 which we limit ourselves to for LFs computed with statistical
 subtraction.

\subsection{A Non-parametric Estimate of the LF}
\label{nonparam_sec}

 We first measure the LF of every cluster by simply binning the sample
 into absolute magnitudes and counting the number of galaxies in each
 bin.  As is done in previous works, we exclude the Brightest Cluster
 Galaxy (BCG) and galaxies brighter than the BCG from the LF
 computation.  The properties of the EDisCS BCGs haven been presented
 separately in \citet{Whiley08}.

 For \lfzp\ the error bars in each bin represent the Poisson errors on
 the retained galaxies, computed using the formulae
 of \citet{Gehrels86}.  For \lfss\ the best-fit LF is that derived
 using the subtraction over the whole CFDF.  There are two sources of
 error that contribute to \lfss.  The first source is the Poisson
 error on the number of galaxies in each cluster field retained as
 members.  The second source of error originates in the uncertain
 background measurement, which we determine using Monte Carlo
 realizations for small sub-tiles of the CFDF in estimating the
 field (see \S\ref{statsub_membsel}).  In this case we computed the LF
 for each Monte Carlo realization of the subtraction and took the 68\%
 confidence intervals of the resultant LFs as an estimate of the
 error.  This error was then added in quadrature to the Poisson error
 to achieve a total error.

 In constructing the LF for each cluster there are two issues to
 consider, the detection limit in observed total magnitudes and the
 corresponding limit in absolute magnitude.  As described in W05 we
 establish our completeness in observed I-band magnitude in an
 empirical way by comparing our number counts to those from much
 deeper surveys (see W05, Fig. 1).  There is ample evidence that the
 intrinsic slope of the I-band number counts is a rising power law at
 faint magnitudes (e.g. Metcalfe et al. 2001; Heidt et al. 2003) and
 we define our completeness as the magnitude at which our observed
 number counts in total magnitudes deviate from a power law defined by
 the deeper observations.  There are two reasons this is reasonable.
 First, the number counts contributed by the cluster at faint
 magnitudes is much smaller than the contribution by the field.  This
 is evidenced by the fact that $80-90\%$ of galaxies are rejected by
 statistical subtraction at $I_{\rm tot} < 24.9$ \citep{Pello09}.
 Also, the slope of our number counts is parallel to that from deeper
 fields at $22<I_{\rm tot}<24$ for the high-z clusters and $23<I_{\rm
   tot}<24$ for the low-z clusters, where we expect the cluster to no
 longer contribute significantly to the counts.  For this reason we
 feel that our faint counts can be directly compared to that of the
 field.  Second, our total magnitudes (which include an aperture
 correction) result in a rapid drop off in the number counts at faint
 magnitudes.  This is not seen in surveys that measure magnitudes
 without an aperture correction but is a direct result that we count
 for a minimal amount of missing flux in our faintest galaxies (see
 \citet{Labbe03} for a more detailed explanation).  \citet{Labbe03}
 also showed that a limit defined in this way corresponds to a near
 perfect detection probability.  Because this is a rather conservative
 estimate of our completeness the signal-to-noise is still high
 (typically $>10$; W05) all the way down to our detection limit,
 allowing the robust computation of magnitudes and colors.

 Once we have established our completeness limit in observed magnitude
 we translate this, for every rest-frame filter, into an absolute
 magnitude limit that is the most conservative (i.e. brightest) given
 the whole range of possible galaxy SEDs.  If a redshifted rest-frame
 filter for a given cluster redshift is blueward of the observed
 \mi-band the brightest limit corresponds to that computed using a 10
 Myr old single age population with solar metallicity and a
 \citet{Sal55} IMF.  This is perhaps overly conservative for red
 galaxies, but results in the most conservative limit for the whole
 catalog, so that we are equally complete at all galaxy colors.  For
 redshifted rest-frame filters redward of the observed \mi-band we
 used Elliptical template from \citet{CWW80}.

 We also created composite LFs for subsamples split by redshift and
 cluster velocity dispersion.  We created the composite and its error
 using the method of \citet{Colless89}, which was also discussed in
 detail in \citet{Popesso05}.  With this method, the composite LF at
 every magnitude represents the mean fraction of galaxies compared to
 the number in a normalization region.  We choose the normalization
 region to be all magnitudes brighter than the brightest completeness
 limit that all clusters in that subsample have in common.

 When creating the composite clusters, we correct them for passive
 evolution to the mean redshift for that subsample.  As we will
 describe in subsequent sections, only the LF for red cluster galaxies
 can be robustly determined and we concentrate mostly on those for the
 rest of the paper.  DL07 showed that the colors of the red sequence
 can be well fit by a passively evolving model with $z_{form}\sim
 2-3$.  We correct the rest-frame magnitudes using a $z_{form}=2$
 single stellar population (SSP) Bruzual \& Charlot (2003; hereafter
 BC03) model with $Z=2.5Z_{\odot}$.  In practice, this small evolution
 correction does not change the binned LF with respect to that
 computed with no correction.  This is because the amount of evolution
 from each cluster to the center of its redshift bin is significantly
 smaller than the 0.5 magnitude bin size used in constructing the LF.
 For the same reason the exact choice of model used makes little
 difference in the resulting composite LF.

 We compute the LF in two different physical radii, $r<0.75$Mpc and
 $r<0.5$\rtwo, where \rtwo\ is defined as the radius within
 which the density is 200 times the critical density:

\begin{equation}
	R_{200}=1.73~\frac{\sigma}{1000~{\rm km~s^{-1}}}\frac{1}{\sqrt{\Omega_\Lambda+\Omega_0(1+z)^3}}~h_{100}^{-1}{\rm~Mpc}
\label{r200_eq}
\end{equation}

where $\sigma$ is the cluster velocity dispersion \citep{Finn05}.  The
area defined by these two radii is entirely contained within the
EDisCS fields for all but one of our clusters (CL1227.9--1138) for
which we take only the inscribed area into account when performing the
statistical subtraction\footnote{Using the EDisCS data it was realized
  that the LCDCS BCG candidate for CL1227.9-1138 was not the actual
  BCG.  The true BCG is located significantly offcenter in our FORS
  data, resulting in the loss of area.}.  For this cluster the lack of
data for $\sim 50\%$ of the galaxies within 0.5\rtwo\ and $\sim 60\%$
within 0.75Mpc should not bias the values of \mstar\ but will result
in a larger error bar on that value.  For only two of the most massive
clusters, CL1216.8-1201 and CL1232.5-1250, is 0.5\rtwo\ larger than
0.75Mpc.  Our conclusions are insensitive to the exact choice of radii
and unless otherwise stated we will use $r<0.75$Mpc as it is most
always the larger of the two and hence will produce the highest S/N
LF.

\subsection{Schechter Function Fits}
\label{schechter_sec}

 We fit \citet{Schechter76} functions to the binned LFs in each
 cluster.  To fit we created a coarse grid in the three fitted
 parameters, i.e. \phistar, \mstar, and $\alpha$.  We calculated
 the \chisq\ value at each grid point and took the best-fit solution
 as an initial guess for the parameters.  We then re-fit the
 parameters with a narrower range and a finer sampling in the
 parameter space.  We determined the formal uncertainty on each
 parameter by first converting the \chisq\ at each grid point into a
 probability via $P_{{\rm \phi^\star}, M^{\star}, \alpha} =
 e^{-\chi^2/2}$ and then by marginalizing the probability along the
 other two parameters to obtain a probability distribution for the
 parameter in question.  We then measured the limits in this parameter
 that enclosed 68\% of the probability as the 1-sigma formal error
 bar.

 To assess the reliability of such fits we created a set of mock
 binned luminosity functions by randomly drawing from a set of input
 values, i.e.  the number of galaxies, \mstar, and $\alpha$.  The
 errors on each mock LF were Poisson errors on the number of galaxies
 in each bin.  For a given set of parameters we created 100 mock
 realizations of that LF and fit each realization using the procedure
 above, and over the absolute magnitude range present in our data.
 While all three Schechter parameters are highly degenerate, we found
 that the most poorly constrained parameter was $\alpha$ followed by
 \mstar.  The ability to retrieve the parameters was also dependent on
 the input value of $\alpha$, since steeper (more negative) $\alpha$'s
 produced more biased answers.  For the red galaxies to which we limit
 our analyses (see subsequent sections) $\alpha > -0.6$ and the bias
 produced by a steep slope is not severe.  Nonetheless, through these
 simulations we found that it was impossible to constrain all three
 parameters simultaneously using the data from an individual cluster,
 or even from a composite luminosity function of only a few clusters.
 We did find however, that we could constrain all three simultaneously
 if we fit a LF with characteristics akin to the composite LF of the
 entire EDisCS sample, split into two bins of redshift.  We therefore
 derive $\alpha$ and its uncertainty for the entire EDisCS sample for
 each band in each redshift bin and use that $\alpha$ when fitting the
 individual and stacked luminosity functions when split by velocity
 dispersion.  Even when fitting to the whole sample, however, the
 uncertainties on $\alpha$ are non-negligible.  To account for this
 uncertainty in the fitting of individual clusters or subsamples of
 the EDisCS clusters, we fit the Schechter function to the data 100
 times, with $\alpha$ fixed each time but drawn randomly from a
 Gaussian with a mean and sigma taken from the fit to the total
 stacked cluster sample.  The 68\% confidence interval in the
 distribution of \mstar\ from these 100 iterations was then added in
 quadrature to the formal uncertainties, derived with a fixed
 $\alpha$, to derive the total uncertainty in \mstar.  This may
 overestimate the error in \mstar\ as it includes some of the sampling
 error twice.

\subsection{Splitting LFs by color}

 We divide our sample by \vi\ color into red sequence galaxies and
 bluer galaxies.  For each cluster we fit the zeropoint of the
 color-magnitude relation (CMR) in \vi\ assuming a fixed slope of
 -0.09 and using the outlier resistant Biweight estimator
 \citep{Beers90} for the zeropoint.  In performing the fit we only use
 spectroscopically confirmed cluster members with no emission lines.
 This was the same method as used by DL07.  We give the best-fit
 zeropoints in Table~\ref{cmrfit_tab} for the 16 clusters for which a
 robust LF determination is possible (see
 \S\ref{results_sec})\footnote{Our values are given at \mitot$=0$
   whereas those from DL07 were given at an apparent magnitude that
   corresponds to \absmag{V}$=-20$ when evolution corrected to $z=0$.
   DL07 also use Vega magnitudes.}.  A relatively constant slope of
 the CMR can be understood if the slope is primarily a result of a
 metallicity trend with magnitude (e.g. Kodama \& Arimoto 1997) among
 galaxies with a uniformly old age \citep{Bower92}, at least among
 bright galaxies.  As shown in, e.g. \citet{Kodama97} and
 \citet{Bower98}, the rate of change of color with time is insensitive
 to metallicity, so using the local value for the CMR slope with our
 intermediate redshift clusters is a reasonable assumption.  As in
 DL07, we select red sequence galaxies as those within $\pm 0.3$
 magnitudes of the best-fit CMR.  This is a compromise between the
 completeness and purity of our red sequence sample.  By allowing our
 color cut to extend below the CMR, we ensure that we do not miss red
 galaxies that are slightly bluer than the CMR, but also increase the
 possibility that there may be some blue galaxy contamination at
 fainter magnitudes where our photometric errors increase.  We tested
 the sensitivity of our results were to the exact form of our red
 sequence select in two ways.  First, we varied the width of our
 selection slice by $\pm 0.05$ mag.  This corresponds to the $\approx
 0.1$ magnitude error in \vi\ for galaxies at the EDisCS magnitude
 limit \citep{White05}.  Second, we selected all galaxies redward of
 the CMR and then reflected them across the CMR.  This latter method
 is similar to what is used by \citet{Gilbank08} and insures high
 sample purity at the risk of missing intrinsically bluer/younger
 galaxies still formally on the red sequence.  In all cases the we
 find that the LFs with these different methods are consistent to
 within 1-sigma, indicating that our results are robust against
 variations in the red sequence selection.  We believe that this must
 be partly true due to our conservative magnitude limit and extremely
 deep VLT photometry.

 For each of the samples split by color we compute the individual and
 composite LFs as described above.  As shown in DL04 and DL07, it is
 also important when establishing the effective magnitude limit on the
 red sequence to take into account that the S/N of the color
 measurement of galaxies becomes worse for redder galaxies at a fixed
 \mitot\ (see Fig. 1 of DL07).  We take this into account when
 determining our completeness limit and find that we may be missing
 some red sequence galaxies in the $24.4<I<24.9$ magnitude bin.
 Although our \lfzp\ estimates for the high-z clusters are computed to
 $I=24.9$, all of the trends described in this paper are completely
 dominated by effects in the bins at $I<24.4$.  We therefore do not
 worry about this minor incompleteness in our last bin.

\subsection{A Comparison Between Methods}
\label{methodcomp_sec}

\begin{figure*}
\epsscale{1.2}
\plotone{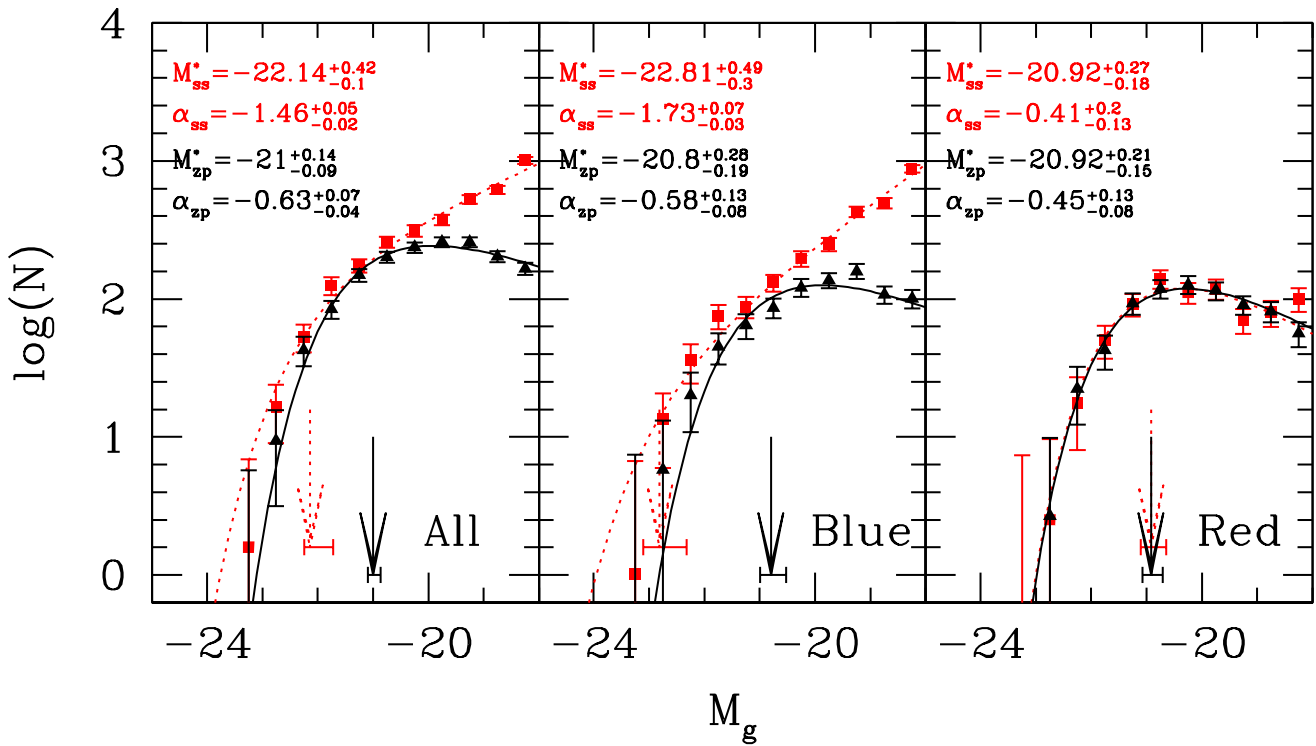}
\vspace{-3in}
\caption {
 The \magrest{g}-band composite luminosity functions of EDisCS cluster
 galaxies.  The left panel is for all galaxies, regardless of color.
 The middle panel is for blue galaxies and the right panel is for red
 galaxies (see text for definition of colors).  The squares show the
 LF determined using statistical subtraction \lfss\ and the triangles
 show the LF determined using photometric redshifts \lfzp.  The solid
 and dotted curves show the best-fit Schechter function fits to \lfzp\
 and \lfss\ respectively and the vertical arrows of the same line type
 show the corresponding best-fit values of \mstar.  The horizontal
 error bars at the base of the arrows give the 68\% confidence limits in
 \mstar.  When including all galaxies \lfss\ has a steeper faint end
 slope and a larger number of bright galaxies than \lfzp.  These
 difference can be traced to the blue galaxies.  Both techniques give
 identical results for the red galaxies.}
\label{all_stack_fig}
\end{figure*}

 We assess the robustness of our LFs by comparing \lfzp\ and \lfss.
 In Figure~\ref{all_stack_fig} we show the composite LF of all EDisCS
 clusters as computed with the two methods.  The \lfss\ of all
 galaxies has a steeper faint end slope and an overabundance of bright
 galaxies compared to \lfzp.  This same behavior is apparent, albeit
 at lower significance, in all the composite and individual LFs.  We
 also compute the LFs separately for blue and red galaxies and plot
 these in the middle and right panels of Fig~\ref{all_stack_fig}
 respectively.  It is obvious from this figure that the discrepancy
 only exists for the blue galaxies.  In contrast, \lfss\ and \lfzp\
 agree completely for red galaxies, as was found in DL07.

 There are at least two possible reasons for the large difference in
 the faint end slope between the two techniques that only manifests
 itself for blue galaxies.  First, the effectiveness of \lfss\ is
 critically dependent on the validity of the field counts used to make
 the statistical subtraction.  The faint-end slope of the blue number
 counts is in general steep (e.g. Koo 1986) and we have checked that
 the faint-end slope of the counts in the CFDF is significantly
 steeper for blue than for red galaxies.  Because the faint-end slope
 of the blue galaxy counts is so steep, the faint-end slope of the
 cluster LF is critically dependent on the exact value of the slope.
 Specifically, the faint-end slope of the counts in the comparison
 field needs to be the same as the faint-end slope of the counts for
 field galaxies in the cluster field.  If there are slight differences
 in the way that magnitudes are measured between the field and cluster
 catalogs, an incorrect faint-end cluster LF can be measured.  Indeed,
 although AUTO magnitudes are used for both the cluster and CFDF
 catalogs the seeing of the CFDF catalog is $\sim 1.5-2$ times worse
 than that of the EDisCS catalogs and there has been no attempt to
 match SExtractor catalog parameters.  As a result, magnitude
 dependent differences in the AUTO magnitudes could be present between
 the two catalogs and this could cause the very steep faint end slope
 of \lfss\ for blue galaxies.  We have checked that a magnitude
 independent change in the CFDF magnitudes of up to 0.2 magnitudes has
 no appreciable effect on the faint end slope but have not explored
 more complicated magnitude dependent effects.  We conclude that
 differences in the way the two surveys measure magnitudes makes it
 difficult to measure the faint-end \lfss\ for blue galaxies, where
 the magnitude measurement of faint galaxies is so crucial.  The
 faint-end slope of red field galaxies in the CFDF is much shallower
 and small errors in the magnitude measurements will not lead to as
 large of errors in the LF.

 Another possible reason for the difference between \lfss\ and
 \lfzp\ for blue galaxies, specifically the large difference in the
 faint end slope, may come from limitations in the photometric
 redshift techniques.  For the spectroscopic sample we verified that
 the photometric redshifts performed similarly for red and blue
 galaxies.  Unfortunately given the spectroscopic magnitude limit we
 were not able to verify how the photometric redshifts performed at
 faint magnitudes.  In general, the performance of photometric
 redshift codes depends on the S/N of the flux measurements since a
 higher S/N measurement allows for a better localization of the
 features (e.g. the 4000\AA\ break) used to determine the redshift.
 For galaxies with weaker intrinsic features in their SEDs, e.g. blue
 galaxies, the photometric S/N must be higher to yield a comparable
 redshift accuracy as for galaxies with stronger features, e.g. red
 galaxies with strong 4000\AA\ breaks.  Since we determine cluster
 membership by integrating \pz, a poorer constraint on \zp\ with a
 correspondingly broader \pz\ will result in a \pclust\ that may fall
 below the \pthresh\ value that was calibrated for brighter galaxies.
 As an additional complication, the slope of the blue star-forming
 galaxy sequence (the blue ``cloud'') is such that faint blue galaxies
 are typically bluer than bright blue galaxies, meaning that the
 photometric redshifts will perform correspondingly worse.  To assess
 whether this effect could cause the downturn on the faint-end
 \lfzp\ for blue galaxies we examined the dependence of the \zp\ 68\%
 confidence limits on \absmag{g} for blue and red galaxies with
 $z_{phot}=z_{clust}\pm 0.05$.  For red galaxies the internal
 \zp\ errors remain small and increase only slowly with \absmag{g}.
 For blue galaxies, however, the internal errors rise more rapidly
 with increasing \absmag{g} and there is a population of blue galaxies
 with very large errors.  Both the blue galaxies with very large
 errors and those on the upper envelope of the main error-magnitude
 relation are flagged as interlopers by the photometric redshifts.
 The absolute magnitude where this increase in the \zp\ uncertainties
 of blue galaxies occurs coincides with the magnitude where the faint
 end slopes of \lfss\ and \lfzp\ start to diverge.  The difficulty in
 using \zp\ to establish membership at faint magnitudes is explored
 further in \citet{Pello09}.  It may be that the best way to study
 blue galaxies with photometric techniques is by using a combination
 of statistical background subtraction and photometric redshift
 membership techniques, such that the photometric redshifts are used
 as a first-pass membership method and the statistical background
 subtraction is then used to subtract off any residual (e.g. Kodama et
 al. 2001; Tanaka et al. 2005).  In practice this will require either
 a large field sample with identical photometry (and hence photometric
 redshift performance) as the target field or a cluster image with a
 wide enough area to have minimal contamination from the cluster at
 the outskirts of the image.

 As mentioned, these two problems in determining the faint end should
 not be (and apparently are not) as severe for red galaxies as for
 blue.  Photometric redshifts seem to perform better for red
 galaxies than for blue, at least in the realm of decreasing
 photometric S/N.  
 The source of the discrepancy between \lfss\ and \lfzp\ at the bright
 end is not as clear.  The CFDF appears to be slightly underdense with
 respect to the FORS Deep Field \citep{Heidt03} and the COMBO-17
 number counts from The Chandra Deep Field South \citep{Wolf04}, which
 would serve to increase the \lfss\ value for EDisCS.  Also, despite
 our best efforts at calibration of \zp\ for bright sources from the
 spectroscopic sample, the photometric redshifts may reject a slightly
 larger number of blue members than red members, which would push
 \lfzp\ down.  In the end we must conclude that the determination of
 the blue-galaxy cluster LF is not robust when only using photometric
 redshifts or statistical subtraction.

 The red galaxy LFs, however, agree astonishingly well, indicating
 that the red galaxy LF is robust to the exact method used.  We
 therefore limit most of our subsequent analysis to the red galaxies
 only.

\subsection{The Local Luminosity Function}
\label{local_LF_sec}

 To measure evolution in the LF it is important to have an appropriate
 local sample.  For many parameters of the galaxy population, e.g. the
 star-forming fraction \citep{Poggianti06} and the early-type fraction
 \citep{Desai07}, there is a strong dependence on $\sigma$ at
 intermediate and high redshift, implying that evolution can only be
 measured in samples matched in velocity dispersion.  No dependence of
 the LF of all cluster galaxies on $\sigma$ has been found at low
 redshift \citep{Propris03} but we wish to test this for red sequence
 galaxies specifically at intermediate and high redshift.  For our
 purposes we therefore require a local sample that has the same range
 in $\sigma$ as our sample and allows for the computation of a
 luminosity function just for red sequence galaxies.  It is also
 desirable that enough local clusters be used so as to average over
 cluster-to-cluster variations and minimize the uncertainties in the
 local anchor of any evolutionary trends.  Finally, it is advantageous
 if the local LF has been computed in multiple bands, to allow the
 measurement of wavelength dependent evolution.  \citet{Propris03},
 \citet{Popesso05}, and \citet{Popesso06} computed composite, high S/N
 LFs from the 2dFGRS and SDSS respectively.  \citet{Propris03} compute
 their LFs only in the $b_j$-band and do not compute them as a
 function of galaxy color.  \citet{Popesso06} presented composite LFs
 for X-ray selected clusters in multiple bands and as a function of
 galaxy color; however, we choose to construct our own SDSS luminosity
 function, for the following reasons.  The sample of (Popesso et
 al. 2005;2006) is x-ray selected, which may cause biases in the
 comparison of the local sample to the EDisCS sample, which is
 optically selected.  Second, \citet{Popesso06} split their LFs by
 color, but not in an analogous way to the EDisCS sample, which again
 complicates the comparison to our results.  Finally, the raw LFs from
 \citet{Popesso06} are not published, but only the two-component
 Schechter fits, which also complicates the comparison to our LFs.

 Our cluster sample is a subset of the sample presented
  in \citet{Linden07}.  This parent sample was selected from the C4
  catalog of \citet{Miller05}, but employs improved algorithms to
  identify the BCG and measure the velocity dispersion. We limit our
  analysis to clusters at $z \le 0.06$, to ensure that the individual
  cluster LFs are complete down to the passively evolved limit of the
  EDisCS clusters (see below), which results in a sample of 167
  clusters.  With this redshift cut-off we can limit our analysis to
  galaxies with $r<20$, where the star/galaxy separation is still
  robust and where colors can be robustly determined.  We used a
  global field sample drawn from the SDSS DR4 catalog and use
  the \textit{model} magnitudes to measure colors and \textit{cmodel}
  magnitudes to measure the total magnitude.  Colors are measured with
  the \textit{model} magnitude, since that measure adopts the same
  aperture in different bands.  The \textit{cmodel} magnitudes fit a
  linear combination of an exponential and de Vaucouleurs profile to
  each galaxy and integrate the combination of these two to derive a
  total magnitude.  It is well known that the Petrosian magnitude of
  SDSS misses flux, especially for early type galaxies with de
  Vaucouleurs profiles and \textit{cmodel} magnitudes should be closer
  to total.

 We isolated cluster members using a statistical subtraction technique
 similar in principle to what was used for the EDisCS clusters, but
 with some significant modifications.  Since photometry of identical
 depth and bandpass coverage was available for both our local cluster
 and ``field'' samples in multiple bands, we performed our statistical
 subtraction in 4 dimensions, using bins in \rs-magnitude as well as
 bins in $g-r$, $r-i$, and $i-z$, and using 0.5 bins in magnitude and
 0.3 magnitude bins in color.  This technique has two main advantages
 over the 2D, i.e. magnitude and single color, subtraction used in
 EDisCS.  Although interloper galaxies may have identical colors to
 cluster members in one color, as more colors are considered, it is
 increasingly difficult to mimic the colors of galaxies at the cluster
 redshift, thereby increasing the contrast of the cluster against the
 background and reducing the number of contaminating galaxies.  In
 addition, because we require that retained galaxies have colors
 matching that of cluster members from \gs\ all the way to \zs, it is
 possible to robustly determine rest-frame magnitudes for any bands in
 between, e.g. \magrest{g}, \magrest{r}, and \magrest{i}.  Following
 \citet{Pimbblet02} here we also expand the bins in magnitude first
 and then in color if the number of expected field galaxies exceeds
 the number of candidate members.  We determine rest-frame magnitudes
 using the kcorrect software v4.1.4\citep{Blanton03}.  For each
 cluster we performed 100 Monte Carlo realizations of the subtraction,
 where each iteration used a new set of random numbers which were then
 compared against the membership probabilities to determine cluster
 membership.  The distribution of the LFs for the full Monte Carlo
 simulation was used measure the uncertainties in the LF.

 To isolate the red sequence in the SDSS clusters we performed an
 outlier-resistant fit the to CMR of the composite SDSS cluster
 population using rest-frame magnitudes and colors and only fitting
 spectroscopically confirmed members with no
 \ha\ emission\footnote{The presence of \ha\ was indicated by a
   measurement of \ha\ with S/N(\ha)$>5$.  The \ha\ measurements were
   obtained from
   \texttt{http://www.mpa-garching.mpg.de/SDSS/DR4/raw\_data.html} as
   computed by \citet{Brinchmann04}}.  Evolution corrections to the
 mean redshift for red galaxies are very small over our redshift range
 and do not make a difference when fitting the CMR or deriving the LF.
 Due to the large number of galaxies over a large range in absolute
 magnitude, we were able to simultaneously fit the slope and zeropoint
 of the relation.  For measuring the \magrest{g} and \magrest{r} LFs
 we fit the CMR in \colrest{g}{r} vs. \magrest{g} and \colrest{g}{r}
 vs. \magrest{r} respectively.  For measuring the \magrest{i} LF we
 fit the CMR in \colrest{r}{i} vs. \magrest{i}.  Similar as to what
 was done for the EDisCS sample, we then classified as red galaxies
 every galaxy within a stripe centered on the CMR in the color and
 magnitude used above for each rest-frame band.  The width of this
 stripe in each color was chosen to correspond to $\Delta
 (U-V)_{rest}=0.3$ at $z=0.6$ assuming that the scatter in the CMR is
 due entirely to age.

 We created individual and composite SDSS cluster LFs using identical
 procedures as with the EDisCS clusters, i.e. binning the individual
 clusters in absolute magnitude and creating the composite
 following \citet{Colless89}.  For every Monte Carlo realization of
 the subtraction we computed the Poisson uncertainty on the LF.  As
 with EDisCS the LFs were computed within $r<0.75$Mpc and
 $r<0.5$\rtwo\ (as computed using Eq.~\ref{r200_eq}).  To determine
 the absolute magnitude limit down to which we construct the LF while
 probing the same galaxies as in the EDisCS clusters, we took the
 absolute magnitude limit for our highest redshift EDisCS cluster and
 corrected it for passive evolution down to $z=0.06$ using a
 $Z=2.5Z_\odot$ BC03 SSP model with $z_{form}=2$.  For example,
 the \magrest{g}-band limit of -18.5 for the highest redshift EDisCS
 cluster corresponds to a \magrest{g}-band limit of -17.5 for the SDSS
 clusters.  The $r<20$ apparent magnitude selection is deep enough
 such that the absolute magnitude limit is the more restrictive cut
 for all of our SDSS clusters.

 We also split our SDSS clusters by $\sigma$ to match the velocity
 dispersion bins in the EDisCS sample.  We defined the SDSS $\sigma$
 threshold taking into account the redshift evolution in $\sigma$ that
 is predicted from the growth of the cluster dark matter halos over
 time.  Using the expected mass accretion history of halos
 \citep{Bower91,Lacey93}, \citet{Poggianti06} used the results of
 \citet{Wechsler02} and \citet{Bullock01} to show that clusters with
 $\sigma=600$~km/s at $z=0.6$ will grow into clusters with
 $\sigma=700$~km/s by $z=0$ (see Fig. 19 of \citet{Jensen08}).  We
 therefore divide our SDSS sample into high and low velocity
 dispersion bins using $\sigma=700 {\rm km s^{-1}}$ as the divider.
 There are 159 clusters in the low $\sigma$ bin and 8 in the high
 $\sigma$ bin.  The mean and median uncertainty in the SDSS velocity
 dispersions is $\pm 62$ and $\pm 55$~km/s respectively.

\begin{figure}
\epsscale{1.2}
\plotone{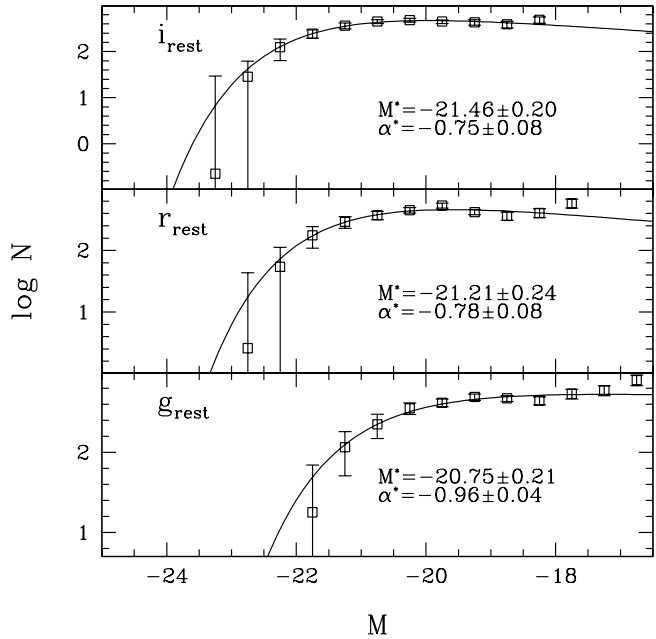}
\caption {
The rest-frame optical composite luminosity functions of red sequence
galaxies for 167 clusters at $z<0.06$ from the SDSS.  The data points,
error bars, and solid line represent one of the Monte Carlo iterations
of the background subtraction, the Poisson errors, and the
corresponding Schechter function fit.  The parameters listed are the
mean over the full set of Monte Carlo realizations and the error bars
on the parameters are the quadrature sum of the Poisson errors and
those from our Monte Carlo simulation of the background subtraction
errors.}
\label{lfsdss_all_fig}
\end{figure}

 We fit each Monte Carlo realization of the composite SDSS LF with a
 Schechter function, allowing all three parameters to vary.  The
 best-fit parameters come from the mean of the realizations and the
 uncertainties on the parameters come from the 68\% confidence
 intervals of distribution from all of the Monte Carlo realizations
 summed in quadrature with the mean formal fit errors.  In
 Figure~\ref{lfsdss_all_fig} we show the SDSS LFs for red sequence
 galaxies and the corresponding Schechter function fits in each band
 for the whole cluster sample.  The Schechter fit parameters are also
 given in Table~\ref{sdss_lfparam_tab}.

\section{Results}
\label{results_sec}

 In this section we present the LFs for red-sequence galaxies in
 the EDisCS clusters.  As we discussed \S\ref{methodcomp_sec} we will
 limit our analysis to the red sequence LF.  Throughout we will
 use \lfzp\ since it gives us access to a larger range of rest-frame
 wavelengths and allows us to go 0.5 magnitudes deeper in our high
 redshift clusters (see \S\ref{restlum_sec}).  We will first present
 the individual cluster LFs and then the composite LFs split by
 redshift and velocity dispersion.  In all cases we do not show
 results for CL1119.3--1129 since this cluster has no NIR data and
 hence has poorly constrained photometric redshifts.  We exclude the
 CL1103.7-1245a and CL1103.7-1245b clusters at $z=0.70$ and 0.63 since
 these clusters overlap on the sky and are too close in redshift to be
 decomposed with the photometric redshifts.  We also exclude
 CL1103.7-1245 since its redshift ($z=0.96$) is too high for our
 imaging to probe far enough down the LF.  

\subsection{Individual Cluster LFs}

\begin{figure*}
\epsscale{1.2}
\plotone{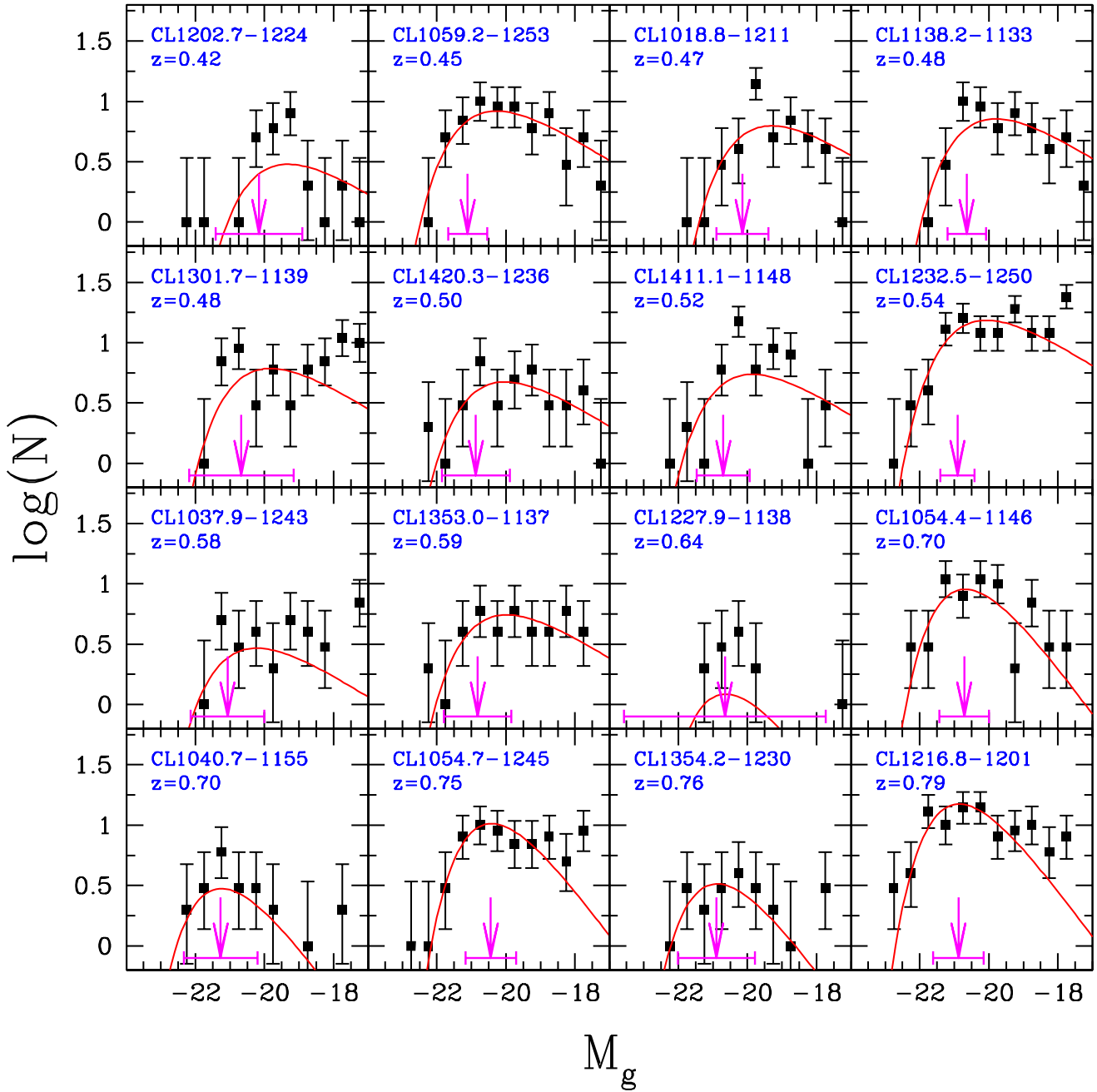}
\caption {
The \magrest{g}-band luminosity functions of red sequence galaxies in
the 16 EDisCS clusters for which \lfzp\ could be determined.  The
clusters increase in redshift to the right and down.  The solid curve
gives the best-fit Schechter function with a slope fixed to the values
fit to the EDisCS composite LF in the corresponding redshift bin.
The arrow and associated error bar indicates the fitted value of \mstar
and its 68\% confidence limits.}
\label{all_indiv_fig}
\end{figure*}

 In Figure~\ref{all_indiv_fig} we present the red-sequence \lfzp\ for
 the remaining 16 EDisCS clusters.  The LF of these clusters in
 the \gs, \rs, and \is-bands is given in the appendix.  In all cases
 we fix the faint-end slope to the value determined from the composite
 LF in the relevant redshift range.  In general the Schechter function
 fits are good with only a few clusters having measured LFs that are
 of too poor quality to obtain a reasonable fit, e.g. CL1227.9-1138.
 In most of the remaining cases it appears that the slope determined
 from the composite LF is an acceptable fit to the individual
 clusters, indicating that a universal LF for red-sequence galaxies is
 possibly in place at these redshifts.  There are, however, some
 exceptions, e.g. CL1301.7-1139 and CL1037.9-1243, where the composite
 faint-end slope does not seem to adequately represent the cluster LF.
 There is no significant observed trend of \mstar\ with cluster
 redshift.  This is in contradiction to the simple expectation that
 red sequence galaxies will be brighter in the past due to passive
 evolution.  This will be addressed in the coming sections.

\subsection{Composite Cluster LFs}

 As shown in Figure~\ref{all_indiv_fig}, the signal-to-noise of the
 individual LFs are too low to make any conclusions about trends with
 redshift or velocity dispersion.  For this reason we create composite
 clusters splitting the sample into two bins at $z=0.6$ and in each
 redshift bin into two bins of velocity dispersion at $\sigma=600{\rm
 km s^{-1}}$.  In all cases we have used the faint-end slope as
 determined from the composite in the same redshift bin.  The
 composite LFs are given in Table~\ref{ediscs_lfstack_tab} and the
 Schechter function parameters for all of the composite LFs are give
 in Table~\ref{ediscs_lfparam_tab}.

\begin{figure*}
\epsscale{1.0}
\plotone{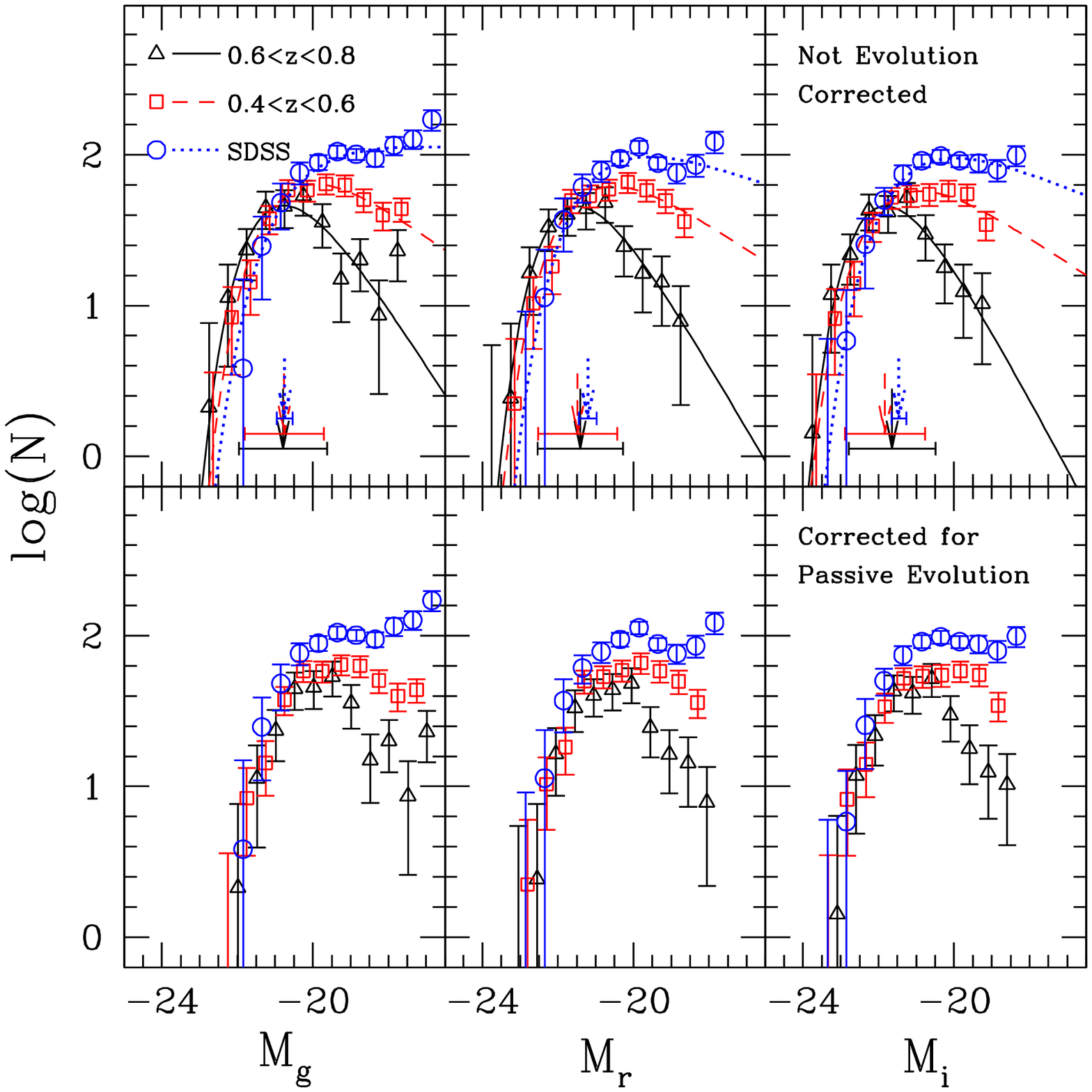}
%\vspace{-2.5in}
\caption { The composite rest-frame \gs, \rs, and \is\ LFs of red
  sequence EDisCS and SDSS cluster galaxies split into bins of
  redshift.  The composites at $0.6<z<0.8$ and $0.4<z<0.6$ have 6 and
  10 clusters respectively.  All plotted LFs and their fits have been
  scaled so that the total luminosity density of the best-fits are
  equal.  \textit{Top row:} The LFs are presented without any
  evolution correction.  Vertical arrows and horizontal error bars
  give the value of \mstar\ and the 68\% confidence interval.  The
  top, middle, and bottom arrow correspond to the SDSS, intermediate
  redshift EDisCS clusters, and high redshift EDisCS clusters
  respectively. The points for the two EDisCS LFs have been offset
  slightly in magnitude for plotting clarity.  \textit{Bottom row:}
  The LFs have all been corrected for passive evolution to the mean
  redshift of the SDSS clusters.  All symbols are as in the top panel.
  The luminous red sequence galaxies are all in place out to $z=0.8$
  but the fainter red sequence galaxies built up dramatically over
  time.}
\label{comp_zsplit_fig}
\end{figure*}

\subsubsection{redshift evolution}
\label{zsplit_sec}

 In the top panel of Figure~\ref{comp_zsplit_fig} we show the
 composite LFs split by redshift at $z=0.6$.  All LFs are only plotted
 as faint as they are complete and, for plotting purposes, are all
 normalized to have the same total integrated luminosity.  In the
 bottom panel we have corrected the LFs for the passive evolution
 expected for a population that formed all of its stars at $z=2$.
 This formation redshift provides a good fit to the color evolution of
 the bright red sequence galaxies in the EDisCS clusters (DL07).  At
 the bright end we do not have high enough signal-to-noise to
 constrain the detailed evolution precisely.  Within the considerable
 uncertainties, however, it appears that the LFs at the bright end are
 all similar after applying the correction for passive evolution.
 This indicates that the bright galaxy population was mostly in place
 in clusters at $z<0.8$.
 
 In contrast to the bright end, there is dramatic evolution in the
 faint-end slope of the LF from $z=0.8$ to $z=0$, in the sense that
 the number of faint galaxies per luminous galaxy are increasing with
 decreasing redshift.  Other works over recent years have found
 similar results (DL04; Kodama et al. 2004; Goto et al. 2005; Tanaka et
 al. 2005; DL07; Stott et al. 2007; Tanaka et al. 2007; Gilbank et
 al. 2008), although \citet{Andreon06} finds no such
 result.  \citet{Stott07} and \citet{Gilbank08} have large samples of
 clusters but we measure the LF to one magnitude fainter than they do
 at all redshifts.  DL04 and DL07 use the same data as in this analysis,
 but concentrate on the evolution in \nlf, as defined using a passively
 evolving luminosity threshold.  Our results using an independent
 analysis are consistent at the 1-sigma level with those from DL04 and
 DL07.

 We also fit Schechter functions to the LFs at different redshifts and
 plot the resulting fits in the top panel of
 Figure~\ref{comp_zsplit_fig}.  As seen in
 Table~\ref{ediscs_lfparam_tab} within the EDisCS sample there is
 strong evolution in $\alpha$ at the $0.5-0.7$ level at a significance
 of $4-5$ sigma (depending on the filter).  The strong evolution
 continues down to the SDSS sample with even higher significance.  At the
 same time there is no evidence for evolution in \mstar\ with
 redshift.  At first glance this seems strange since the bright
 galaxies are already in place in the clusters and appear to be
 brighter in the past.  However, the evolution in \mstar\ only tracks
 the luminosity evolution of the whole galaxy population if all the
 galaxies evolve in the same way.  Given the rapid evolution at the
 faint end, this is definitely not the case.  The highly degenerate
 nature of $\alpha$ and \mstar\ make it hard to interpret them simply
 as evolution of any given part of the galaxy population.  As a side
 note, the lack of evolution in \mstar\ that we observe is also
 consistent at the $1-2\sigma$ level with the results
 of \citet{Tanaka05}, who also find a strongly evolving faint end.

 Stronger constraints on the joint evolution of the Schechter function
 parameters would be enabled most effectively by higher
 signal-to-noise at the bright end, as our sample at fainter
 magnitudes is statistically quite robust.  The small number of
 bright galaxies is a direct consequence of our limited number of
 clusters and of their modest mass (or richness).  The best way to
 improve the constraints on the Schechter function parameters will be
 to construct LFs for much larger samples of clusters.

 In \S\ref{form_epoch_sec} we will discuss this further and present
 implications for the formation histories of red sequence galaxies in
 clusters.

\subsubsection{sample split by velocity dispersion}
\label{sigsplit_sec}

\begin{figure*}
\epsscale{1.0}
\plotone{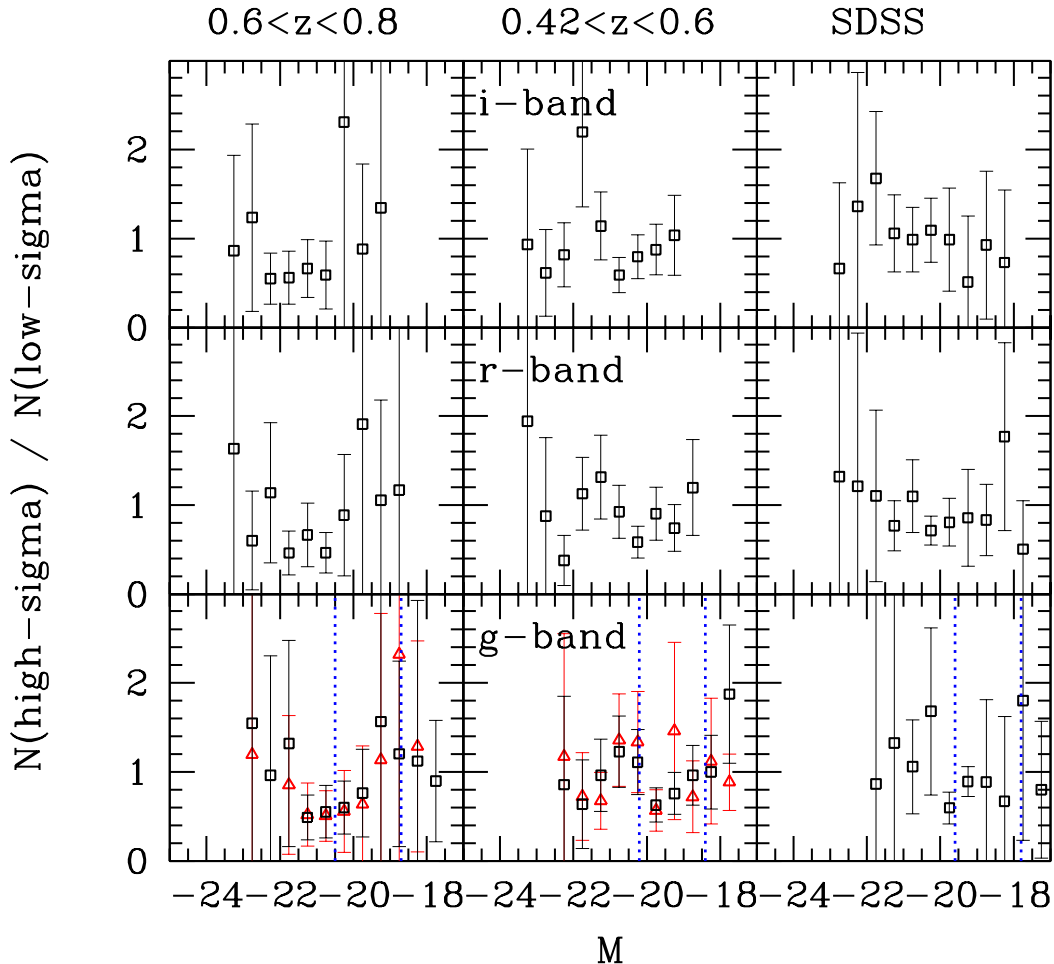}
\vspace{-0.5in}
\caption {
The ratio of the LF of red sequence galaxies in clusters of different
velocity dispersion.  Large differences in the shapes of the LFs will
appear as significant deviations from a horizontal line in this plot.
Each column indicates a different redshift bin and each row represents
a different rest-frame band.  The ratio has been normalized to a mean
of unity to demonstrate relative differences with magnitude.  For
the \gs-band LF for EDisCS clusters we show the ratio of \lfss\ as red
triangles to compare to the ratio of \lfzp\ given as black squares.
All EDisCS points for the \rs\ and
\is-bands are using \lfzp.  The vertical dotted lines in the bottom
row give the division between bright and faint galaxies that
correspond to those used in \citet{DeLucia07}.}
\label{comp_sigsplit_fig}
\end{figure*}

\begin{figure}
\epsscale{1.0}
\plotone{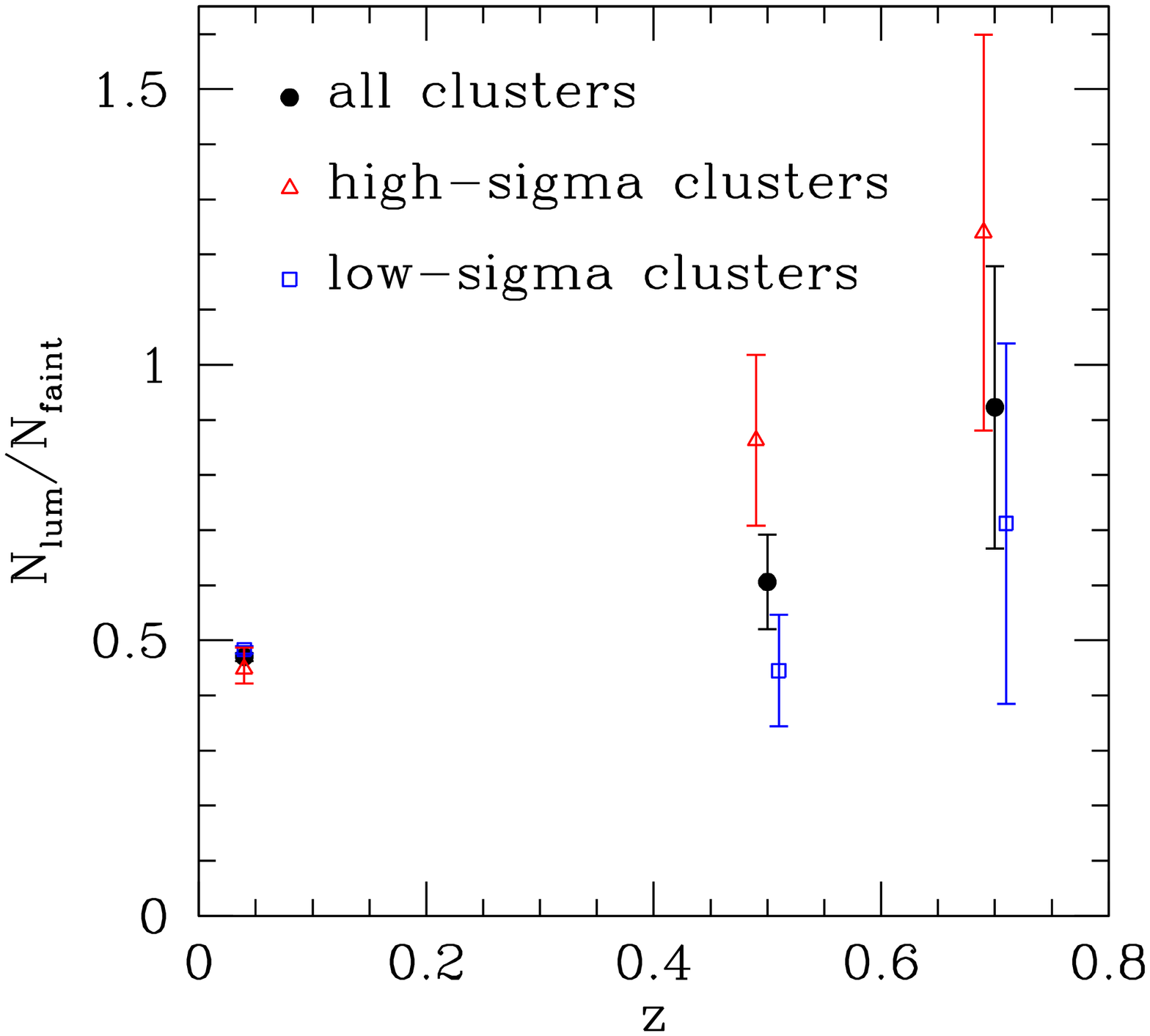}
%\vspace{-0.5in}
\caption {
The ratio of luminous-to-faint red sequence galaxies in clusters at
$z<0.8$.  The dividing line between luminous and faint galaxies has
been corrected for passive evolution and is similar to that used in
\citet{DeLucia07}.  There is a  trend that the high
velocity dispersion clusters evolve quicker than the lower velocity
dispersion clusters and there is no difference with velocity
dispersion in the SDSS clusters.}
\label{lum_faint_fig}
\end{figure}

 We also examined the differences between the red sequence LFs when we
 split our sample by cluster velocity dispersion at $\sigma=600$~km/s.
 The SDSS sample was split at $\sigma=700$~km/s corresponding to the
 expected growth of clusters over time (see \S\ref{local_LF_sec}.)  To
 assess the differences we take the ratio of the LFs in different
 $\sigma$ bins and show the ratio in Figure~\ref{comp_sigsplit_fig}.
 In this figure trends of the ratio with magnitude illustrate
 differences between the high and low velocity dispersion clusters.
 There are no major trends with magnitude but in the $0.42<z<0.6$
 redshift bin the low velocity dispersion clusters have systematically
 more faint galaxies relative to bright galaxies, than high velocity
 dispersion clusters.  Still, formally all of these relations are
 statistically consistent with a constant.  As part of their analysis
 of \nlf\ DL07 and \citet{Gilbank08} also divided their samples by
 velocity dispersion and cluster richness respectively.  For reference
 we draw vertical lines in the bottom panel of
 Figure~\ref{comp_sigsplit_fig} that correspond to the dividing lines
 between luminous and faint galaxies from DL07.  Calculating \nlf\ as
 in DL07 we find entirely consistent results for the EDisCS sample and
 plot them in Figure~\ref{lum_faint_fig}.  For SDSS clusters we find
 no significant difference between the LFs for clusters of different
 velocity dispersions, whereas DL07 found a small
 difference\footnote{DL07 split their SDSS sample at 600~km/s but
   found no difference when splitting at 700~km/s, although they only
   had four clusters in their $\sigma>700$~km/s bin.}.  Our SDSS
 cluster sample is about twice as large as that of DL07, uses a more
 sophisticated method for the background subtraction, and uses model
 magnitudes instead of Petrosian magnitudes.  Because of the
 improvements to our SDSS LF with respect that in DL07 we believe our
 result although the conclusions are in any case not sensitive to the
 difference.  In addition the lack of a dependence of \nlf\ on
 $\sigma$ for local clusters agrees with the results from
 \citet{Propris03} who found that their $b_J$ LFs were consistent at
 the 2$\sigma$ level for clusters with $\sigma$ greater and less than
 800~km/s.  It must be noted, however, that we are only considering
 the red sequence cluster LF and therefore it is not possible to
 directly compare with the results of \citet{Propris03}.  Taken
 together, our results imply that the higher velocity dispersion
 clusters evolved at a faster rate than the low velocity dispersion
 clusters.  A dependence on cluster velocity dispersion is also seen
 by \citet{Poggianti06} who find that the fraction of star-forming
 members in clusters evolves most quickly in clusters with velocity
 dispersions similar to those for the most massive 50\% of the EDisCS
 clusters.

 \citet{Gilbank08} analyzed the dependence of \nlf\ on cluster
 richness and also found that richer clusters evolve quicker than
 clusters with low richness.  They also find that low-richness
 clusters at $0.4<z<0.6$ have a higher \nlf\ than high-richness
 clusters.  This is in the opposite sense of the trend found by DL07
 and our work, in that higher velocity dispersion clusters have a
 higher \nlf\ than low velocity dispersion clusters.  While this
 difference in our work is not very significant, the difference
 between our results and those of \citet{Gilbank08} seem to be as they
 go in opposite directions.  Unfortunately it is difficult to make a
 straightforward comparison between our results as \citet{Gilbank08}
 split their clusters at a richness that corresponds to $\sim
 400$~km/s, below which we only have four clusters.  We therefore have
 almost no constraint on the behavior of \nlf\ for these very poor
 systems.  To resolve this discrepancy will require computing \nlf\ in
 the Gilbank sample with our velocity dispersion cut.  Another
 difference in \citet{Gilbank08} with respect to our work is that
 those authors use a color cut that isolates only the red side of the
 red sequence, whereas our color cut encompasses a band centered on
 the peak of the red sequence.  It is possible, therefore, that our
 cut has a different contribution from blue galaxies at the blue side
 of the red sequence.  It is not clear, however, which cut is more
 physically meaningful.  While the \citet{Gilbank08} cut only isolates
 galaxies with the reddest colors it may miss galaxies that have been
 most recently added to the red sequence and therefore those that have
 the youngest stellar population ages and bluest colors.  At present
 we cannot determine the nature of the discrepancy as a cut only on
 the red side of the red sequence in our data would substantially
 reduce the significance which  we can measure \nlf.

 We do not have enough galaxies to reliably fit $\alpha$ and \mstar\
 independently for our composite LFs split by velocity dispersion.
 Because these two parameters are highly degenerate and there are
 indications that \nlf\ is different in the different bins of $\sigma$
 we therefore do not attempt to interpret the Schechter function fits
 split by $\sigma$.

\subsubsection{radial gradients}

 We examined whether there was any evidence for radial gradients in
 the LF.  We do not have enough galaxies to split our clusters up into
 annuli so instead we compared the LFs within 0.75 and 0.5~Mpc, with
 the acknowledgment that these two are correlated.  For both the
 clusters at $0.4<z<0.6$ and those at $0.6<z<0.8$ the galaxies at
 $0.5~{\rm Mpc}<r<0.75~{\rm Mpc}$ make up $\sim 30\%$ of the galaxies
 at $r<0.75~{\rm Mpc}$.  For the EDisCS clusters we made a similar
 comparison as in Figure~\ref{comp_sigsplit_fig} and find that there
 are no significant differences in the LFs at different radii.  As
 expected from this, there are also no significant differences in the
 Schechter function parameters.  These results are true for all
 rest-frame filters.  In their study of local clusters,
 \citet{Paolillo01} find results consistent with ours but for the
 whole galaxy population instead of just for those on the red
 sequence.  \citet{Popesso06} measure the red sequence LF at different
 radii in local clusters and find that there are radial trends in the
 LF, but primarily at magnitudes fainter than what we probe with our
 data.  At magnitudes corresponding to those we probe they find only a
 weak dependence of the LF with cluster radius, which is entirely
 compatible with our lower precision measurement of the radial trends.
 \citet{Hansen09} measure the LF of red galaxies in different radial
 bins in many more clusters than \citet{Popesso06} but to
 significantly brighter magnitude limits and also find no radial
 dependence in the shape of the LF.

\begin{figure}
\epsscale{1.2}
\plotone{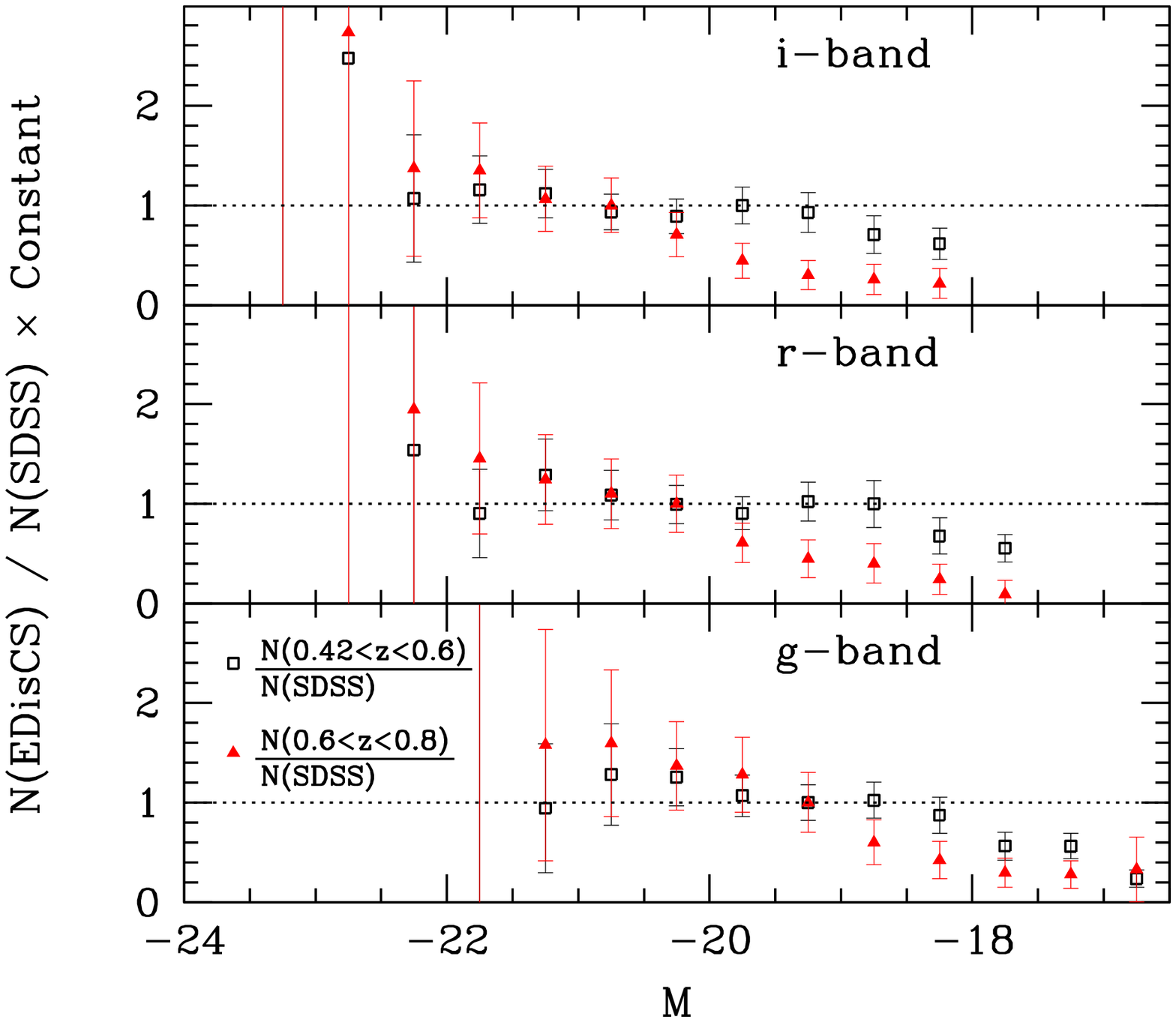}
%\vspace{-2.5in}
\caption {
 The ratio of the EDisCS composite red sequence LFs to those from the
 SDSS in the rest-frame \gs, \rs, and \is-bands.  In all cases the
 EDisCS LFs have been corrected for passive evolution expected for a
 population that formed its stars at $z=2$.  The ratio has also been
 scaled to have a median of unity to allow comparisons of the
 intermediate and high redshift EDisCS clusters.  The horizontal lines
 is drawn at unity to guide the eye.  In all three bands the bright
 end is consistent with a constant value, which appears to turn over
 at fainter magnitudes.  The turnover magnitude evolves to fainter
 magnitudes at lower redshifts.}
\label{rat_zsplit_fig}
\end{figure}

\section{Discussion}

\subsection{Mass dependent build-up of red sequence cluster galaxies}
\label{form_epoch_sec}

 As shown in \S\ref{zsplit_sec} and Figure~\ref{comp_zsplit_fig} there
 is dramatic evolution in the LFs of red sequence galaxies from
 $z=0.8$ to $z=0$.  At the same time this evolution appears to proceed
 at different rates for galaxies of different luminosities.  Red
 sequence galaxies at $L\gtrsim {\rm L}^\star$ appear to be in place
 at $z\sim 0.8$ but the fainter galaxies were added to the red
 sequence at much later times.  As a different way to visualize this
 we plot in Figure~\ref{rat_zsplit_fig} the ratio of the
 passive-evolution-corrected EDisCS LFs to those from the SDSS.  At
 bright magnitudes both LFs are consistent with a constant ratio,
 implying that the EDisCS and SDSS LFs have the same shape.  At
 fainter magnitudes the ratio then decreases.  In the intermediate
 redshift clusters it appears that there is a threshold magnitude,
 fainter than which there is a deficit of red sequence galaxies with
 respect to the SDSS clusters and above which there appears to be a
 constant ratio.  For the high redshift clusters there is not enough
 signal-to-noise to determine if a threshold magnitude exists or if
 there is simply a monotonic trend to fainter magnitudes.  In any case
 it is clear that the magnitude brighter than which the red sequence
 is in place is fainter at lower redshifts.  This is reminiscent of
 the results by \citet{Bundy06} who show that there is a threshold
 stellar mass above which star formation appears to be quenched in
 field galaxies, and that this threshold evolves to lower masses at
 lower redshifts.  However, cluster-associated processes for quenching
 star formation, e.g. ram-pressure stripping,
 strangulation/starvation, or harassment, are different from the
 mass-dependent quenching that may be present in the field.  It may be
 that the evolving mass threshold in the field is imprinted on the
 cluster red sequence in the form of infalling red galaxies whose star
 formation is halted for reasons not associated with the cluster.  We
 cannot, however, directly confirm that hypothesis with our data.  It
 will be very interesting in the future to compare the mass functions
 of cluster and field galaxies to see if there are differences between
 the cluster and field.

 These results imply that the bright cluster galaxies were already in
 place and on the red sequence by $z=0.8$.  This is consistent with
 the evolution of the colors of these galaxies, which is fit well by a
 SSP with $2<z_{form}<3$ (Standford et al. 1998; Holden et al. 2004;
 DL07).  It is also perfectly consistent with the formation redshifts
 found by fundamental plane studies (e.g. van Dokkum \& Franx 2001;
 van der Wel et al. 2005; J{\o}rgensen et al. 2006; van Dokkum \& van
 der Marel 2007; Saglia et al. in preparation).  \citet{Barger98}
 found no formal evidence of evolution in their \mk-band LF for
 morphologically classified early type galaxies, but had large enough
 error bars to be compatible with the expected amount of passive
 evolution over their redshift interval.  They did perform an analysis
 of the surface brightness evolution, however, and found the amount of
 fading expected from passive evolution.  Later works performed more
 extensive LF analyses using \mk-band LF \citep{depropris99}, and in
 the rest-frame NIR \citep{Andreon06,depropris07} and find evolution
 in \mstar\ consistent with passive evolution and $z_{form}>1.5$ when
 fit to galaxies with a faint limit $1-3$ magnitudes brighter than
 ours.  This may imply that the red sequence galaxies dominate the
 galaxy mass function of clusters even out to $z\sim 0.8$, such that
 the total LF of bright galaxies in the rest-frame NIR is really
 dominated by red sequence galaxies.  We will address this directly in
 our future analysis of the galaxy stellar mass function in our
 clusters (Arag\'on-Salamanca et al. in preparation).

 In previous works, the high inferred formation redshift for bright
 red galaxies has been used by many LF studies to imply that the
 population of all red sequence cluster galaxies was already in place
 at $z>1$.  We, however, have shown that there is indeed a
 differential build-up of the red sequence with redshift, such that
 fainter galaxies were added to the red sequence at lower redshifts
 than brighter galaxies.  This may in fact be consistent with claims
 that the early type fraction increased significantly at $z<1$
 (e.g. Dressler et al. 1997; Fasano et al. 2000; Smith et al. 2005;
 Postman et al. 2005; Desai et al. 07) implying a late buildup of
 significant fractions of the red galaxy population.  This is explored
 in more detail in \citet{Sanchez09}, which examines the joint
 evolution of ages, metallicities, and morphologies of galaxies on the
 red sequence.

\subsection{Comparison to field}
\label{field_sec}

 We compare our red-sequence LF to two independently computed red
 sequence LFs for the field\footnote{There are other measures of the
   early-type galaxy LF in the field at $z=1$ (e.g. Zucca et al. 2006;
   Scarlata et al. 2007) but we choose not to compare directly with
   these surveys for various reasons.  For example, \citet{Zucca06}
   identify red galaxies by spectroscopic type and it is not clear how
   that corresponds to our selection by color.  \citet{Scarlata07} do
   not aperture correct their magnitude measures and their LFs are 1-2
   magnitudes shallower than ours, while not having a higher
   signal-to-noise than the NDWFS.}.  The first field LF is that
 published by \citet{Brown07} in the NOAO Deep Wide Field Survey
 (NDWFS).  The NDWFS LF is computed over 6.96 deg$^2$.  It is somewhat
 shallower than the EDisCS survey but the LF has excellent
 signal-to-noise at both the bright and faint-end of the LF.
 Magnitudes in the NDWFS were measured in circular apertures of
 $r=4\farcs0$ and had an aperture correction derived from simulations.
 Although our surveys have substantially different depth and image
 quality, we have checked that our total magnitude estimates are
 compatible with those in the NDWFS.  We compared the aperture
 corrections we apply to our AUTO magnitudes in EDisCS to the
 difference between the AUTO magnitudes of NDWFS galaxies and their
 total magnitude estimate \citep{Brown07}.  This difference is
 similar, indicating that our correction brings our total magnitudes
 into rough agreement with theirs.  Both our survey and the NDWFS
 select red galaxies in similar ways, implying that our results can be
 directly compared.

For our second comparison we determine the LF for red sequence
galaxies using a modified version of the data and LF technique
presented in \citet{Marchesini07}.  The original data are comprised of
multi-band optical through NIR photometry over 6 dispersed fields,
with a total area of 355 arcmin$^2$. The fields are comprised of the
Hubble Deep Field South, The Chandra Deep Field South, and the four
fields from the Deep NIR Multi-Wavelength Survey by Yale-Chile (MUSYC;
Quadri et al. 2007).  The main modification that we have made to the
data presented in \citet{Marchesini07} is that we have included
photometry past the \mk-band to 8.0$\mu$m using \textit{Spitzer}
observations (Wuyts et al. 2008; Marchesini et al. 2009).  We have
also determined the photometric redshifts using a different code
(EAZY; Brammer et al. 2008).  The data are \mk-selected but we used
the observed $I-K$ colors of galaxies at $z<1$ in the EDisCS \mi-band
selected catalog to verify that we can detect all red sequence
galaxies in the \mk-band data at $z<1$.  Therefore we can use the
\mk-band selected data to construct a pseudo \mi-band selected red
sequence sample analogous to that for EDisCS.  Total magnitudes in the
MUSYC survey are constructed using SExtractor AUTO apertures with
aperture corrections derived in an identical way as our own.  The
MUSYC red sequence definition is identical to our own and that from
\citet{Brown07} is very similar, ensuring that we are selecting red
sequence galaxies in the same way.

The NDWFS red galaxy LFs are consistent with those from COMBO-17
\citep{Bell04} and DEEP2 \citep{Willmer06,Faber07} but are derived
over a much larger area and have a very well determined bright end.
The NDWFS LF is only computed in the rest-frame \mb-band and so we
compare it to the EDisCS LF also computed in the rest-frame \mb-band.
The MUSYC LFs are statistically consistent with the NDWFS LFs at
bright magnitudes but go significantly deeper and have LFs in the the
\gs, \rs, and \is-bands.  In both field samples, the Schechter
function fits to the field data have been performed using a
maximum-likelihood technique with $\alpha$ as a free parameter and the
plotted points are the $1/V_{\rm max}$ measurements of the LF which
include the contribution from field-to-field variance for the MUSYC
data.

\begin{figure}
\epsscale{1.2}
\plotone{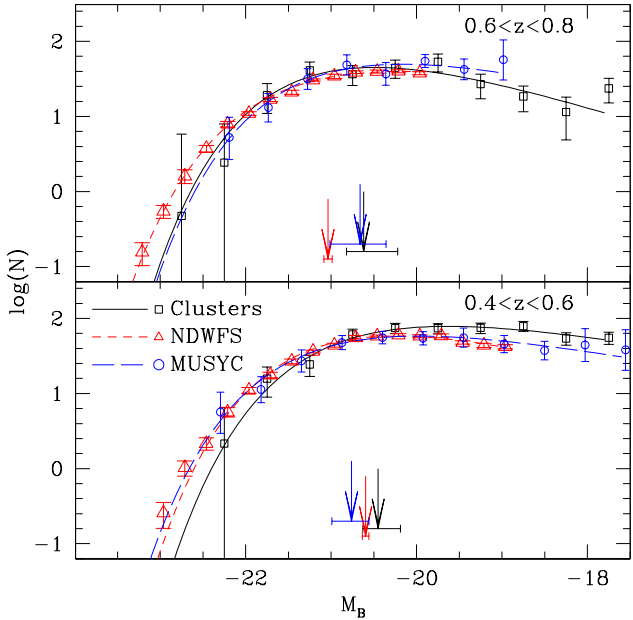}
%\vspace{-2.5in}
\caption {
A comparison of the \mb-band LFs and Schechter function fits for the
composite LF of red sequence galaxies in EDisCS clusters and the LF of
red sequence galaxies in the field as determined from the NDWFS (Brown
et al. 2007) and from MUSYC, in two redshift bins.  The vertical arrows
give the values of \mstar\ and its 68\% confidence limits.  The upper,
middle, and lower arrow refer to the MUSYC, EDisCS, and NDWFS LFs
respectively.  We have used the NDWFS Schechter fits that allow
$\alpha$ to vary.  The field LFs have been scaled vertically to have
the same total luminosity density as the EDisCS LF. }
\label{ediscs_ndwfs_comp_fig}
\end{figure}

\begin{figure}
\epsscale{1.2}
\plotone{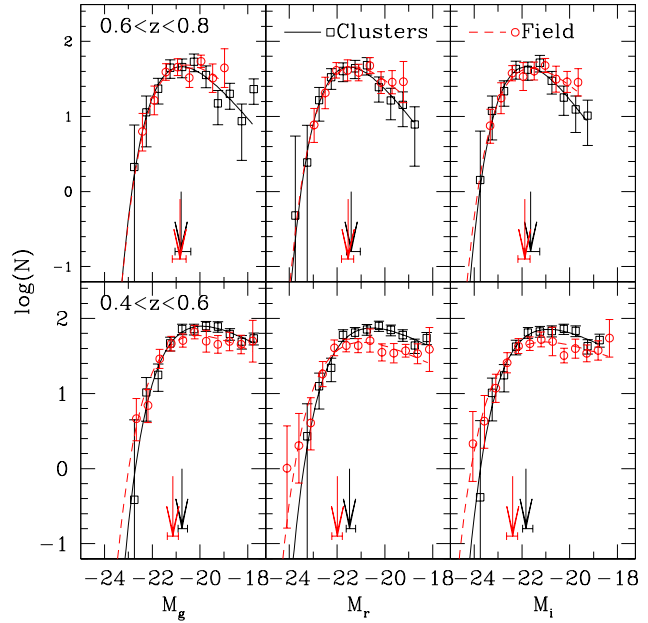}
%\vspace{-2.5in}
\caption {
A comparison of the \gs, \rs, and \is-band LFs and Schechter function
fits for the composite LF of red sequence galaxies in EDisCS clusters
and the LF of red sequence galaxies in the field (as determined from
the MUSYC survey), in two redshift bins.  The vertical arrows give the
values of \mstar\ and its 68\% confidence limits.  The upper and lower
arrow refer to the EDisCS and MUSYC survey respectively.  The field LF
has been scaled vertically to have the same total luminosity density
as the EDisCS LF.}
\label{ediscs_musyc_comp_fig}
\end{figure}

 We compare the field LFs to those from EDisCS in
 Figure~\ref{ediscs_ndwfs_comp_fig} and \ref{ediscs_musyc_comp_fig}.
 In both figures for display purposes we have normalized the different
 LFs to have the same integrated luminosity.  The bright ends of the
 field and clusters are consistent in all cases.  At fainter
 magnitudes the cluster LF appears be be over abundant compared to the
 field at $0.4<z<0.6$ but this trend reverses itself at $0.6<z<0.8$.
 The reversal is most apparent at the reddest rest-frame wavelengths
 (the right two columns of Figure~\ref{ediscs_musyc_comp_fig}), where
 the MUSYC LF extends to similar depths as the EDisCS LF.

 With regards to the best-fit Schechter function parameters, those
 from the two field surveys agree with each other at better than
 1.5-sigma.  It appears that the field value of \mstar\ is
 systematically brighter than for the clusters in both redshift bins,
 but only at the $\sim 2$-sigma level.  At $0.4<z<0.6$ the faint end
 slope of the clusters is steeper than for the field at the $\sim
 2$-sigma level.  At $0.6<z<0.8$ however, this has reversed, with the
 field having an $\alpha$ that is $\sim 2$ sigma steeper than for
 clusters.  The change in the relative slopes is interesting as, in
 addition to the dependence of \nlf\ evolution on cluster velocity
 dispersion presented in \S\ref{sigsplit_sec}, it provides further
 indication that the buildup of the red sequence happens at different
 rates in different environments.  Further it appears that the
 population of the faint-end of the red sequence happens more quickly
 in clusters than in the field.  This could occur if the main channels
 for the population of the red sequence in the field occur at a
 roughly constant or decreasing rate with increasing cosmic time, as
 may be the case for AGN feedback (e.g. Croton et al. 2006) or
 galaxy-galaxy mergers (e.g. Lotz et al. 2008), while quenching
 processes associated with clusters become more efficient with
 increasing time.  This is plausible in a $\Lambda$CDM universe where
 cluster assembly proceeds most rapidly at late times, implying that
 the processes associated with quenching in clusters would also become
 more efficient as one moves towards lower redshift.  At face value,
 this is consistent with the results of \citet{Desai07}, who found
 that the S0 fraction in clusters from EDisCS and \citet{Fasano00} was
 changing slowly at $z>0.5$ but increased much more rapidly starting
 at $z=0.4-0.5$.  While, the exact time at which the red sequence
 assembly in clusters and the field cross and the time at which the S0
 buildup becomes significant are not very well determined it is
 possible that the difference in redshift is real.  If it is we may
 speculate that the $\sim 0.7$ Gyr difference in time between $z=0.6$
 and $z=0.5$ may reflect an intrinsic delay between morphological
 transformation and the truncation of the SFR.  As an example of such
 a scenario, the SFR may have been truncated during one pass through
 the cluster, while the morphological transformation may have required
 repeated cluster passages to build up the bulge \citep{Christlein04}.

It is worth noting that \citet{GB08} compile many different
measurements of \nlf\ in the field and cluster and find that
\nlf\ evolves more quickly in the field than in the cluster.  It is
difficult to directly compare our results as we do not directly
compare \nlf\ between the field and cluster.  %\citet{GB08} assume the
%same definition of bright and faint galaxies as in DL04 but some of
%the cluster and field studies do not have data down to the faint
%magnitude limit of DL04, in which case only the Schechter function
%fits are integrated past the limit of the data to derive \nlf.  This
%further complicates the interpretation of the \citet{GB08} results in
%comparison to ours.  
We also find that the difference between the
field and cluster is most dramatic at red rest-frame wavelengths,
whereas \citet{GB08} measured their \nlf\ in the rest-frame \mv-band.
It is also curious that the field \nlf\ in \citet{GB08} is observed to
evolve quicker than for clusters but in DL07 and in this work,
high-mass clusters are observed to evolve quicker in \nlf\ than
low-mass clusters (see \S\ref{sigsplit_sec} for a discussion of these
results).  All we can say with some certainty is that the bright ends
of the field and cluster measured LFs agree within our, admittedly,
large errorbars but that the faint ends do not, that this disagreement
increases towards redder rest-frame bands, and that there seems to be
a tentative indication that the direction of this disagreement changes
over the EDisCS redshift range.

 It is also interesting to discuss our results in light of similar
 cluster vs. field comparisons in the local Universe.  Our finding
 that the faint-end slope in clusters is steeper than in the field at
 $0.4<z<0.6$ is in qualitative agreement with the local results using
 spectroscopically defined non-starforming galaxies from 2dF
 \citep{Propris03}.  However, \citet{Propris03} found that the
 clusters have a brighter \mstar\ than the field, which may be
 inconsistent with our results at $0.4<z<0.6$.  This may imply the
 presence of relative evolution in the bright end of the LF between
 the field and the cluster but it is important to keep in mind that
 our composite cluster LF is noisy at the bright end and that the
 local studies used spectroscopic techniques to identify galaxies with
 no star formation.  Keeping this in mind, as we showed in
 Figure~\ref{comp_zsplit_fig} the bright end of the measured EDisCS
 LFs are consistent with pure passive evolution, which would argue
 against a significant increase in the cluster red sequence population
 at the luminous end toward lower redshift.  One possible explanation
 is that red galaxies in the field are younger than those in clusters,
 and will therefore fade by a larger amount toward lower redshift.  A
 useful check of this comes from Fundamental Plane studies.
 \citet{vanderwel05} have shown that the evolution in \mllam{B} for
 massive galaxies in the cluster and field is $\Delta ln (M/L_B) =
 (-1.12\pm0.06)z$ and $(-1.2\pm0.18)z$ respectively.  These correspond
 to $\Delta M_B=(-1.22\pm0.06)z$ and $(-1.30\pm0.20)z$ for cluster and
 field galaxies respectively.  The lack of a difference between the LF
 evolution of red galaxies in the field and in clusters agrees with
 \citet{dokkum07} who have shown that massive ellipticals in clusters
 and the field have luminosity weighted ages that are within $4.1\%$
 ($\approx 0.4$ Gyr) of each other.  \citet{Gebhardt03}, however, find
 a $\sim 2$ Gyr difference in the age of cluster and field ellipticals
 but the analysis of \citet{dokkum07} involves more sophisticated
 dynamical modeling and also corrects for selection effects such as,
 ``progenitor bias'' \citep{dokkum01}.  In any case, this small
 difference between clusters and the field is not enough to explain
 the difference that we see with respect to \citet{Propris03}.
 Another explanation must therefore be found to explain the apparent
 difference between the \mstar\ in the field and clusters at
 intermediate and low redshifts and what it implies for the relative
 evolution of bright ellipticals in these two extremes of environment.
 To better study the relative evolution of the bright end of the LF in
 the field and in the cluster it will be necessary to construct red
 sequence LFs for much larger samples of clusters.

\subsection{The integrated growth of the red sequence}
\label{ml_sec}

\begin{figure}
\epsscale{1.2}
\plotone{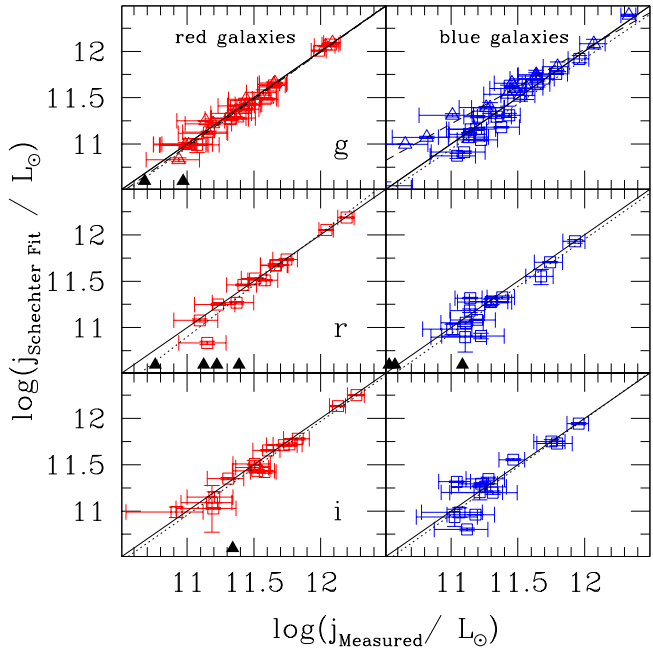}
%\vspace{-2.5in}
\caption {
 A comparison of $j$ for galaxies in EDisCS clusters as computed by
 integrating the measured LF and by integrating the Schechter function
 fits to each cluster.  The left-hand column is for red galaxies and
 the right-hand column is for blue galaxies.  The open squares show
 the measured values with error bars on each quantity for the integral
 of \lfzp\ and the open triangles (only shown in the top panels -
 the \gs-band) are for the integral of \lfss.  The vertical error bars
 are usually smaller than the points.  The solid triangles on the
 bottom of each panel show the clusters with poor Schechter function
 fits.  The solid diagonal line shows the one-to-one relation.  The
 dotted and dashed lines show the least squares fit to the data
 for \lfzp\ and \lfss\ respectively.  The measured LFs seem to have
 converged for all the red LFs and for the blue \lfzp.  We are missing
 some light for the faintest clusters when integrating \lfss.  }
\label{intcomp_fig}
\end{figure}

 In this subsection we explore the buildup of the total amount of
 light on the red sequence as the clusters evolve from $z\sim0.4-0.8$
 to $z\sim0$.  We start by measuring the total light on the cluster
 red sequence, \jcrs, by integrating the measured LFs in each cluster.
 We do this both at $r<0.75$~Mpc and at $r<0.5$\rtwo, which scales
 with \mtwo.  We present the results here for $r<0.75$~Mpc.  The
 choice of aperture for the \jcrs\ computation is somewhat arbitrary,
 but we note that the results using $r<0.5$\rtwo\ are consistent with
 those computed using $r<0.75$~Mpc but the trends are not as
 significant.  In the left-hand column of Figure~\ref{intcomp_fig} we
 compare the total light on the red sequence in each EDisCS cluster
 calculated both using the measured LF down to the magnitude limit of
 each cluster and using the complete integral of the best-fit
 Schechter function.  The right-hand column is the same for blue
 galaxies, which we discuss later in this section.  There are 3-4
 clusters (with the number depending on the rest-frame bandpass) which
 have low measured luminosity and as a consequence also did not have
 well constrained Schechter fits (e.g. CL1227.9-1138 in
 Figure~\ref{all_indiv_fig}).  For the rest of the clusters, however,
 the two measures of \jcrs\ correlate very highly, with a mean offset
 of $3-5\%$ and a scatter of $9\%$.  This demonstrates that our
 observations go deep enough to directly probe almost all of the light
 on the red sequence.  Although the formal errors on the integral of
 the Schechter functions are much smaller than for the measured LF
 integrals, this is primarily because of our assumption of a
 parametric form for the LF and because we assume a well constrained
 faint-end slope, as determined by a fit to the composite LF in each
 redshift bin.  To be as conservative as possible we therefore use
 \jcrs\ derived from the measured LFs for the rest of this discussion.
 This has the added advantage of allowing us to include some of the
 lowest luminosity clusters that had poorly constrained Schechter
 function fits.

 To perform a consistent comparison between low and high redshift
 clusters, and to mitigate any secondary dependences on cluster mass
 we compared our clusters to those in the local universe as a function
 of cluster mass.  We calculate the mass, \mtwo, from the velocity
 dispersion following \citet{Finn05}:

\begin{equation}
{\rm M_{200}} = 1.2\times 10^{15}  (\frac{\sigma}{1000 {\rm km s^{-1}}} )^3\frac{1}{\sqrt{(\Omega_\Lambda + \Omega_M(1 + z)^3)}} h_{100}^{-1} M_\odot.
\end{equation}

\begin{figure}
\epsscale{1.2}
\plotone{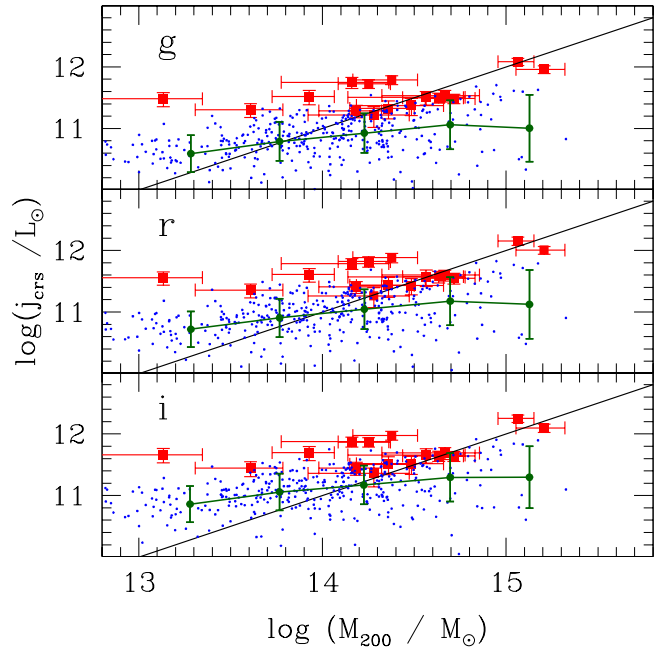}
%\vspace{-2.5in}
\caption {
 A comparison of \mtwo\ vs. \jcrs\ in EDisCS clusters to that from
 clusters in the SDSS, in the rest-frame \gs, \rs, and \is-bands.  The
 large solid squares show the EDisCS clusters, where the horizontal
 error bars show the 68\% confidence intervals on \mtwo\ that stem from
 the uncertainty in the velocity dispersions.  The small dots show the
 values for the SDSS clusters.  The large circles are the geometric
 mean of the individual SDSS \jcrs\ values in different mass bins.
 The solid vertical error bars on the SDSS points show the geometric
 standard deviation in each mass bin.  The diagonal black line shows
 the slope that galaxies will lie on if they have constant \lmtwo.
 The SDSS clusters deviate significantly from this relation indicating
 a real trend of \lmtwo\ with \mtwo.  }
\label{m_l_fig}
\end{figure}

 In Figure~\ref{m_l_fig} we plot \jcrs\ vs. \mtwo\ for the EDisCS
 clusters and for the SDSS clusters.  Even at a fixed mass in the SDSS
 sample there is significant scatter in integrated luminosity.  This
 is intrinsic scatter in the cluster population as $\sim 95\%$ of the
 SDSS clusters have lower than 20\% error on the integrated
 luminosity.  Although we have a large sample of SDSS clusters, the
 size of the local sample, especially at the massive end is the
 dominant uncertainty in the following analysis.  The EDisCS clusters
 lie at brighter luminosities than the SDSS clusters at the same
 cluster mass.  At a basic level this is expected because the red
 galaxies will fade as a function of time, moving the EDisCS points in
 the direction of the SDSS locus.  We will quantify this evolution
 below.

 It is also clear that the SDSS clusters deviate significantly from
 the one-to-one relation between \mtwo\ and \jcrs
 (i.e. constant \lmtwo), which indicates that there is a residual
 dependence of \lmtwo\ on \mtwo.  In this case it appears that more
 massive clusters have a lower \lmtwo\ than less massive clusters.  We
 have confirmed that this trend is not due to the uncertainties on the
 velocity dispersion for SDSS clusters.  A similar deviation from the
 one-to-one relation between \jcrs\ and \mtwo\ exists for the EDisCS
 clusters, such that clusters with low \mtwo\ are brighter than the
 one-to-one relation, but it is not clear how robust this is given the
 small numbers of very low mass clusters.

\begin{figure}
\epsscale{1.2}
%\epsscale{0.75}
\plotone{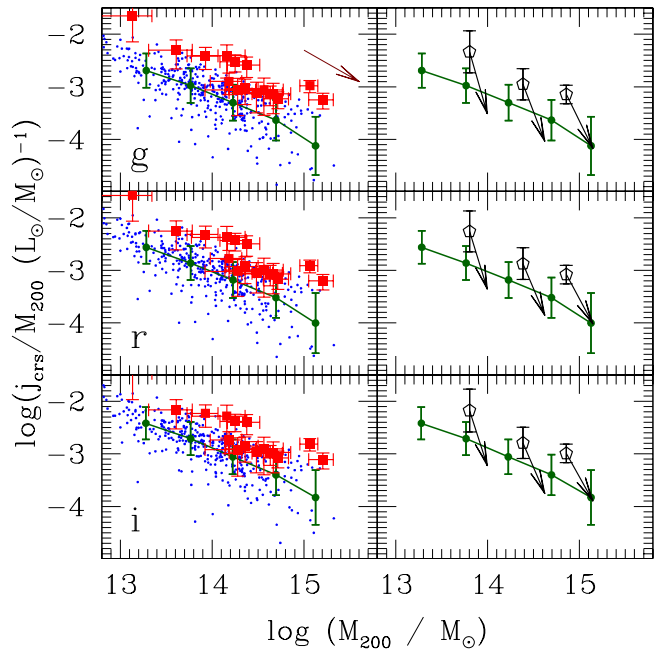}
%\vspace{-2.5in}
\caption {
 A comparison of \lmtwo\ vs \mtwo\ for red sequence galaxies in EDisCS
 clusters to that from clusters in the SDSS, in the rest-frame \gs,
 \rs, and \is-bands.  (\textit{left column}) The large solid squares
 show the EDisCS clusters, where the vertical and horizontal error bars
 show the 68\% confidence intervals on \lmtwo\ and the errors in
 \mtwo\ that stem from the uncertainty in the velocity dispersions,
 respectively.  The small dots show the values for the SDSS clusters.
 The large circles are the geometric mean of the individual SDSS
 \lmtwo\ values in different mass bins.  The solid vertical error bars
 show the geometric standard deviation.  Note that the errors in
 \lmtwo\ and \mtwo\ are correlated.  The arrow in the upper left panel
 demonstrates the change in \lmtwo\ that results from a factor of four
 change in \mtwo.  (\textit{right column}) The circles and error bars
 are the same as in the left column.  The large open pentagons show
 the geometric mean of the EDisCS clusters in three mass bins chosen
 to contain roughly equal numbers of clusters, with the vertical
 error bars showing the geometric standard deviation for each value.
 The horizontal position of each pentagon is determined by the
 geometric mean of the masses for the EDisCS clusters in that mass
 bin.  The arrows show the expected evolution from \zclust\ to $z=0$
 for a model in which the cluster mass increases by accretion but
 where no new galaxies are added to the red sequence and where those
 that are already present at the epoch of observation fade passively
 as SSPs with $z_f=2$.}
\label{ml_m_fig}
\end{figure}

\begin{figure}
\epsscale{1.2}
\plotone{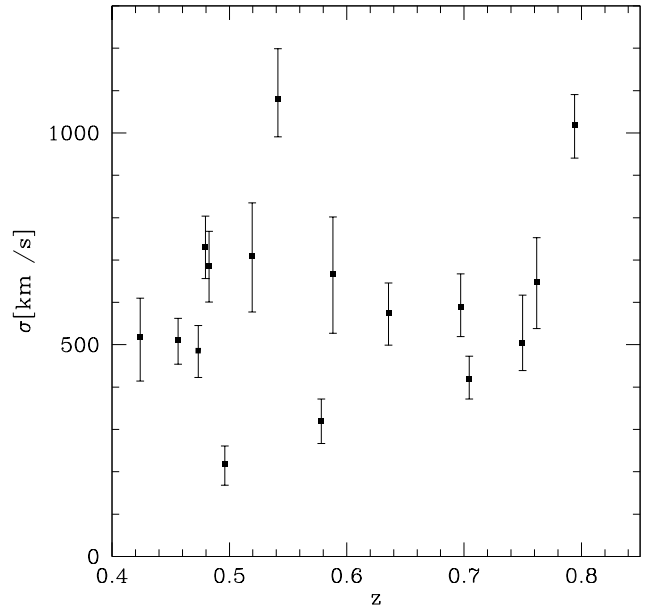}
%\vspace{-2.5in}
\caption {
 The velocity dispersions for our 16 clusters as a function of
 redshift.  A Spearman Rank test gives a 4\% probability that sigma is
 correlated with redshift.  }
\label{sigz_fig}
\end{figure}

 In Figure~\ref{ml_m_fig} we plot \lmtwo\ vs. \mtwo\ for the EDisCS
 and SDSS clusters.  In the left column we plot the individual values
 and in the right column we show the geometric mean \lmtwo\ of the
 EDisCS clusters in three mass bins and compare them to the SDSS
 clusters.  The mass bins were chosen to contain approximately similar
 numbers of objects.  We do not have enough clusters to bin both in
 mass and in redshift, but there is no dependence of sigma and
 redshift in our sample (see Figure~\ref{sigz_fig}) and a Spearman
 rank test gives only a 4\% probability that the two values are
 correlated, i.e. our mass bins should not be affected by secondary
 correlation of cluster mass with redshift.  Again, as in
 Figure~\ref{m_l_fig} it is clear that the \lmtwo\ values for the
 EDisCS clusters are larger than for the SDSS clusters.

\begin{figure}
\epsscale{1.2}
\plotone{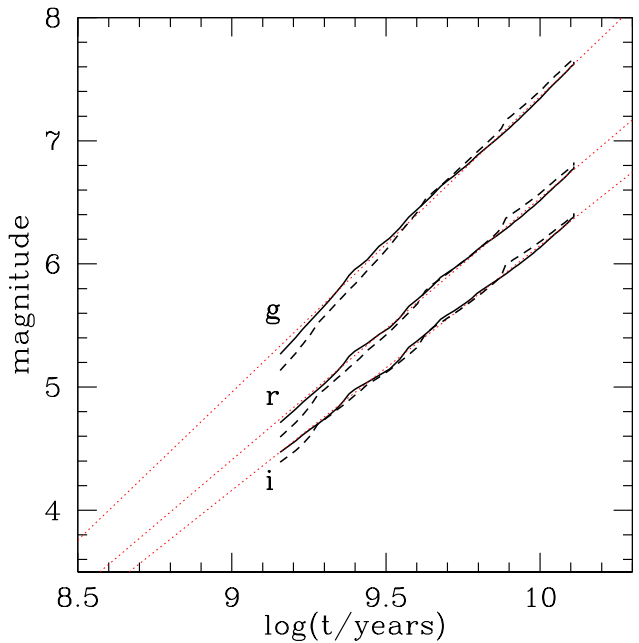}
%\vspace{-2.5in}
\caption { The fading as a function of time for simple stellar
  populations in the \gs, \rs, and \is-bands.  The zeropoint of the
  y-axis is arbitrary.  The thick solid line and thick dashed lines
  show the evolution in magnitude for SSP models with Z=Z$_\odot$ and
  2.5Z$_\odot$ respectively.  The thin dotted line is the linear fit
  to the solar metallicity model.  The 2.5Z$_\odot$ model is brighter
  in the mean by 0.2-0.4 magnitudes at these ages and we have
  subtracted out the difference to highlight the similar slopes.  At
  $(t)>1.4$ Gyr the fading in magnitude is well approximated by a
  linear relation in magnitude and log$(t)$.  }
\label{magfade_fig}
\end{figure}

\begin{figure}
\epsscale{1.2}
\plotone{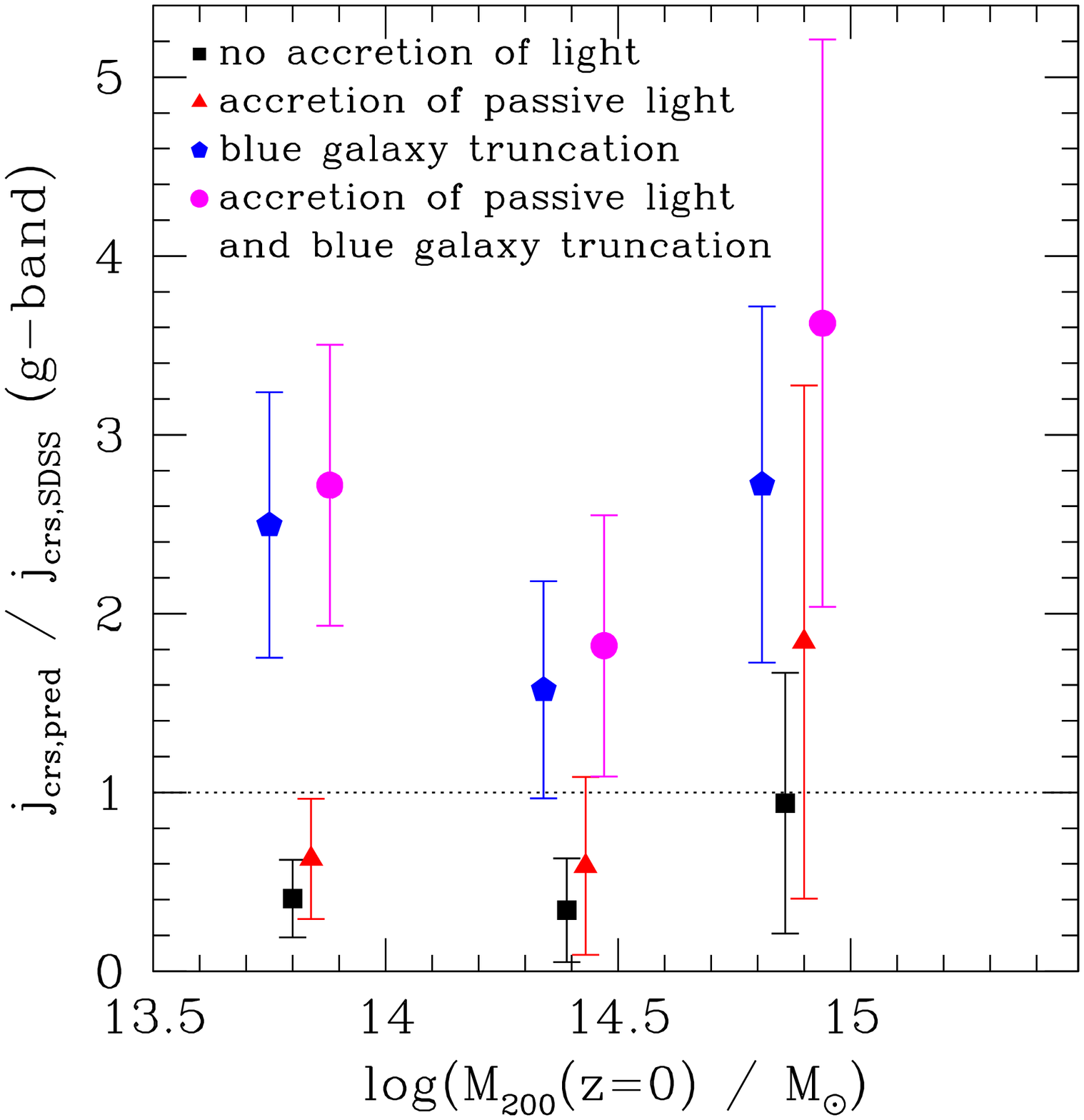}
%\vspace{-2.5in}
\caption {
 The ratio of the predicted red sequence \gs-band luminosity, \jpred,
 of EDisCS clusters at $z=0$ to the measured \jcrs\ from SDSS
 clusters, as a function of cluster mass.  We show the predicted
 values for 4 toy models described in the text.  The squares show a
 model in which only mass is added to the clusters but in which no new
 galaxies are added to the red sequence and those that exist at the
 epoch of observation evolve passively.  The triangles show what
 happens when galaxies are added onto the red sequence with the same
 \mtotl\ and its evolution as cluster galaxies.  The pentagons show a
 model in which no additional red galaxies are accreted into the
 cluster but in which the blue galaxies in the clusters at $z=0.6$ are
 assumed to have constant SFRs prior to the epoch of observation but
 subsequently have their SFR truncated and evolve passively
 thereafter.  The circles show a model in which both the blue galaxies
 in the clusters are added to the red sequence and in which red
 passively evolving galaxies are accreted from the field.  In all
 models the error bars account for the dispersion in \jcrs\ values in
 the EDisCS data.  In the third and fourth models the error bars also
 account for the different \jbg\ for different membership techniques
 and for the range of possible SFH.  In each mass bin the points have
 been offset in mass for clarity.  }
\label{extra_lum_fig}
\end{figure}

 Even though the EDisCS clusters have higher \jcrs\ for their
 \mtwo\ at $z\sim 0.6$ they must by definition evolve by $z\sim0$ onto
 the local relation defined by the SDSS clusters.  To explore what
 this constraint implies for the build-up of light (or mass) on the
 red sequence we have constructed a set of 4 toy models which we
 describe below.  In all models we assume that two processes are
 universally present.  First, the mass of clusters will grow via
 accretion of matter from the surrounding cosmic web.  Second, the
 galaxies on the red sequence at the epoch of observation will only
 fade as a function of time, i.e. we assume that galaxies on the red
 sequence are passive and stay on the red sequence.  We account for
 the growth of clusters by tracking the median growth in mass for
 clusters of a certain mass using the results of \citet{Wechsler02}
 and \citet{Bullock01}\footnote{We computed this using programs
   obtained from
   \texttt{http://www.physics.uci.edu/\~{}bullock/CVIR/}}$^,$\footnote{Note
   that our highest mass clusters will evolve into objects that are so
   rare as not to be present at all in the SDSS C4 sample.  This is
   also the fate of most of the numerous massive X-ray clusters at
   $z>0.5$ that are found in the literature.}.  To estimate the fading
 of cluster red sequence galaxies we fit a linear relation to the
 change of magnitude as a function of log(time) for each bandpass.  In
 Figure~\ref{magfade_fig} we show the fading in magnitude for models
 of two different metallicities\citep{BC03}.  A linear fading of
 magnitude with log(age) is a very good approximation for SSPs with
 age$>1.4$Gyr, as was shown originally by \citet{Tinsley80}.  Models
 with Z=Z$_\odot$ and 2.5Z$_\odot$ differ in the amount of fading by
 $<0.13$ magnitudes over the $5.7$Gyr period from $z=0.6$ to $z=0$.
 We adopt the Z=Z$_\odot$ model but our results are insensitive to
 this choice.  In tracking the fading stellar populations we evaluate
 whether galaxies fade below the absolute magnitude limit that we
 adopt for the SDSS.  The first, simplest, and most unrealistic toy
 model is one in which only mass is accreted but no light is added to
 the red sequence of the cluster.  The light in the cluster therefore
 decreases by pure fading.  In the right-hand panel of
 Figure~\ref{ml_m_fig} we demonstrate how this model would cause the
 EDisCS clusters to evolve from \zclust\ down to $z=0$.  This model
 results in predicted \lmtwo\ values at $z=0$ that are too low
 compared to the SDSS.  In Figure~\ref{extra_lum_fig} we demonstrate
 this in another way by plotting the ratio of the predicted $z=0$
 \gs-band red sequence luminosity, \jpred, to the observed SDSS red
 sequence luminosity \jsdss.  From this figure it is clear that this
 first model results in predicted $z=0$ luminosities that are too
 faint by a factor of $\sim 1-3$ depending on the mass range, although
 formally the required mass growth implied in all three mass bins are
 consistent with each other.  The red sequence light in clusters must
 therefore grow by a similar factor from $z\sim 0.6$ to the $z\sim 0$.
 For clusters with \mtwo$<10^{14.6}$\msol the required growth is a
 factor of 2.5--3.  Because light and mass are proportional for red
 sequence galaxies, this also implies that the stellar mass on the red
 sequence needs to therefore grow by a factor of $1-3$ with a growth
 of 2.5--3 required at \mtwo$<10^{14.6}$\msol.  That the largest
 implied growth may come from clusters of low to intermediate masses
 may be consistent with the results of \citet{Poggianti06}, who find
 that the star-forming fraction in clusters at $z<0.8$ evolves most
 rapidly at intermediate cluster velocity dispersions (\mtwo) but
 evolves little at the highest velocity dispersions.

 To explore different scenarios for how additional light (and stellar
 mass) may be added to the red sequence we therefore consider a second
 model in which the cluster mass growth is accompanied by the
 accretion of stars with the same $M_{\rm tot}/L$ and its expected
 evolution, where $M_{\rm tot}/L$ is derived for each cluster
 individually using the total light in red sequence cluster galaxies.
 In other words, this conservative model assumes that the only
 galaxies added to the red sequence are red sequence field galaxies
 with the same ages and SFHs as cluster ellipticals and with the same
 proportion of light relative to the total mass.  This model would be
 consistent, for example, with the very small age differences ($\sim
 4\%$) found between cluster and field ellipticals by
 \citet{dokkum07}.  Assuming all galaxies that fall in to clusters at
 $z<0.6$ end up on the red sequence at $z=0$ this model represents the
 minimum amount of light that can be added to the cluster red sequence
 by accretion from the field.  The results of this model are shown as
 the solid triangles in Figure~\ref{extra_lum_fig}.  As expected these
 models yield higher predicted $z=0$ luminosities for the EDisCS
 clusters, and are more consistent with the expectations for SDSS
 clusters.

 It is interesting to discuss the predictions of this model in
 relation to the roughly factor of two growth in mass on the red
 sequence inferred from field studies \cite{Bell04,Faber07,Brown07}.
 In interpreting this it is important to remember that the ``field''
 surveys contain a range of environments, including moderate mass
 clusters for the largest surveys like the NDWFS and extending to
 massive groups for the MUSYC, DEEP-2, and COMBO-17 surveys.  The
 observed mass growth on the red sequence in field surveys represents
 the actual transformation of blue galaxies to red galaxies.
 Clusters, on the other hand, are growing their total mass by a factor
 of $\sim 2$ at $z<0.6$ and, at least partly, will be increasing the
 total amount of red light merely via the accretion of red galaxies
 from the field as in model 2 (above).  Thus, the total increase that
 we infer in the mass on the red sequence is consistent with only a
 moderate additional transformation of blue to red galaxies.
 Nonetheless, our value for the required mass growth is still rather
 uncertain due to the significant intrinsic scatter both in the SDSS
 and EDisCS cluster \lmtwo\ values.  Also, our middle and lowest mass
 bins (\mtwo$<10^{14.6}$\msol) imply a factor of $\sim 2.5-3$ growth
 in the red sequence stellar mass at $z<0.8$, which may exceed the
 observed growth in the field and the expected total mass growth in
 clusters.  As we show in \S\ref{field_sec} there is differential
 evolution in the shapes of the LFs in the field and clusters and this
 does imply that clusters evolve more rapidly in the number of
 galaxies on the red sequence.  There is also evidence from
 \citet{Poggianti06} that star forming galaxies are being truncated
 preferentially in cluster environments but \citet{Finn08} find that
 the decline in the SFRs of cluster galaxies at $z<0.8$ is comparable
 to that in the field.  While it is difficult to draw precise
 conclusions about the necessary mass transfer to the red sequence, it
 is clear that a pure passive fading of the cluster red sequence seen
 at $z= 0.4-0.8$ will result in clusters that are systematically too
 faint compared to those seen locally.  Within the large uncertainties
 in the model predictions there is no significant trend with cluster
 mass, which is also true for each of the following models.

 As a third model we calculate how much light is added to the EDisCS
 cluster red sequences by $z=0$ by all of the blue galaxies in the
 clusters at $z\sim0.4-0.8$, with no additional infall of red galaxies
 (pentagons; Figure~\ref{extra_lum_fig}).  In calculating this
 estimate we take into account the uncertainties resulting from the
 differences between \lfzp\ and \lfss\ that we determined in
 \S\ref{methodcomp_sec}.  Indeed, the total luminosity of blue
 galaxies in the \gs-band, \jbg, determined from \lfss\ ranges from
 0.7 to 5.7 times larger than \jbg\ determined from \lfzp, with a
 median of 1.8, as calculated over all clusters.  In addition, as is
 shown in the right-hand column of Figure~\ref{intcomp_fig} the blue
 galaxy \gs-band \lfss\ has not converged completely for low
 luminosity clusters due to the very steep faint-end slope.
 Nonetheless, for most of the clusters the missing light below our
 magnitude limit is small and \jbg\ as derived from \lfss\ and
 \lfzp\ should still bracket the true value of \jbg.  For every
 cluster we therefore use the mean of \jbg\ as determined from
 \lfss\ and \lfzp\ and the values for the two methods as an indication
 of the uncertainty in \jbg.  Because we can only calculate both
 \lfss\ and \lfzp\ for the \gs-band (see \S\ref{restlum_sec}) we limit
 our analyses for the following models to that single bandpass.  We
 then assume that the galaxies in each cluster have been forming stars
 constantly prior to the epoch of observation and that they have
 achieved solar metallicity by the time they are observed.  We assume
 that they have $A_V=1$, which corresponds to $A_g=1.17$.  Subsequent
 to observation we assume that these galaxies continue forming stars
 for 1 Gyr before abruptly ceasing their star formation and losing
 their dust.  Except for the extra extinction, this is similar to the
 delayed truncation model of DL07 and is consistent with the evolution
 in the shape of the LF from $z=0.8$ to 0.  For every cluster we
 calculate the expected luminosity contribution that these blue
 galaxies make to \jpred\ assuming that the galaxies were forming
 stars for 3--6 Gyr prior to observation.  As we can see in
 Figure~\ref{extra_lum_fig} this simple model over predicts the amount
 of light on the red sequence in local clusters, especially so for the
 highest and lowest mass bins.

 In reality, galaxies must be accreted onto the cluster over time and
 we therefore consider a fourth model in which both old ellipticals
 are accreted onto the cluster (model 2) and blue galaxies within the
 cluster are transformed (model 3).  Being the sum of models 2 and 3,
 this fourth model naturally also over predicts the predicted light in
 local clusters, by a factor of 1.8--3.6.  In understanding why this
 last model over predicts the local luminosity of clusters it is
 important first to remember that these models are in many ways very
 conservative in how much light is added to the red sequence by $z=0$.
 The amount of mass that our clusters accrete is determined by
 ${\rm \Lambda}$CDM and for this accreted mass we add the smallest
 possible amount of light to the red sequence by only accreting old
 galaxies at the same \lmtwo\ as the cluster.  Accreting galaxies that
 are not as old as cluster ellipticals (e.g. blue galaxies) will
 increase the luminosities of the clusters by $z=0$ with respect to
 our fourth model.  In addition, there is little way to avoid the
 transformation of blue cluster galaxies at $z=0.6$ to red and dead
 ones by $z=0$ so our third model should also be valid.  Because we
 accrete as little light as possible for the expected mass accretion,
 our fourth model can be thought of a lower limit on the amount of
 light added to the cluster (but see below).  In light of these rather
 conservative assumptions it is perhaps puzzling that the simple model
 nonetheless produces too much cluster red sequence light at $z=0$.
 As a note, the amount by which model 3 and 4 over predict local
 luminosities of clusters when computing \jcrs\ at $r<0.5$\rtwo\ is
 consistent with, but not as large as when computing \jcrs\ at
 $r<0.75$~Mpc.

 There are at least three possible resolutions to this apparent
 discrepancy.  First, it may be that clusters accreted significant
 light in galaxies that never enter the red sequence by $z=0$,
 i.e. from star-forming galaxies that have not had time to cease star
 formation and redden since they fell in.  In our SDSS sample we also
 computed the blue luminosity function and find that $26\pm3\%$ of the
 total cluster light comes from galaxies bluer than the red sequence.
 Reducing the contribution to the final red sequence luminosity by
 this amount corrects for our assumption that all accreted galaxies
 end up on the red sequence.  This brings the fourth model into closer
 agreement with local cluster luminosities but still systematically
 over predicts them.

 A second possibility is that blue star-forming galaxies at $z=0.6$
 that enter the red sequence at $z=0$ may still be dust enshrouded,
 lowering their total observed luminosity, something not encapsulated
 in our simple models.  Indeed \citet{Wolf05} find that roughly $1/3$
 of red sequence galaxies in the Abell 901/902 supercluster are dusty
 star forming galaxies with $\langle A_V\rangle \approx 0.6$.  If the
 same fraction of dusty red galaxies is present in all clusters, and
 under the simplifying assumption that the fraction of dusty galaxies
 is independent of galaxy luminosity, then this extinction changes
 the \jcrs\ by a factor of 0.87.  Taken alone this is obviously too
 small to make a significant contribution to reconciling model 4 with
 the local \jcrs\ values.  However, combining this with the amount of
 light from accreted galaxies that never make it onto the red sequence
 (see previous paragraph) would change \jpred\ by a factor of 0.64,
 which still results in \jpred\ values that are systematically a
 factor of 1.15--2.3 too high compared to local clusters, but are
 consistent within the 68\% confidence limits.

 Thirdly, it is possible that a substantial fraction of stars / mass
 is in cluster components other than red sequence galaxies, and is
 thus neglected in our measurement of the LF.  For example, our LFs
 exclude the BCG and the intracluster light (ICL) which could harbor a
 significant fraction of the total stellar mass and may evolve
 differently than the red galaxies.  Indeed, cluster galaxies are
 expected to sink to the cluster center and merge with the BCG
 \citep[``cannibalism''][]{ostriker75,white76}, a phenomenon that has
 been observed in some low redshift BCGs\citep{Lauer88}.  Cannibalism
 may be especially relevant for the most massive cluster galaxies, for
 which the dynamical friction timescale is similar to the timescale we
 are probing here (a few Gyrs). Hence, the extra mass in stars our toy
 model predicts could simply have been accreted onto the BCGs.
 However, \citet{Whiley08} found that the properties of the EDisCS and
 SDSS BCGs are consistent with passive evolution since $z\sim2$,
 implying no appreciable BCG mass growth. This result seems at odds
 with the factor 3--4 mass growth of BCGs predicted by simulations
 \citep{DeLucia07b}. One needs to keep in mind, however, that
 \citet{Whiley08} considered only the central 37~kpc of each BCG and
 that any mass accreted in mergers must predominantly be accreted onto
 the envelope of the BCG and/or the ICL.  This is also a viable
 possibility to explain the discrepancy between our toy model
 prediction for the mass on the red sequence: rather than remaining in
 the galaxies we see in the cluster at $z\sim0.6$, a significant
 fraction of old stars could be part of the ICL at $z=0$, and are thus
 not accounted for in our LFs.  There are various mechanisms by which
 cluster galaxies may get disrupted and lose their stars to the ICL:
 e.g. stripping due to tidal forces in the cluster gravitational field
 \citep{Merritt84}, or galaxy harassment \citep{Farouki81,Moore96}.
 Indeed, the mass in the ICL has been measured in a number of
 low-redshift clusters, e.g. \citet{Gonzalez07} find that the
 BCG$+$ICL contribute 20\%$-$40\% of the stellar light within
 $R_{500}$ in clusters of the range in velocity dispersion we are
 considering here. This result is consistent with that of
 \citet{Zibetti05}, who found that the BCG$+$ICL contribute $\sim$30\%
 of the stellar light in stacked SDSS clusters. Furthermore, the color
 of the ICL is comparable to, or even slightly redder than, the total
 color of cluster galaxies \citet{Zibetti05}, implying that the ICL
 may have originated from red sequence galaxies.  \citet{Gonzalez05}
 use the position angle and ellipticity for a set of lower redshift
 clusters to decompose the BCG and ICL and find that 80\% of the light
 on average comes from the ICL.  We do not know the ICL contribution
 in our high redshift clusters, but if we assume that 20\% of the
 light that would be on the red sequence at $z=0$ ends up in the
 BCG$+$ICL this would move \jpred\ for the fourth model even further
 into agreement with local values, implying that no new processes are
 needed to reconcile the red sequence luminosities of clusters at
 $0.4<z<0.8$ with those locally.

 \citet{Gonzalez07} find that the fraction of light in the BCG+ICL
 decreases with increasing cluster velocity dispersion.  \citet{Lin04}
 measure a BCG magnitude that may include some ICL and they find that
 the luminosity fraction also decreases with increasing cluster mass.
 In both cases the trends have a large scatter at velocity dispersions
 corresponding to our clusters and it is possible that a trend is
 present in our data but masked by the large scatter within our own
 sample.  If such a trend exists it is possible that the BCG and ICL
 build up at different rates and with a different response to the
 accretion history of the cluster.  Hopefully more progress will be
 made with future high resolution simulations of clusters in a
 cosmological context, and direct measurements of the high redshift
 ICL.

\section{Summary and Conclusions}

 In this paper we have measured the rest-frame optical LFs for 16
 clusters at $0.4<z<0.8$ that are drawn from the ESO Distant Cluster
 Survey (EDisCS).  These clusters have a range in velocity dispersions
 and, in contrast to massive x-ray selected high redshift clusters,
 are the progenitors of ``typical'' clusters in the local
 Universe \citep{Jensen08}.

 We determined membership for our clusters using a
 photometric-redshift-based technique and one based on statistical
 background subtraction.  From a detailed comparison of these two
 methods we concluded that the LF could only be robustly determined
 for red sequence galaxies and that the two methods resulted in very
 different \mstar\ and $\alpha$ values for blue galaxies.  We
 therefore focus on the LFs of red sequence galaxies.  We computed
 individual LFs for our clusters and composite LFs for the whole
 sample as well as for sample split into subsets by redshift at
 $z=0.6$ and velocity dispersion at $\sigma=600$~km/s.  For the
 individual and composite LFs we fit Schechter functions, where we
 fixed the faint-end slope to the value determined from the fit to the
 composite of all EDisCS clusters in two different redshift bins.

 As a low redshift comparison sample we used a cluster catalog drawn
 from the SDSS and calculate the composite LF and its Schechter
 function fit as for the EDisCS clusters.  When splitting the SDSS
 sample into bins of velocity dispersion we take into account the
 average mass growth in clusters as expected from numerical
 simulations.  In this way we can compare clusters at high redshift to
 (representatives of) their likely descendants at low redshift,
 something that has not been possible with previous LF studies that
 either concentrated on very massive high redshift clusters - whose
 descendants would be largely absent from local volumes - or that had
 no velocity dispersion information for the clusters.

 We measure significant evolution in the LF of cluster red sequence
 galaxies at $z<0.8$.  In detail the LFs show evolution in the bright
 end consistent with passive evolution but show a dramatic increase in
 the number of faint galaxies relative to bright ones toward lower
 redshifts, both within our own survey and when compared to the SDSS
 cluster sample.  As a simple characterization of this evolution we
 measure the ratio of luminous to faint galaxies as in
 \citet{DeLucia07} and find similar results.  We also measure the
 build-up of the red sequence as a more detailed function of magnitude
 and find tentative evidence for an evolving magnitude threshold
 brighter than which the LF is in place w.r.t. the local LF.  It is
 not clear if this evolving magnitude threshold is in any way related
 to the evolving mass threshold seen in field samples, above which
 star formation is truncated (e.g. Bundy et al. 2006), or if it
 corresponds to a different cluster related quenching mechanism.
 Indeed, the late build-up of the faint red sequence in clusters may
 also be related to the increase in the S0 fraction seen toward lower
 redshift.

 We perform Schechter function fits to our LFs and find significant
 evolution in $\alpha$ but no evolution in \mstar, despite finding
 that the measured LFs at the bright end are consistent with passive
 evolution.  This highlights the complications of using \mstar\ as a
 measure of the evolution in the luminosity of the galaxy population
 as a whole when $\alpha$ is also simultaneously undergoing strong
 evolution.  In our case it must be that luminosities of the whole
 galaxy population are not evolving in lock step.

 We split our sample into two bins of velocity dispersion and find only
 small differences in the detailed LFs, although we find the same
 result as DL07 that the ratio of luminous to faint galaxies is higher,
 and evolves more quickly, in clusters of higher velocity dispersion.
 We find indistinguishable luminous-to-faint ratios for SDSS clusters
 of different velocity dispersion, similar to that found by
 \citet{Propris03}.

 We looked for radial trends by examining the EDisCS LFs computed at
 $r<0.75Mpc$ and $r<0.5Mpc$ and find no difference in either
 \mstar\ or $\alpha$ for red sequence galaxies.  This comparison,
 however, is uncertain as the two radial bins are highly correlated.

 Using the field LF of red sequence galaxies measured from the
 NDWFS \citep{Brown07} and MUSYC we compared our cluster LFs to the
 coeval field LF for similarly selected galaxies.  At $0.6<z<0.8$ the
 field has more faint galaxies relative to bright ones than the
 clusters but at $0.4<z<0.6$ this has reversed, with the clusters
 having more faint galaxies than the field.  This epoch is similar to
 that in which the buildup of the S0 population in clusters starts to
 become significant \citep{Desai07}.  Combined with the more rapid
 evolution of \nlf\ for high velocity dispersion clusters, the
 different rates of evolution in the LFs imply that dense environments
 are more efficient than the field at adding galaxies to the red
 sequence at $z<1$.  These trends in the ratio of luminous to faint
 galaxies are reflected in the Schechter function fits.  At both
 redshifts the EDisCS LF has a more negative $\alpha$ than the field
 but a slightly fainter \mstar.  While the former agrees with local
 cluster-field comparisons from 2dF, the latter disagrees.
 Discovering the cause of this discrepancy in the relative \mstar\
 values will require larger samples of clusters at intermediate
 redshift to increase the signal-to-noise of the LF at high
 luminosities.

 To constrain different mechanisms for building up the red sequence
 galaxy population we measure the total red sequence light in the
 EDisCS and quantify its evolution w.r.t. clusters from SDSS.
 Clusters at high redshift are overluminous compared to their likely
 local descendants.  Once passive fading is
 accounted for it appears that the clusters are a factor of
 $1-3$ underluminous compared to the local clusters that are their
 likely descendants.  Since light traces stellar mass on the red
 sequence, this implies that the mass on the red sequence in clusters
 must grow by a factor of $1-3$ at $z<0.8$, with most of the
 growth occurring at the faint end of the LF where we directly witness
 strong evolution.  This is a similar amount of growth in the red
 sequence as inferred from studies of the field
 LF \citep{Bell04,Faber07,Brown07} and indicates that the additional
 transformation of blue galaxies to red galaxies in clusters may be
 modest.  However, due to the significant uncertainties we cannot
 determine if the total amount of mass added depends on cluster
 velocity dispersion.  Evidence for environmental effects come
 predominantly in the dependence of the shape evolution of the LF on
 cluster velocity dispersion and in differences between the cluster
 and field.

 To explore what physical mechanisms may be driving the assembly of
 the red sequence we explore a set of simple toy models that
 incorporate many of the processes that should add galaxies to the red
 sequence in clusters in a conservative manner.  Accounting for all
 necessary processes we find that these models overpredict the light
 in local clusters.  The model predictions can be reconciled with the
 data by a combination of three previously known processes: blue
 galaxies that have fallen in since $z<0.6$ but are not on the red
 sequence at $z=0$, attenuation of light on the red sequence by dust
 extinction, and the transfer of stars from galaxies to the diffuse
 cluster light via tidal stripping.

 The results presented in this paper were only made possible with a
 large sample of clusters that span a range of redshift and velocity
 dispersion, that have accurately measured velocity dispersions, and
 that have deep multi-wavelength photometry over a significant
 fraction of the virial radius.  Our analysis was limited in large
 part by our reliance on photometric methods to isolate cluster
 members, and by the limited number of clusters in our sample.  To
 improve upon the analysis several ingredients are needed.  First, we
 need more clusters over a large range of mass, to confirm the mass
 dependent assembly of the red sequence.  Second, we need to move to
 larger radii so that we can probe out past \rtwo\ and thereby include
 many more galaxies in our sample.  Third, we need increased
 spectroscopy of blue cluster members to allow a robust luminosity
 function determination for galaxies of all colors.  This last element
 would also be assisted if we had wide fields as we would then be able
 to do statistical background subtraction using ``field'' samples with
 identical photometry and roughly co-spatial on the sky, thus
 bypassing many of the problems inherent in using external fields for
 background estimation.

\acknowledgements

GR would like to thank Casey Papovich, Jennifer Lotz, Daniel
Eisenstein, and Mark Dickinson for useful discussions during the
writing of this paper.  GR would also like to acknowledge the support
of the Leo Goldberg Fellowship during his time at NOAO.  The Dark
Cosmology Centre is funded by the Danish National Research Foundation.

\ifemulate
	\begin{deluxetable*}{cccc}
\else
	\begin{deluxetable}{cccc}
\fi
\tablecaption{Zeropoints of fits to red sequence}
\tablewidth{0pt}
\tablehead{\colhead{${\rm Cluster}$} & \colhead{$z$} & \colhead{${\rm ZP}_{V-I;I}$} & \colhead{$\sigma_{\rm ZP}$}\\
\colhead{$$} & \colhead{$$} & \colhead{${\rm mag}$} & \colhead{${\rm mag}$}}
\startdata
CL1018.8-1211 & $ 0.47 $ & $  3.45 $ & $  0.10 $ \\ 
CL1037.9-1243 & $ 0.58 $ & $  3.65 $ & $  0.08 $ \\ 
CL1040.7-1155 & $ 0.70 $ & $  3.81 $ & $  0.09 $ \\ 
CL1054.4-1146 & $ 0.70 $ & $  3.86 $ & $  0.06 $ \\ 
CL1054.7-1245 & $ 0.75 $ & $  4.08 $ & $  0.11 $ \\ 
CL1059.2-1253 & $ 0.46 $ & $  3.42 $ & $  0.07 $ \\ 
CL1138.2-1133 & $ 0.48 $ & $  3.44 $ & $  0.11 $ \\ 
CL1202.7-1224 & $ 0.42 $ & $  3.34 $ & $  0.08 $ \\ 
CL1216.8-1201 & $ 0.79 $ & $  4.00 $ & $  0.11 $ \\ 
CL1227.9-1138 & $ 0.64 $ & $  3.77 $ & $  0.06 $ \\ 
CL1232.5-1250 & $ 0.54 $ & $  3.58 $ & $  0.16 $ \\ 
CL1301.7-1139 & $ 0.48 $ & $  3.49 $ & $  0.12 $ \\ 
CL1353.0-1137 & $ 0.59 $ & $  3.67 $ & $  0.12 $ \\ 
CL1354.2-1230 & $ 0.76 $ & $  4.03 $ & $  0.02 $ \\ 
CL1411.1-1148 & $ 0.52 $ & $  3.50 $ & $  0.15 $ \\ 
CL1420.3-1236 & $ 0.50 $ & $  3.51 $ & $  0.09 $ \\ 
\enddata
\label{cmrfit_tab}
\tablecomments {Zeropoints of $V-I$ vs. $I_{\rm tot}$ color magnitude relation are calculated for spectroscopically defined non-starforming galaxies.  These are defined where \mitot$=0$, which differs from the definition of \citet{DeLucia07}.}
\ifemulate
	\end{deluxetable*}
\else
	\end{deluxetable}
\fi

\ifemulate
	\begin{deluxetable*}{cccc}
\else
	\begin{deluxetable}{cccc}
\fi
\tablecaption{SDSS LF Schechter Function Parameters}
\tablewidth{0pt}
\tablehead{\colhead{${\rm filter}$} & \colhead{$\sigma_{clust}$} & \colhead{$M_\star$} & \colhead{$\alpha$}\\
\colhead{$$} & \colhead{$$} & \colhead{$M-5~{\rm log}~h_{70}$} & \colhead{$$}}
\startdata
g & $ {\rm all~clusters} $ & $ -20.75_{-0.03}^{+0.03} $ & $ -0.96_{-0.01}^{+0.00} $ \\ 
g & $ {\rm \geq 700~km/s} $ & $ -20.54_{-0.03}^{+0.04} $ & $ -0.92_{-0.04}^{+0.03} $ \\ 
g & $ {\rm < 700~km/s} $ & $ -20.88_{-0.06}^{+0.06} $ & $ -1.00_{-0.02}^{+0.02} $ \\ 
r & $ {\rm all~clusters} $ & $ -21.21_{-0.04}^{+0.05} $ & $ -0.78_{-0.02}^{+0.03} $ \\ 
r & $ {\rm \geq 700~km/s} $ & $ -21.38_{-0.06}^{+0.07} $ & $ -0.82_{-0.05}^{+0.04} $ \\ 
r & $ {\rm < 700~km/s} $ & $ -21.20_{-0.05}^{+0.05} $ & $ -0.78_{-0.03}^{+0.02} $ \\ 
i & $ {\rm all~clusters} $ & $ -21.46_{-0.04}^{+0.03} $ & $ -0.75_{-0.01}^{+0.02} $ \\ 
i & $ {\rm \geq 700~km/s} $ & $ -21.07_{-0.10}^{+0.09} $ & $ -0.43_{-0.13}^{+0.14} $ \\ 
i & $ {\rm < 700~km/s} $ & $ -21.53_{-0.03}^{+0.04} $ & $ -0.78_{-0.01}^{+0.02} $ \\ 
\enddata
\label{sdss_lfparam_tab}
\ifemulate
	\end{deluxetable*}
\else
	\end{deluxetable}
\fi

\ifemulate
	\clearpage
	\begin{landscape}
	\begin{deluxetable*}{cccccccc}
\else
	\begin{deluxetable}{cccccccc}
	\setlength{\tabcolsep}{0.05in}
	\rotate
	\tablecolumns{8}
	\tabletypesize{\scriptsize}
\fi
\tablewidth{0pt}
\tablecaption{ED${\rm is}$CS Composite LFs}
\tablehead{\colhead{$$} & \colhead{$0.4<z<0.8$} & \colhead{$0.4<z<0.6$} & \colhead{$0.4<z<0.6$} & \colhead{$0.4<z<0.6$} & \colhead{$0.6<z<0.8$} & \colhead{$0.6<z<0.8$} & \colhead{$0.6<z<0.8$}\\
\colhead{$$} & \colhead{${\rm all~clusters}$} & \colhead{${\rm all~clusters}$} & \colhead{$\geq 600~km/s$} & \colhead{$< 600~km/s$} & \colhead{${\rm all~clusters}$} & \colhead{$\geq 600~km/s$} & \colhead{$< 600~km/s$}\\
\colhead{$M$} & \colhead{$\Phi$} & \colhead{$\Phi$} & \colhead{$\Phi$} & \colhead{$\Phi$} & \colhead{$\Phi$} & \colhead{$\Phi$} & \colhead{$\Phi$}}
\startdata
\cutinhead{$ g $-band}
$ -24.5<M_g\leq -24.0 $& $ < 7.03 $& $ < 4.05 $& $ < 2.83 $& $ < 2.52 $& $ < 5.37 $& $ < 2.99 $& $ < 3.46 $\\ 
$ -24.0<M_g\leq -23.5 $& $ < 7.03 $& $ < 4.05 $& $ < 2.83 $& $ < 2.52 $& $ < 5.37 $& $ < 2.99 $& $ < 3.46 $\\ 
$ -23.5<M_g\leq -23.0 $& $ < 7.03 $& $ < 4.05 $& $ < 2.83 $& $ < 2.52 $& $ < 5.37 $& $ < 2.99 $& $ < 3.46 $\\ 
$ -23.0<M_g\leq -22.5 $& $  2.69\pm 7.19 $& $  0.38\pm 4.08 $& $  0.48\pm 2.89 $& $ < 2.52 $& $  2.12\pm 5.54 $& $  1.84\pm 3.26 $& $  0.50\pm 3.51 $\\ 
$ -22.5<M_g\leq -22.0 $& $ 22.23\pm 9.94 $& $ 10.32\pm 6.04 $& $  6.05\pm 4.48 $& $  4.11\pm 3.65 $& $ 11.35\pm 7.42 $& $  8.19\pm 5.27 $& $  3.56\pm 4.41 $\\ 
$ -22.0<M_g\leq -21.5 $& $ 42.49\pm11.70 $& $ 17.71\pm 7.01 $& $  8.77\pm 4.93 $& $  8.01\pm 4.34 $& $ 23.46\pm 8.79 $& $ 19.22\pm 7.25 $& $  6.08\pm 4.82 $\\ 
$ -21.5<M_g\leq -21.0 $& $ 92.96\pm16.40 $& $ 46.53\pm 9.82 $& $ 28.96\pm 7.32 $& $ 17.51\pm 5.91 $& $ 44.55\pm12.33 $& $ 22.58\pm 7.12 $& $ 19.26\pm 7.87 $\\ 
$ -21.0<M_g\leq -20.5 $& $ 118.55\pm17.77 $& $ 72.31\pm11.43 $& $ 50.47\pm 9.45 $& $ 23.94\pm 6.42 $& $ 45.52\pm12.86 $& $ 24.81\pm 7.68 $& $ 18.72\pm 8.13 $\\ 
$ -20.5<M_g\leq -20.0 $& $ 127.00\pm18.88 $& $ 71.93\pm11.61 $& $ 48.00\pm 9.24 $& $ 25.14\pm 6.70 $& $ 53.58\pm14.01 $& $ 30.55\pm 8.49 $& $ 21.29\pm 8.82 $\\ 
$ -20.0<M_g\leq -19.5 $& $ 114.53\pm17.10 $& $ 79.37\pm12.13 $& $ 39.01\pm 8.61 $& $ 36.09\pm 7.48 $& $ 35.69\pm11.54 $& $ 23.12\pm 7.66 $& $ 12.65\pm 7.06 $\\ 
$ -19.5<M_g\leq -19.0 $& $ 90.43\pm14.04 $& $ 77.84\pm12.23 $& $ 42.72\pm 8.59 $& $ 32.73\pm 7.58 $& $ 14.98\pm 7.20 $& $ 13.02\pm 6.28 $& $  3.47\pm 3.77 $\\ 
$ -19.0<M_g\leq -18.5 $& $ 81.18\pm13.30 $& $ 62.33\pm10.94 $& $ 38.80\pm 8.47 $& $ 23.46\pm 6.44 $& $ 20.01\pm 7.57 $& $ 15.85\pm 5.94 $& $  5.49\pm 4.28 $\\ 
$ -18.5<M_g\leq -18.0 $& $ 56.21\pm11.72 $& $ 49.15\pm10.17 $& $ 31.17\pm 7.81 $& $ 18.16\pm 6.01 $& $  8.65\pm 6.05 $& $  6.66\pm 4.04 $& $  2.48\pm 3.69 $\\ 
$ -18.0<M_g\leq -17.5 $& $ 76.73\pm13.09 $& $ 54.11\pm 9.60 $& $ 44.58\pm 8.53 $& $ 13.84\pm 5.06 $& $ 23.09\pm 8.57 $& $ 16.16\pm 7.10 $& $  7.52\pm 4.68 $\\ 
\cutinhead{$ r $-band}
$ -24.5<M_r\leq -24.0 $& $ < 6.60 $& $ < 4.00 $& $ < 2.94 $& $ < 2.45 $& $ < 4.94 $& $ < 3.06 $& $ < 3.02 $\\ 
$ -24.0<M_r\leq -23.5 $& $  0.53\pm 6.64 $& $ < 4.00 $& $ < 2.94 $& $ < 2.45 $& $  0.48\pm 4.98 $& $  0.63\pm 3.16 $& $ < 3.02 $\\ 
$ -23.5<M_r\leq -23.0 $& $  5.19\pm 7.12 $& $  2.70\pm 4.57 $& $  2.28\pm 3.78 $& $  0.66\pm 2.59 $& $  2.44\pm 5.19 $& $  2.20\pm 3.47 $& $  0.52\pm 3.09 $\\ 
$ -23.0<M_r\leq -22.5 $& $ 29.88\pm10.40 $& $ 12.48\pm 6.26 $& $  7.59\pm 4.85 $& $  4.85\pm 3.72 $& $ 16.49\pm 7.82 $& $  9.76\pm 5.42 $& $  6.25\pm 4.60 $\\ 
$ -22.5<M_r\leq -22.0 $& $ 57.23\pm13.33 $& $ 21.98\pm 7.59 $& $  8.03\pm 4.82 $& $ 11.83\pm 4.97 $& $ 33.23\pm10.24 $& $ 26.48\pm 8.38 $& $  8.99\pm 5.53 $\\ 
$ -22.0<M_r\leq -21.5 $& $ 101.63\pm15.94 $& $ 60.24\pm10.61 $& $ 41.29\pm 8.93 $& $ 20.51\pm 5.93 $& $ 40.30\pm11.25 $& $ 20.49\pm 7.03 $& $ 17.07\pm 6.85 $\\ 
$ -21.5<M_r\leq -21.0 $& $ 109.67\pm16.65 $& $ 64.73\pm10.85 $& $ 47.36\pm 9.09 $& $ 20.18\pm 6.09 $& $ 43.92\pm11.93 $& $ 27.43\pm 8.44 $& $ 15.90\pm 6.96 $\\ 
$ -21.0<M_r\leq -20.5 $& $ 118.39\pm17.46 $& $ 71.33\pm11.50 $& $ 44.50\pm 9.25 $& $ 26.99\pm 6.67 $& $ 48.12\pm12.62 $& $ 24.58\pm 7.26 $& $ 20.33\pm 7.86 $\\ 
$ -20.5<M_r\leq -20.0 $& $ 101.41\pm15.16 $& $ 79.59\pm12.29 $& $ 38.57\pm 8.79 $& $ 36.96\pm 7.66 $& $ 24.60\pm 9.03 $& $ 17.66\pm 6.85 $& $  7.69\pm 5.10 $\\ 
$ -20.0<M_r\leq -19.5 $& $ 83.02\pm13.36 $& $ 70.37\pm11.54 $& $ 43.40\pm 8.84 $& $ 26.93\pm 6.92 $& $ 16.33\pm 7.32 $& $ 15.44\pm 7.17 $& $  3.13\pm 3.36 $\\ 
$ -19.5<M_r\leq -19.0 $& $ 71.75\pm12.58 $& $ 60.04\pm10.72 $& $ 33.40\pm 8.13 $& $ 25.23\pm 6.47 $& $ 14.26\pm 6.93 $& $ 11.03\pm 5.59 $& $  4.04\pm 3.78 $\\ 
$ -19.0<M_r\leq -18.5 $& $ 50.30\pm10.80 $& $ 43.58\pm 9.34 $& $ 30.64\pm 7.60 $& $ 14.35\pm 5.37 $& $  7.83\pm 5.65 $& $  6.30\pm 4.08 $& $  2.08\pm 3.26 $\\ 
$ -18.5<M_r\leq -18.0 $& ...& $ 54.94\pm10.16 $& $ 37.95\pm 8.13 $& $ 18.51\pm 5.92 $& ...& ...& ...\\ 
$ -18.0<M_r\leq -17.5 $& ...& ...& ...& ...& ...& ...& ...\\ 
\cutinhead{$ i $-band}
$ -24.5<M_i\leq -24.0 $& $ < 6.45 $& $ < 3.90 $& $ < 2.77 $& $ < 2.51 $& $ < 4.84 $& $ < 3.00 $& $ < 3.12 $\\ 
$ -24.0<M_i\leq -23.5 $& $  1.93\pm 6.55 $& $  0.41\pm 3.94 $& $  0.50\pm 2.85 $& $ < 2.51 $& $  1.42\pm 4.94 $& $  1.86\pm 3.27 $& $ < 3.12 $\\ 
$ -23.5<M_i\leq -23.0 $& $ 22.48\pm 9.41 $& $ 10.20\pm 5.87 $& $  6.02\pm 4.41 $& $  4.14\pm 3.65 $& $ 11.82\pm 6.97 $& $  8.18\pm 5.27 $& $  3.89\pm 4.09 $\\ 
$ -23.0<M_i\leq -22.5 $& $ 40.15\pm10.90 $& $ 17.43\pm 6.89 $& $  8.15\pm 4.80 $& $  8.49\pm 4.47 $& $ 21.73\pm 8.01 $& $ 17.50\pm 6.63 $& $  5.84\pm 4.45 $\\ 
$ -22.5<M_i\leq -22.0 $& $ 87.44\pm15.39 $& $ 42.04\pm 9.32 $& $ 23.16\pm 6.73 $& $ 18.15\pm 5.93 $& $ 42.87\pm11.48 $& $ 23.65\pm 7.69 $& $ 17.65\pm 7.16 $\\ 
$ -22.0<M_i\leq -21.5 $& $ 108.61\pm16.03 $& $ 64.97\pm10.42 $& $ 54.26\pm 9.53 $& $ 15.84\pm 5.39 $& $ 41.78\pm11.44 $& $ 23.35\pm 7.12 $& $ 17.12\pm 7.36 $\\ 
$ -21.5<M_i\leq -21.0 $& $ 119.74\pm17.39 $& $ 67.00\pm10.95 $& $ 43.61\pm 8.72 $& $ 24.50\pm 6.54 $& $ 52.02\pm12.84 $& $ 31.93\pm 8.93 $& $ 19.85\pm 7.98 $\\ 
$ -21.0<M_i\leq -20.5 $& $ 97.43\pm15.34 $& $ 68.10\pm11.41 $& $ 31.05\pm 7.82 $& $ 33.68\pm 7.46 $& $ 29.81\pm 9.92 $& $ 17.13\pm 6.08 $& $ 11.93\pm 6.41 $\\ 
$ -20.5<M_i\leq -20.0 $& $ 88.31\pm13.66 $& $ 72.39\pm11.42 $& $ 39.31\pm 8.33 $& $ 31.64\pm 7.23 $& $ 17.87\pm 7.56 $& $ 17.38\pm 7.66 $& $  3.10\pm 3.45 $\\ 
$ -20.0<M_i\leq -19.5 $& $ 78.96\pm12.73 $& $ 68.22\pm11.14 $& $ 39.00\pm 8.16 $& $ 28.52\pm 7.03 $& $ 12.40\pm 6.29 $& $  8.59\pm 4.32 $& $  4.01\pm 3.86 $\\ 
$ -19.5<M_i\leq -19.0 $& $ 51.95\pm11.05 $& $ 42.79\pm 9.16 $& $ 26.64\pm 7.08 $& $ 16.45\pm 5.59 $& $ 10.27\pm 6.18 $& $  8.44\pm 5.24 $& $  2.59\pm 3.40 $\\ 
$ -19.0<M_i\leq -18.5 $& ...& $ 51.44\pm10.03 $& $ 32.30\pm 7.75 $& $ 19.60\pm 6.12 $& ...& ...& $  4.01\pm 3.86 $\\ 
$ -18.5<M_i\leq -18.0 $& ...& ...& ...& ...& ...& ...& ...\\ 
$ -18.0<M_i\leq -17.5 $& ...& ...& ...& ...& ...& ...& ...\\ 
\enddata
\label{ediscs_lfstack_tab}
\tablecomments {Magnitudes are given in units of $M-5~{\rm log}~h_{70}.$  Composite LFs are only given for magnitudes brighter than which all clusters in each redshift and velocity bin are complete.}
\ifemulate
	\end{deluxetable*}
	\clearpage
	\end{landscape}
\else
	\end{deluxetable}
\fi

\ifemulate
	\begin{deluxetable*}{cccccccc}
\else
	\begin{deluxetable}{cccccccc}
\fi
\tablecaption{Schechter Function Parameters for EDisCS Composite LFs}
\tablewidth{0pt}
\tablehead{\colhead{${\rm redshift}$} & \colhead{$\sigma_{clust}$} & \colhead{$M^\star_g$} & \colhead{$\alpha_g$} & \colhead{$M^\star_r$} & \colhead{$\alpha_r$} & \colhead{$M^\star_i$} & \colhead{$\alpha_i$}\\
\colhead{$$} & \colhead{$$} & \colhead{$M-5~{\rm log}~h_{70}$} & \colhead{$$} & \colhead{$M-5~{\rm log}~h_{70}$} & \colhead{$$} & \colhead{$M-5~{\rm log}~h_{70}$} & \colhead{$$}}
\startdata
$ 0.4<z<0.8 $ & $ {\rm all~clusters} $ & $ -20.92^{+0.21}_{-0.15} $ & $ -0.45^{+0.13}_{-0.08} $ & $ -21.51^{+0.23}_{-0.14} $ & $ -0.36^{+0.16}_{-0.08} $ & $ -21.80^{+0.22}_{-0.17} $ & $ -0.34^{+0.16}_{-0.10} $ \\ 
$ 0.4<z<0.6 $ & $ {\rm all~clusters} $ & $ -20.76^{+0.24}_{-0.16} $ & $ -0.54^{+0.13}_{-0.08} $ & $ -21.48^{+0.26}_{-0.14} $ & $ -0.58^{+0.13}_{-0.06} $ & $ -21.83^{+0.28}_{-0.14} $ & $ -0.58^{+0.15}_{-0.06} $ \\ 
$ 0.4<z<0.6 $ & $ {\rm \geq 600~km/s} $ & $ -20.73^{+0.14}_{-0.10} $ & ...\tablenotemark{b} & $ -21.38^{+0.12}_{-0.10} $ & ...\tablenotemark{b} & $ -21.81^{+0.12}_{-0.12} $ & ...\tablenotemark{b} \\ 
$ 0.4<z<0.6 $ & $ {\rm < 600~km/s} $ & $ -20.80^{+0.20}_{-0.14} $ & ...\tablenotemark{b} & $ -21.50^{+0.20}_{-0.14} $ & ...\tablenotemark{b} & $ -21.75^{+0.22}_{-0.16} $ & ...\tablenotemark{b} \\ 
$ 0.6<z<0.8 $ & $ {\rm all~clusters} $ & $ -20.79^{+0.40}_{-0.26} $ & $ -0.02^{+0.41}_{-0.18} $ & $ -21.41^{+0.38}_{-0.20} $ & $ 0.08^{+0.41}_{-0.15} $ & $ -21.64^{+0.38}_{-0.24} $ & $ 0.17^{+0.42}_{-0.20} $ \\ 
$ 0.6<z<0.8 $ & $ {\rm \geq 600~km/s} $ & $ -20.77^{+0.18}_{-0.14} $ & ...\tablenotemark{a} & $ -21.37^{+0.18}_{-0.14} $ & ...\tablenotemark{a} & $ -21.67^{+0.18}_{-0.14} $ & ...\tablenotemark{a} \\ 
$ 0.6<z<0.8 $ & $ {\rm < 600~km/s} $ & $ -20.81^{+0.26}_{-0.22} $ & ...\tablenotemark{a} & $ -21.46^{+0.22}_{-0.20} $ & ...\tablenotemark{a} & $ -21.66^{+0.22}_{-0.18} $ & ...\tablenotemark{a} \\ 
\enddata
\label{ediscs_lfparam_tab}
\tablenotetext{a}{Uses distribution of $\alpha$ determined from fits to all clusters at $0.4<z<0.6$.}
\tablenotetext{b}{Uses distribution of $\alpha$ determined from fits to all clusters at $0.6<z<0.8$.}
\ifemulate
	\end{deluxetable*}
\else
	\end{deluxetable}
\fi

\clearpage
\appendix

\section{Rest-frame optical luminosity functions of ED${\rm is}$CS clusters}
\ifemulate
	\begin{landscape}
	\begin{deluxetable*}{ccccccccccc}
\else
	\begin{deluxetable}{ccccccccccc}
	\setlength{\tabcolsep}{0.05in}
	\rotate
	\tabletypesize{\scriptsize}
\fi
\tablewidth{0pt}
\tablecaption{Rest-frame $ g $-band LFs for ED${\rm is}$CS clusters at $ 0.4 < z < 0.6 $}
\tablehead{\colhead{$M_g$} & \colhead{$\Phi_{\rm cl1018.8-1211}$} & \colhead{$\Phi_{\rm cl1037.9-1243}$} & \colhead{$\Phi_{\rm cl1059.2-1253}$} & \colhead{$\Phi_{\rm cl1138.2-1133}$} & \colhead{$\Phi_{\rm cl1202.7-1224}$} & \colhead{$\Phi_{\rm cl1232.5-1250}$} & \colhead{$\Phi_{\rm cl1301.7-1139}$} & \colhead{$\Phi_{\rm cl1353.0-1137}$} & \colhead{$\Phi_{\rm cl1411.1-1148}$} & \colhead{$\Phi_{\rm cl1420.3-1236}$}\\
\colhead{$M-5~{\rm log}~h_{70}$} & \colhead{$$} & \colhead{$$} & \colhead{$$} & \colhead{$$} & \colhead{$$} & \colhead{$$} & \colhead{$$} & \colhead{$$} & \colhead{$$} & \colhead{$$}}
\startdata
$ -24.5<M_g\leq -24.0 $& $ < 2.00$& $ < 2.00$& $ < 2.00$& $ < 2.00$& $ < 2.00$& $ < 2.00$& $ < 2.00$& $ < 2.00$& $ < 2.00$& $ < 2.00$ \\ 
$ -24.0<M_g\leq -23.5 $& $ < 2.00$& $ < 2.00$& $ < 2.00$& $ < 2.00$& $ < 2.00$& $ < 2.00$& $ < 2.00$& $ < 2.00$& $ < 2.00$& $ < 2.00$ \\ 
$ -23.5<M_g\leq -23.0 $& $ < 2.00$& $ < 2.00$& $ < 2.00$& $ < 2.00$& $ < 2.00$& $ < 2.00$& $ < 2.00$& $ < 2.00$& $ < 2.00$& $ < 2.00$ \\ 
$ -23.0<M_g\leq -22.5 $& $ < 2.00$& $ < 2.00$& $ < 2.00$& $ < 2.00$& $ < 2.00$& $ 1^{+2.41}_{-0.83} $& $ < 2.00$& $ < 2.00$& $ < 2.00$& $ < 2.00$ \\ 
$ -22.5<M_g\leq -22.0 $& $ < 2.00$& $ < 2.00$& $ 1^{+2.41}_{-0.83} $& $ < 2.00$& $ 1^{+2.41}_{-0.83} $& $ 3^{+3.00}_{-1.63} $& $ < 2.00$& $ 2^{+2.73}_{-1.29} $& $ 1^{+2.41}_{-0.83} $& $ 2^{+2.73}_{-1.29} $ \\ 
$ -22.0<M_g\leq -21.5 $& $ 1^{+2.41}_{-0.83} $& $ 1^{+2.41}_{-0.83} $& $ 5^{+3.45}_{-2.15} $& $ 1^{+2.41}_{-0.83} $& $ 1^{+2.41}_{-0.83} $& $ 4^{+3.24}_{-1.91} $& $ 1^{+2.41}_{-0.83} $& $ 1^{+2.41}_{-0.83} $& $ 2^{+2.73}_{-1.29} $& $ 1^{+2.41}_{-0.83} $ \\ 
$ -21.5<M_g\leq -21.0 $& $ 1^{+2.41}_{-0.83} $& $ 5^{+3.45}_{-2.15} $& $ 7^{+3.83}_{-2.58} $& $ 3^{+3.00}_{-1.63} $& $ < 2.00$& $ 13^{+4.74}_{-3.55} $& $ 7^{+3.83}_{-2.58} $& $ 4^{+3.24}_{-1.91} $& $ 1^{+2.41}_{-0.83} $& $ 3^{+3.00}_{-1.63} $ \\ 
$ -21.0<M_g\leq -20.5 $& $ 3^{+3.00}_{-1.63} $& $ 3^{+3.00}_{-1.63} $& $ 10^{+4.32}_{-3.10} $& $ 10^{+4.32}_{-3.10} $& $ 1^{+2.41}_{-0.83} $& $ 16^{+5.12}_{-3.95} $& $ 9^{+4.16}_{-2.94} $& $ 6^{+3.65}_{-2.37} $& $ 6^{+3.65}_{-2.37} $& $ 7^{+3.83}_{-2.58} $ \\ 
$ -20.5<M_g\leq -20.0 $& $ 4^{+3.24}_{-1.91} $& $ 4^{+3.24}_{-1.91} $& $ 9^{+4.16}_{-2.94} $& $ 9^{+4.16}_{-2.94} $& $ 5^{+3.45}_{-2.15} $& $ 12^{+4.61}_{-3.41} $& $ 3^{+3.00}_{-1.63} $& $ 4^{+3.24}_{-1.91} $& $ 15^{+5.00}_{-3.83} $& $ 3^{+3.00}_{-1.63} $ \\ 
$ -20.0<M_g\leq -19.5 $& $ 14^{+4.87}_{-3.69} $& $ 2^{+2.73}_{-1.29} $& $ 9^{+4.16}_{-2.94} $& $ 6^{+3.65}_{-2.37} $& $ 6^{+3.65}_{-2.37} $& $ 12^{+4.61}_{-3.41} $& $ 6^{+3.65}_{-2.37} $& $ 6^{+3.65}_{-2.37} $& $ 6^{+3.65}_{-2.37} $& $ 5^{+3.45}_{-2.15} $ \\ 
$ -19.5<M_g\leq -19.0 $& $ 5^{+3.45}_{-2.15} $& $ 5^{+3.45}_{-2.15} $& $ 6^{+3.65}_{-2.37} $& $ 8^{+4.00}_{-2.76} $& $ 8^{+4.00}_{-2.76} $& $ 19^{+5.47}_{-4.32} $& $ 3^{+3.00}_{-1.63} $& $ 4^{+3.24}_{-1.91} $& $ 9^{+4.16}_{-2.94} $& $ 6^{+3.65}_{-2.37} $ \\ 
$ -19.0<M_g\leq -18.5 $& $ 7^{+3.83}_{-2.58} $& $ 4^{+3.24}_{-1.91} $& $ 8^{+4.00}_{-2.76} $& $ 6^{+3.65}_{-2.37} $& $ 2^{+2.73}_{-1.29} $& $ 12^{+4.61}_{-3.41} $& $ 6^{+3.65}_{-2.37} $& $ 4^{+3.24}_{-1.91} $& $ 8^{+4.00}_{-2.76} $& $ 3^{+3.00}_{-1.63} $ \\ 
$ -18.5<M_g\leq -18.0 $& $ 5^{+3.45}_{-2.15} $& $ 3^{+3.00}_{-1.63} $& $ 3^{+3.00}_{-1.63} $& $ 4^{+3.24}_{-1.91} $& $ 1^{+2.41}_{-0.83} $& $ 12^{+4.61}_{-3.41} $& $ 7^{+3.83}_{-2.58} $& $ 6^{+3.65}_{-2.37} $& $ 1^{+2.41}_{-0.83} $& $ 3^{+3.00}_{-1.63} $ \\ 
$ -18.0<M_g\leq -17.5 $& $ 4^{+3.24}_{-1.91} $& $ < 2.00$& $ 5^{+3.45}_{-2.15} $& $ 5^{+3.45}_{-2.15} $& $ 2^{+2.73}_{-1.29} $& $ 24^{+6.00}_{-4.86} $& $ 11^{+4.46}_{-3.26} $& $ 4^{+3.24}_{-1.91} $& $ 3^{+3.00}_{-1.63} $& $ 4^{+3.24}_{-1.91} $ \\ 
$ -17.5<M_g\leq -17.0 $& $ 1^{+2.41}_{-0.83} $& $ 7^{+3.83}_{-2.58} $& $ 2^{+2.73}_{-1.29} $& $ 2^{+2.73}_{-1.29} $& $ 1^{+2.41}_{-0.83} $& ...& $ 10^{+4.32}_{-3.10} $& ...& ...& $ 1^{+2.41}_{-0.83} $ \\ 
$ -17.0<M_g\leq -16.5 $& ...& ...& ...& $ 2^{+2.73}_{-1.29} $& ...& ...& ...& ...& ...& ... \\ 
\enddata
\tablecomments{$\Phi$ gives the number of galaxies per magnitude bin for each cluster and does not depend on $ h_{70}$.}
\ifemulate
	\end{deluxetable*}
	\clearpage
	\end{landscape}
\else
	\end{deluxetable}
\fi

\ifemulate
	\clearpage
	\begin{landscape}
	\begin{deluxetable*}{ccccccc}
\else
	\begin{deluxetable}{ccccccc}
	\setlength{\tabcolsep}{0.05in}
	\rotate
	\tabletypesize{\scriptsize}
\fi
\tablewidth{0pt}
\tablecaption{Rest-frame $ g $-band LFs for ED${\rm is}$CS clusters at $ 0.6 < z < 0.8 $}
\tablehead{\colhead{$M_g$} & \colhead{$\Phi_{\rm cl1040.7-1155}$} & \colhead{$\Phi_{\rm cl1054.4-1146}$} & \colhead{$\Phi_{\rm cl1054.7-1245}$} & \colhead{$\Phi_{\rm cl1216.8-1201}$} & \colhead{$\Phi_{\rm cl1227.9-1138}$} & \colhead{$\Phi_{\rm cl1354.2-1230}$}\\
\colhead{$M-5~{\rm log}~h_{70}$} & \colhead{$$} & \colhead{$$} & \colhead{$$} & \colhead{$$} & \colhead{$$} & \colhead{$$}}
\startdata
$ -24.5<M_g\leq -24.0 $& $ < 2.00$& $ < 2.00$& $ < 2.00$& $ < 2.00$& $ < 2.00$& $ < 2.00$ \\ 
$ -24.0<M_g\leq -23.5 $& $ < 2.00$& $ < 2.00$& $ < 2.00$& $ < 2.00$& $ < 2.00$& $ < 2.00$ \\ 
$ -23.5<M_g\leq -23.0 $& $ < 2.00$& $ < 2.00$& $ < 2.00$& $ < 2.00$& $ < 2.00$& $ < 2.00$ \\ 
$ -23.0<M_g\leq -22.5 $& $ < 2.00$& $ < 2.00$& $ 1^{+2.41}_{-0.83} $& $ 3^{+3.00}_{-1.63} $& $ < 2.00$& $ < 2.00$ \\ 
$ -22.5<M_g\leq -22.0 $& $ 2^{+2.73}_{-1.29} $& $ 3^{+3.00}_{-1.63} $& $ 1^{+2.41}_{-0.83} $& $ 4^{+3.24}_{-1.91} $& $ < 2.00$& $ 1^{+2.41}_{-0.83} $ \\ 
$ -22.0<M_g\leq -21.5 $& $ 3^{+3.00}_{-1.63} $& $ 3^{+3.00}_{-1.63} $& $ 3^{+3.00}_{-1.63} $& $ 13^{+4.74}_{-3.55} $& $ < 2.00$& $ 3^{+3.00}_{-1.63} $ \\ 
$ -21.5<M_g\leq -21.0 $& $ 6^{+3.65}_{-2.37} $& $ 11^{+4.46}_{-3.26} $& $ 8^{+4.00}_{-2.76} $& $ 10^{+4.32}_{-3.10} $& $ 2^{+2.73}_{-1.29} $& $ 2^{+2.73}_{-1.29} $ \\ 
$ -21.0<M_g\leq -20.5 $& $ 3^{+3.00}_{-1.63} $& $ 8^{+4.00}_{-2.76} $& $ 10^{+4.32}_{-3.10} $& $ 14^{+4.87}_{-3.69} $& $ 3^{+3.00}_{-1.63} $& $ 3^{+3.00}_{-1.63} $ \\ 
$ -20.5<M_g\leq -20.0 $& $ 3^{+3.00}_{-1.63} $& $ 11^{+4.46}_{-3.26} $& $ 9^{+4.16}_{-2.94} $& $ 14^{+4.87}_{-3.69} $& $ 4^{+3.24}_{-1.91} $& $ 4^{+3.24}_{-1.91} $ \\ 
$ -20.0<M_g\leq -19.5 $& $ 2^{+2.73}_{-1.29} $& $ 10^{+4.32}_{-3.10} $& $ 7^{+3.83}_{-2.58} $& $ 8^{+4.00}_{-2.76} $& $ 2^{+2.73}_{-1.29} $& $ 3^{+3.00}_{-1.63} $ \\ 
$ -19.5<M_g\leq -19.0 $& $ < 2.00$& $ 2^{+2.73}_{-1.29} $& $ 7^{+3.83}_{-2.58} $& $ 9^{+4.16}_{-2.94} $& $ < 2.00$& $ 2^{+2.73}_{-1.29} $ \\ 
$ -19.0<M_g\leq -18.5 $& $ 1^{+2.41}_{-0.83} $& $ 7^{+3.83}_{-2.58} $& $ 8^{+4.00}_{-2.76} $& $ 10^{+4.32}_{-3.10} $& $ < 2.00$& $ 1^{+2.41}_{-0.83} $ \\ 
$ -18.5<M_g\leq -18.0 $& $ < 2.00$& $ 3^{+3.00}_{-1.63} $& $ 5^{+3.45}_{-2.15} $& $ 6^{+3.65}_{-2.37} $& $ < 2.00$& $ < 2.00$ \\ 
$ -18.0<M_g\leq -17.5 $& $ 2^{+2.73}_{-1.29} $& $ 3^{+3.00}_{-1.63} $& $ 9^{+4.16}_{-2.94} $& $ 8^{+4.00}_{-2.76} $& $ < 2.00$& $ 3^{+3.00}_{-1.63} $ \\ 
$ -17.5<M_g\leq -17.0 $& ...& ...& ...& ...& $ 1^{+2.41}_{-0.83} $& ... \\ 
$ -17.0<M_g\leq -16.5 $& ...& ...& ...& ...& ...& ... \\ 
\enddata
\tablecomments{$\Phi$ gives the number of galaxies per magnitude bin for each cluster and does not depend on $ h_{70}$ .  The LF for CL1227.9-1138 has been computed over $<50\%$ of the full cluster area and so must be re-normalized by the full area.}
\ifemulate
	\end{deluxetable*}
	\clearpage
	\end{landscape}
\else
	\end{deluxetable}
\fi

\ifemulate
	\clearpage
	\begin{landscape}
	\begin{deluxetable*}{ccccccccccc}
\else
	\begin{deluxetable}{ccccccccccc}
	\setlength{\tabcolsep}{0.05in}
	\rotate
	\tabletypesize{\scriptsize}
\fi
\tablewidth{0pt}
\tablecaption{Rest-frame $ r $-band LFs for ED${\rm is}$CS clusters at $ 0.4 < z < 0.6 $}
\tablehead{\colhead{$M_r$} & \colhead{$\Phi_{\rm cl1018.8-1211}$} & \colhead{$\Phi_{\rm cl1037.9-1243}$} & \colhead{$\Phi_{\rm cl1059.2-1253}$} & \colhead{$\Phi_{\rm cl1138.2-1133}$} & \colhead{$\Phi_{\rm cl1202.7-1224}$} & \colhead{$\Phi_{\rm cl1232.5-1250}$} & \colhead{$\Phi_{\rm cl1301.7-1139}$} & \colhead{$\Phi_{\rm cl1353.0-1137}$} & \colhead{$\Phi_{\rm cl1411.1-1148}$} & \colhead{$\Phi_{\rm cl1420.3-1236}$}\\
\colhead{$M-5~{\rm log}~h_{70}$} & \colhead{$$} & \colhead{$$} & \colhead{$$} & \colhead{$$} & \colhead{$$} & \colhead{$$} & \colhead{$$} & \colhead{$$} & \colhead{$$} & \colhead{$$}}
\startdata
$ -24.5<M_r\leq -24.0 $& $ < 2.00$& $ < 2.00$& $ < 2.00$& $ < 2.00$& $ < 2.00$& $ < 2.00$& $ < 2.00$& $ < 2.00$& $ < 2.00$& $ < 2.00$ \\ 
$ -24.0<M_r\leq -23.5 $& $ < 2.00$& $ < 2.00$& $ < 2.00$& $ < 2.00$& $ < 2.00$& $ < 2.00$& $ < 2.00$& $ < 2.00$& $ < 2.00$& $ < 2.00$ \\ 
$ -23.5<M_r\leq -23.0 $& $ < 2.00$& $ < 2.00$& $ 1^{+2.41}_{-0.83} $& $ < 2.00$& $ < 2.00$& $ 1^{+2.41}_{-0.83} $& $ < 2.00$& $ 1^{+2.41}_{-0.83} $& $ < 2.00$& $ < 2.00$ \\ 
$ -23.0<M_r\leq -22.5 $& $ < 2.00$& $ < 2.00$& $ 2^{+2.73}_{-1.29} $& $ 1^{+2.41}_{-0.83} $& $ 1^{+2.41}_{-0.83} $& $ 4^{+3.24}_{-1.91} $& $ 1^{+2.41}_{-0.83} $& $ 1^{+2.41}_{-0.83} $& $ 1^{+2.41}_{-0.83} $& $ 2^{+2.73}_{-1.29} $ \\ 
$ -22.5<M_r\leq -22.0 $& $ 2^{+2.73}_{-1.29} $& $ 2^{+2.73}_{-1.29} $& $ 4^{+3.24}_{-1.91} $& $ < 2.00$& $ 1^{+2.41}_{-0.83} $& $ 5^{+3.45}_{-2.15} $& $ 1^{+2.41}_{-0.83} $& $ 1^{+2.41}_{-0.83} $& $ 2^{+2.73}_{-1.29} $& $ 3^{+3.00}_{-1.63} $ \\ 
$ -22.0<M_r\leq -21.5 $& $ 3^{+3.00}_{-1.63} $& $ 4^{+3.24}_{-1.91} $& $ 10^{+4.32}_{-3.10} $& $ 3^{+3.00}_{-1.63} $& $ < 2.00$& $ 13^{+4.74}_{-3.55} $& $ 8^{+4.00}_{-2.76} $& $ 7^{+3.83}_{-2.58} $& $ 5^{+3.45}_{-2.15} $& $ 4^{+3.24}_{-1.91} $ \\ 
$ -21.5<M_r\leq -21.0 $& $ 2^{+2.73}_{-1.29} $& $ 4^{+3.24}_{-1.91} $& $ 7^{+3.83}_{-2.58} $& $ 13^{+4.74}_{-3.55} $& $ 2^{+2.73}_{-1.29} $& $ 17^{+5.24}_{-4.08} $& $ 9^{+4.16}_{-2.94} $& $ 2^{+2.73}_{-1.29} $& $ 5^{+3.45}_{-2.15} $& $ 4^{+3.24}_{-1.91} $ \\ 
$ -21.0<M_r\leq -20.5 $& $ 7^{+3.83}_{-2.58} $& $ 3^{+3.00}_{-1.63} $& $ 10^{+4.32}_{-3.10} $& $ 8^{+4.00}_{-2.76} $& $ 5^{+3.45}_{-2.15} $& $ 10^{+4.32}_{-3.10} $& $ 2^{+2.73}_{-1.29} $& $ 5^{+3.45}_{-2.15} $& $ 13^{+4.74}_{-3.55} $& $ 3^{+3.00}_{-1.63} $ \\ 
$ -20.5<M_r\leq -20.0 $& $ 11^{+4.46}_{-3.26} $& $ 3^{+3.00}_{-1.63} $& $ 7^{+3.83}_{-2.58} $& $ 5^{+3.45}_{-2.15} $& $ 6^{+3.65}_{-2.37} $& $ 12^{+4.61}_{-3.41} $& $ 7^{+3.83}_{-2.58} $& $ 6^{+3.65}_{-2.37} $& $ 5^{+3.45}_{-2.15} $& $ 8^{+4.00}_{-2.76} $ \\ 
$ -20.0<M_r\leq -19.5 $& $ 3^{+3.00}_{-1.63} $& $ 4^{+3.24}_{-1.91} $& $ 6^{+3.65}_{-2.37} $& $ 8^{+4.00}_{-2.76} $& $ 8^{+4.00}_{-2.76} $& $ 19^{+5.47}_{-4.32} $& $ 2^{+2.73}_{-1.29} $& $ 4^{+3.24}_{-1.91} $& $ 10^{+4.32}_{-3.10} $& $ 4^{+3.24}_{-1.91} $ \\ 
$ -19.5<M_r\leq -19.0 $& $ 9^{+4.16}_{-2.94} $& $ 5^{+3.45}_{-2.15} $& $ 8^{+4.00}_{-2.76} $& $ 5^{+3.45}_{-2.15} $& $ 2^{+2.73}_{-1.29} $& $ 11^{+4.46}_{-3.26} $& $ 4^{+3.24}_{-1.91} $& $ 4^{+3.24}_{-1.91} $& $ 6^{+3.65}_{-2.37} $& $ 2^{+2.73}_{-1.29} $ \\ 
$ -19.0<M_r\leq -18.5 $& $ 4^{+3.24}_{-1.91} $& $ 2^{+2.73}_{-1.29} $& $ 3^{+3.00}_{-1.63} $& $ 4^{+3.24}_{-1.91} $& $ < 2.00$& $ 17^{+5.24}_{-4.08} $& $ 7^{+3.83}_{-2.58} $& $ 3^{+3.00}_{-1.63} $& $ 2^{+2.73}_{-1.29} $& $ 3^{+3.00}_{-1.63} $ \\ 
$ -18.5<M_r\leq -18.0 $& $ 3^{+3.00}_{-1.63} $& $ 4^{+3.24}_{-1.91} $& $ 5^{+3.45}_{-2.15} $& $ 3^{+3.00}_{-1.63} $& $ 2^{+2.73}_{-1.29} $& $ 19^{+5.47}_{-4.32} $& $ 12^{+4.61}_{-3.41} $& $ 3^{+3.00}_{-1.63} $& $ 2^{+2.73}_{-1.29} $& $ 4^{+3.24}_{-1.91} $ \\ 
$ -18.0<M_r\leq -17.5 $& $ < 2.00$& $ 2^{+2.73}_{-1.29} $& $ 2^{+2.73}_{-1.29} $& $ 5^{+3.45}_{-2.15} $& $ < 2.00$& ...& $ 10^{+4.32}_{-3.10} $& ...& $ 1^{+2.41}_{-0.83} $& $ 1^{+2.41}_{-0.83} $ \\ 
$ -17.5<M_r\leq -17.0 $& ...& ...& ...& $ 2^{+2.73}_{-1.29} $& $ 1^{+2.41}_{-0.83} $& ...& ...& ...& ...& ... \\ 
$ -17.0<M_r\leq -16.5 $& ...& ...& ...& ...& ...& ...& ...& ...& ...& ... \\ 
\enddata
\tablecomments{$\Phi$ gives the number of galaxies per magnitude bin for each cluster and does not depend on $ h_{70}$.}
\ifemulate
	\end{deluxetable*}
	\clearpage
	\end{landscape}
\else
	\end{deluxetable}
\fi

\ifemulate
	\clearpage
	\begin{landscape}
	\begin{deluxetable*}{ccccccc}
\else
	\begin{deluxetable}{ccccccc}
	\setlength{\tabcolsep}{0.05in}
	\rotate
	\tabletypesize{\scriptsize}
\fi
\tablewidth{0pt}
\tablecaption{Rest-frame $ r $-band LFs for ED${\rm is}$CS clusters at $ 0.6 < z < 0.8 $}
\tablehead{\colhead{$M_r$} & \colhead{$\Phi_{\rm cl1040.7-1155}$} & \colhead{$\Phi_{\rm cl1054.4-1146}$} & \colhead{$\Phi_{\rm cl1054.7-1245}$} & \colhead{$\Phi_{\rm cl1216.8-1201}$} & \colhead{$\Phi_{\rm cl1227.9-1138}$} & \colhead{$\Phi_{\rm cl1354.2-1230}$}\\
\colhead{$M-5~{\rm log}~h_{70}$} & \colhead{$$} & \colhead{$$} & \colhead{$$} & \colhead{$$} & \colhead{$$} & \colhead{$$}}
\startdata
$ -24.5<M_r\leq -24.0 $& $ < 2.00$& $ < 2.00$& $ < 2.00$& $ < 2.00$& $ < 2.00$& $ < 2.00$ \\ 
$ -24.0<M_r\leq -23.5 $& $ < 2.00$& $ < 2.00$& $ < 2.00$& $ 1^{+2.41}_{-0.83} $& $ < 2.00$& $ < 2.00$ \\ 
$ -23.5<M_r\leq -23.0 $& $ < 2.00$& $ 1^{+2.41}_{-0.83} $& $ 1^{+2.41}_{-0.83} $& $ 2^{+2.73}_{-1.29} $& $ < 2.00$& $ < 2.00$ \\ 
$ -23.0<M_r\leq -22.5 $& $ 4^{+3.24}_{-1.91} $& $ 2^{+2.73}_{-1.29} $& $ 1^{+2.41}_{-0.83} $& $ 8^{+4.00}_{-2.76} $& $ < 2.00$& $ 1^{+2.41}_{-0.83} $ \\ 
$ -22.5<M_r\leq -22.0 $& $ 3^{+3.00}_{-1.63} $& $ 8^{+4.00}_{-2.76} $& $ 4^{+3.24}_{-1.91} $& $ 12^{+4.61}_{-3.41} $& $ 1^{+2.41}_{-0.83} $& $ 4^{+3.24}_{-1.91} $ \\ 
$ -22.0<M_r\leq -21.5 $& $ 5^{+3.45}_{-2.15} $& $ 9^{+4.16}_{-2.94} $& $ 9^{+4.16}_{-2.94} $& $ 10^{+4.32}_{-3.10} $& $ 2^{+2.73}_{-1.29} $& $ 2^{+2.73}_{-1.29} $ \\ 
$ -21.5<M_r\leq -21.0 $& $ 2^{+2.73}_{-1.29} $& $ 9^{+4.16}_{-2.94} $& $ 10^{+4.32}_{-3.10} $& $ 12^{+4.61}_{-3.41} $& $ 3^{+3.00}_{-1.63} $& $ 4^{+3.24}_{-1.91} $ \\ 
$ -21.0<M_r\leq -20.5 $& $ 4^{+3.24}_{-1.91} $& $ 10^{+4.32}_{-3.10} $& $ 8^{+4.00}_{-2.76} $& $ 15^{+5.00}_{-3.83} $& $ 4^{+3.24}_{-1.91} $& $ 2^{+2.73}_{-1.29} $ \\ 
$ -20.5<M_r\leq -20.0 $& $ 1^{+2.41}_{-0.83} $& $ 8^{+4.00}_{-2.76} $& $ 7^{+3.83}_{-2.58} $& $ 7^{+3.83}_{-2.58} $& $ 1^{+2.41}_{-0.83} $& $ 2^{+2.73}_{-1.29} $ \\ 
$ -20.0<M_r\leq -19.5 $& $ < 2.00$& $ 2^{+2.73}_{-1.29} $& $ 6^{+3.65}_{-2.37} $& $ 8^{+4.00}_{-2.76} $& $ < 2.00$& $ 3^{+3.00}_{-1.63} $ \\ 
$ -19.5<M_r\leq -19.0 $& $ 1^{+2.41}_{-0.83} $& $ 4^{+3.24}_{-1.91} $& $ 5^{+3.45}_{-2.15} $& $ 7^{+3.83}_{-2.58} $& $ < 2.00$& $ 1^{+2.41}_{-0.83} $ \\ 
$ -19.0<M_r\leq -18.5 $& $ < 2.00$& $ 4^{+3.24}_{-1.91} $& $ 4^{+3.24}_{-1.91} $& $ 4^{+3.24}_{-1.91} $& $ < 2.00$& $ < 2.00$ \\ 
$ -18.5<M_r\leq -18.0 $& $ 1^{+2.41}_{-0.83} $& $ < 2.00$& ...& ...& $ < 2.00$& ... \\ 
$ -18.0<M_r\leq -17.5 $& ...& ...& ...& ...& ...& ... \\ 
$ -17.5<M_r\leq -17.0 $& ...& ...& ...& ...& ...& ... \\ 
$ -17.0<M_r\leq -16.5 $& ...& ...& ...& ...& ...& ... \\ 
\enddata
\tablecomments{$\Phi$ gives the number of galaxies per magnitude bin for each cluster and does not depend on $ h_{70}$ .  The LF for CL1227.9-1138 has been computed over $<50\%$ of the full cluster area and so must be re-normalized by the full area.}
\ifemulate
	\end{deluxetable*}
	\clearpage
	\end{landscape}
\else
	\end{deluxetable}
\fi

\ifemulate
	\clearpage
	\begin{landscape}
	\begin{deluxetable*}{ccccccccccc}
\else
	\begin{deluxetable}{ccccccccccc}
	\setlength{\tabcolsep}{0.05in}
	\rotate
	\tabletypesize{\scriptsize}
\fi
\tablewidth{0pt}
\tablecaption{Rest-frame $ i $-band LFs for ED${\rm is}$CS clusters at $ 0.4 < z < 0.6 $}
\tablehead{\colhead{$M_i$} & \colhead{$\Phi_{\rm cl1018.8-1211}$} & \colhead{$\Phi_{\rm cl1037.9-1243}$} & \colhead{$\Phi_{\rm cl1059.2-1253}$} & \colhead{$\Phi_{\rm cl1138.2-1133}$} & \colhead{$\Phi_{\rm cl1202.7-1224}$} & \colhead{$\Phi_{\rm cl1232.5-1250}$} & \colhead{$\Phi_{\rm cl1301.7-1139}$} & \colhead{$\Phi_{\rm cl1353.0-1137}$} & \colhead{$\Phi_{\rm cl1411.1-1148}$} & \colhead{$\Phi_{\rm cl1420.3-1236}$}\\
\colhead{$M-5~{\rm log}~h_{70}$} & \colhead{$$} & \colhead{$$} & \colhead{$$} & \colhead{$$} & \colhead{$$} & \colhead{$$} & \colhead{$$} & \colhead{$$} & \colhead{$$} & \colhead{$$}}
\startdata
$ -24.5<M_i\leq -24.0 $& $ < 2.00$& $ < 2.00$& $ < 2.00$& $ < 2.00$& $ < 2.00$& $ < 2.00$& $ < 2.00$& $ < 2.00$& $ < 2.00$& $ < 2.00$ \\ 
$ -24.0<M_i\leq -23.5 $& $ < 2.00$& $ < 2.00$& $ < 2.00$& $ < 2.00$& $ < 2.00$& $ 1^{+2.41}_{-0.83} $& $ < 2.00$& $ < 2.00$& $ < 2.00$& $ < 2.00$ \\ 
$ -23.5<M_i\leq -23.0 $& $ < 2.00$& $ < 2.00$& $ 1^{+2.41}_{-0.83} $& $ < 2.00$& $ 1^{+2.41}_{-0.83} $& $ 3^{+3.00}_{-1.63} $& $ < 2.00$& $ 2^{+2.73}_{-1.29} $& $ 1^{+2.41}_{-0.83} $& $ 2^{+2.73}_{-1.29} $ \\ 
$ -23.0<M_i\leq -22.5 $& $ 1^{+2.41}_{-0.83} $& $ 1^{+2.41}_{-0.83} $& $ 4^{+3.24}_{-1.91} $& $ 1^{+2.41}_{-0.83} $& $ 1^{+2.41}_{-0.83} $& $ 3^{+3.00}_{-1.63} $& $ 1^{+2.41}_{-0.83} $& $ 1^{+2.41}_{-0.83} $& $ 2^{+2.73}_{-1.29} $& $ 2^{+2.73}_{-1.29} $ \\ 
$ -22.5<M_i\leq -22.0 $& $ 1^{+2.41}_{-0.83} $& $ 5^{+3.45}_{-2.15} $& $ 8^{+4.00}_{-2.76} $& $ 2^{+2.73}_{-1.29} $& $ < 2.00$& $ 10^{+4.32}_{-3.10} $& $ 5^{+3.45}_{-2.15} $& $ 4^{+3.24}_{-1.91} $& $ 1^{+2.41}_{-0.83} $& $ 3^{+3.00}_{-1.63} $ \\ 
$ -22.0<M_i\leq -21.5 $& $ 3^{+3.00}_{-1.63} $& $ 2^{+2.73}_{-1.29} $& $ 7^{+3.83}_{-2.58} $& $ 12^{+4.61}_{-3.41} $& $ < 2.00$& $ 17^{+5.24}_{-4.08} $& $ 11^{+4.46}_{-3.26} $& $ 5^{+3.45}_{-2.15} $& $ 6^{+3.65}_{-2.37} $& $ 5^{+3.45}_{-2.15} $ \\ 
$ -21.5<M_i\leq -21.0 $& $ 5^{+3.45}_{-2.15} $& $ 3^{+3.00}_{-1.63} $& $ 9^{+4.16}_{-2.94} $& $ 7^{+3.83}_{-2.58} $& $ 4^{+3.24}_{-1.91} $& $ 14^{+4.87}_{-3.69} $& $ 4^{+3.24}_{-1.91} $& $ 4^{+3.24}_{-1.91} $& $ 13^{+4.74}_{-3.55} $& $ 3^{+3.00}_{-1.63} $ \\ 
$ -21.0<M_i\leq -20.5 $& $ 10^{+4.32}_{-3.10} $& $ 4^{+3.24}_{-1.91} $& $ 8^{+4.00}_{-2.76} $& $ 5^{+3.45}_{-2.15} $& $ 7^{+3.83}_{-2.58} $& $ 7^{+3.83}_{-2.58} $& $ 4^{+3.24}_{-1.91} $& $ 5^{+3.45}_{-2.15} $& $ 6^{+3.65}_{-2.37} $& $ 3^{+3.00}_{-1.63} $ \\ 
$ -20.5<M_i\leq -20.0 $& $ 7^{+3.83}_{-2.58} $& $ 2^{+2.73}_{-1.29} $& $ 7^{+3.83}_{-2.58} $& $ 8^{+4.00}_{-2.76} $& $ 7^{+3.83}_{-2.58} $& $ 16^{+5.12}_{-3.95} $& $ 3^{+3.00}_{-1.63} $& $ 5^{+3.45}_{-2.15} $& $ 7^{+3.83}_{-2.58} $& $ 8^{+4.00}_{-2.76} $ \\ 
$ -20.0<M_i\leq -19.5 $& $ 7^{+3.83}_{-2.58} $& $ 5^{+3.45}_{-2.15} $& $ 8^{+4.00}_{-2.76} $& $ 5^{+3.45}_{-2.15} $& $ 4^{+3.24}_{-1.91} $& $ 17^{+5.24}_{-4.08} $& $ 6^{+3.65}_{-2.37} $& $ 3^{+3.00}_{-1.63} $& $ 9^{+4.16}_{-2.94} $& $ 4^{+3.24}_{-1.91} $ \\ 
$ -19.5<M_i\leq -19.0 $& $ 6^{+3.65}_{-2.37} $& $ 2^{+2.73}_{-1.29} $& $ 5^{+3.45}_{-2.15} $& $ 5^{+3.45}_{-2.15} $& $ 1^{+2.41}_{-0.83} $& $ 14^{+4.87}_{-3.69} $& $ 2^{+2.73}_{-1.29} $& $ 4^{+3.24}_{-1.91} $& $ 3^{+3.00}_{-1.63} $& $ 2^{+2.73}_{-1.29} $ \\ 
$ -19.0<M_i\leq -18.5 $& $ 4^{+3.24}_{-1.91} $& $ 3^{+3.00}_{-1.63} $& $ 5^{+3.45}_{-2.15} $& $ 3^{+3.00}_{-1.63} $& $ 2^{+2.73}_{-1.29} $& $ 11^{+4.46}_{-3.26} $& $ 12^{+4.61}_{-3.41} $& $ 4^{+3.24}_{-1.91} $& $ 1^{+2.41}_{-0.83} $& $ 4^{+3.24}_{-1.91} $ \\ 
$ -18.5<M_i\leq -18.0 $& $ < 2.00$& $ < 2.00$& $ 2^{+2.73}_{-1.29} $& $ 4^{+3.24}_{-1.91} $& $ < 2.00$& $ 10^{+4.32}_{-3.10} $& $ 7^{+3.83}_{-2.58} $& ...& $ 3^{+3.00}_{-1.63} $& $ 2^{+2.73}_{-1.29} $ \\ 
$ -18.0<M_i\leq -17.5 $& ...& ...& $ 1^{+2.41}_{-0.83} $& $ 2^{+2.73}_{-1.29} $& $ 1^{+2.41}_{-0.83} $& ...& ...& ...& ...& ... \\ 
$ -17.5<M_i\leq -17.0 $& ...& ...& ...& ...& ...& ...& ...& ...& ...& ... \\ 
$ -17.0<M_i\leq -16.5 $& ...& ...& ...& ...& ...& ...& ...& ...& ...& ... \\ 
\enddata
\tablecomments{$\Phi$ gives the number of galaxies per magnitude bin for each cluster and does not depend on $ h_{70}$.}
\ifemulate
	\end{deluxetable*}
	\clearpage
	\end{landscape}
\else
	\end{deluxetable}
\fi

\ifemulate
	\clearpage
	\begin{landscape}
	\begin{deluxetable*}{ccccccc}
\else
	\begin{deluxetable}{ccccccc}
	\setlength{\tabcolsep}{0.05in}
	\rotate
	\tabletypesize{\scriptsize}
\fi
\tablewidth{0pt}
\tablecaption{Rest-frame $ i $-band LFs for ED${\rm is}$CS clusters at $ 0.6 < z < 0.8 $}
\tablehead{\colhead{$M_i$} & \colhead{$\Phi_{\rm cl1040.7-1155}$} & \colhead{$\Phi_{\rm cl1054.4-1146}$} & \colhead{$\Phi_{\rm cl1054.7-1245}$} & \colhead{$\Phi_{\rm cl1216.8-1201}$} & \colhead{$\Phi_{\rm cl1227.9-1138}$} & \colhead{$\Phi_{\rm cl1354.2-1230}$}\\
\colhead{$M-5~{\rm log}~h_{70}$} & \colhead{$$} & \colhead{$$} & \colhead{$$} & \colhead{$$} & \colhead{$$} & \colhead{$$}}
\startdata
$ -24.5<M_i\leq -24.0 $& $ < 2.00$& $ < 2.00$& $ < 2.00$& $ < 2.00$& $ < 2.00$& $ < 2.00$ \\ 
$ -24.0<M_i\leq -23.5 $& $ < 2.00$& $ < 2.00$& $ < 2.00$& $ 3^{+3.00}_{-1.63} $& $ < 2.00$& $ < 2.00$ \\ 
$ -23.5<M_i\leq -23.0 $& $ 2^{+2.73}_{-1.29} $& $ 3^{+3.00}_{-1.63} $& $ 2^{+2.73}_{-1.29} $& $ 4^{+3.24}_{-1.91} $& $ < 2.00$& $ 1^{+2.41}_{-0.83} $ \\ 
$ -23.0<M_i\leq -22.5 $& $ 3^{+3.00}_{-1.63} $& $ 4^{+3.24}_{-1.91} $& $ 3^{+3.00}_{-1.63} $& $ 13^{+4.74}_{-3.55} $& $ < 2.00$& $ 2^{+2.73}_{-1.29} $ \\ 
$ -22.5<M_i\leq -22.0 $& $ 6^{+3.65}_{-2.37} $& $ 10^{+4.32}_{-3.10} $& $ 7^{+3.83}_{-2.58} $& $ 9^{+4.16}_{-2.94} $& $ 2^{+2.73}_{-1.29} $& $ 3^{+3.00}_{-1.63} $ \\ 
$ -22.0<M_i\leq -21.5 $& $ 3^{+3.00}_{-1.63} $& $ 10^{+4.32}_{-3.10} $& $ 9^{+4.16}_{-2.94} $& $ 13^{+4.74}_{-3.55} $& $ 3^{+3.00}_{-1.63} $& $ 2^{+2.73}_{-1.29} $ \\ 
$ -21.5<M_i\leq -21.0 $& $ 3^{+3.00}_{-1.63} $& $ 9^{+4.16}_{-2.94} $& $ 9^{+4.16}_{-2.94} $& $ 15^{+5.00}_{-3.83} $& $ 4^{+3.24}_{-1.91} $& $ 5^{+3.45}_{-2.15} $ \\ 
$ -21.0<M_i\leq -20.5 $& $ 2^{+2.73}_{-1.29} $& $ 9^{+4.16}_{-2.94} $& $ 7^{+3.83}_{-2.58} $& $ 9^{+4.16}_{-2.94} $& $ 2^{+2.73}_{-1.29} $& $ 1^{+2.41}_{-0.83} $ \\ 
$ -20.5<M_i\leq -20.0 $& $ < 2.00$& $ 2^{+2.73}_{-1.29} $& $ 6^{+3.65}_{-2.37} $& $ 7^{+3.83}_{-2.58} $& $ < 2.00$& $ 4^{+3.24}_{-1.91} $ \\ 
$ -20.0<M_i\leq -19.5 $& $ 1^{+2.41}_{-0.83} $& $ 5^{+3.45}_{-2.15} $& $ 5^{+3.45}_{-2.15} $& $ 6^{+3.65}_{-2.37} $& $ < 2.00$& $ < 2.00$ \\ 
$ -19.5<M_i\leq -19.0 $& $ < 2.00$& $ 2^{+2.73}_{-1.29} $& $ 5^{+3.45}_{-2.15} $& $ 6^{+3.65}_{-2.37} $& $ < 2.00$& $ 1^{+2.41}_{-0.83} $ \\ 
$ -19.0<M_i\leq -18.5 $& $ 1^{+2.41}_{-0.83} $& $ 3^{+3.00}_{-1.63} $& $ 5^{+3.45}_{-2.15} $& ...& $ < 2.00$& ... \\ 
$ -18.5<M_i\leq -18.0 $& ...& ...& ...& ...& $ < 2.00$& ... \\ 
$ -18.0<M_i\leq -17.5 $& ...& ...& ...& ...& ...& ... \\ 
$ -17.5<M_i\leq -17.0 $& ...& ...& ...& ...& ...& ... \\ 
$ -17.0<M_i\leq -16.5 $& ...& ...& ...& ...& ...& ... \\ 
\enddata
\tablecomments{$\Phi$ gives the number of galaxies per magnitude bin for each cluster and does not depend on $ h_{70}$ .  The LF for CL1227.9-1138 has been computed over $<50\%$ of the full cluster area and so must be re-normalized by the full area.}
\ifemulate
	\end{deluxetable*}
	\clearpage
	\end{landscape}
\else
	\end{deluxetable}
\fi

\end{document}